\documentclass{article} 

\usepackage[utf8]{inputenc}         
\usepackage[T1]{fontenc}            
\usepackage{lmodern}                
\usepackage{amsmath, amssymb}       
\usepackage{graphicx}               
\usepackage{caption} 
\usepackage{subcaption}             
\usepackage{booktabs}               
\usepackage[colorlinks=true, linkcolor=blue, citecolor=blue, urlcolor=blue]{hyperref}    
\usepackage{wrapfig}
\usepackage[authoryear]{natbib}                 
\usepackage[margin=1in]{geometry}   
\usepackage{titling}
\usepackage{xcolor}     
\usepackage{svg}
\usepackage{underscore}
\usepackage[cache=false]{minted}
\usepackage{tikz}
\usetikzlibrary{positioning, arrows.meta, shapes.geometric, shapes.misc, calc}

\bibpunct{(}{)}{;}{a}{,}{,}
\definecolor{titlecolor}{rgb}{0.2157, 0.4549, 0.1961}

\usepackage{amsmath,amsfonts}
\usepackage{microtype}
\usepackage[utf8]{inputenc}
\usepackage{graphicx}
\usepackage{subcaption}
\usepackage{booktabs} 
\usepackage{colortbl}
\usepackage{xcolor}

\usepackage{pgfplots,pgfplotstable}
\pgfplotsset{compat=1.18}
\usepgfplotslibrary{polar}
\usepackage{multirow}
\usepackage{adjustbox}  

\usepackage{hyperref}
\usepackage{url}
\usepackage{bbding}
\usepackage{pifont}
\usepackage{wasysym}
\usepackage{amssymb}
\usepackage{mdframed}
\usepackage{longtable}

\usepackage[capitalize,noabbrev]{cleveref}

\hypersetup{
    pdfborderstyle={/S/U/W 1}, 
    linkbordercolor=red,       
    citebordercolor=green,     
    filebordercolor=magenta,   
   urlbordercolor=cyan        
}

\usepackage{enumitem}

\usepackage{interval}
\usepackage{tabularx}
\usepackage{xspace}
\usepackage{mathtools}  
\usepackage{siunitx}    

\usepackage{placeins}
\makeatletter
\renewcommand{\paragraph}{%
  \@startsection{paragraph}{4}%
  {\z@}{0.5em}{-5em}%
  {\normalfont\normalsize\bfseries}%
}

\makeatother
\usepackage{algorithm2e}
\usepackage{amsmath}

\usepackage{multirow}
\usepackage{colortbl}
\usepackage{xcolor}
\usepackage{amssymb}
\definecolor{xl}{HTML}{B0E3E6}
\definecolor{long}{HTML}{FAD9D5}
\definecolor{think}{HTML}{DAE8FC}
\definecolor{chat}{HTML}{D0CEE2}
\definecolor{nvidiaGreen}{RGB}{118,185,0}
\definecolor{qwenPurple}{RGB}{116,81,207}
\definecolor{closedGray}{RGB}{90,90,110}
\usepackage{fancyhdr}               

\newcommand{\yyyymmdd}{\the\year-\the\month-\the\day}

\fancypagestyle{firstpage}{%
  \fancyhf{}%
  \fancyhead[L]{\includegraphics[height=0.7cm]{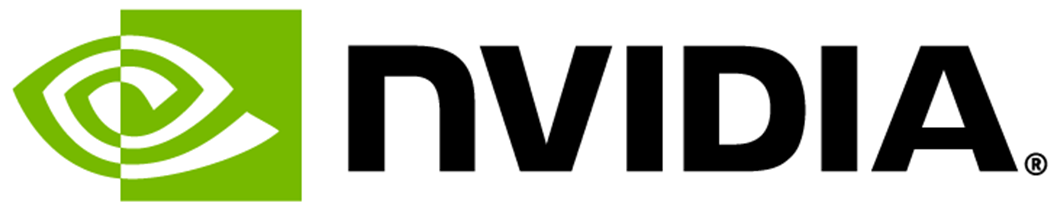}}%
  \fancyhead[R]{\yyyymmdd}%
  \fancyfoot[L]{\footnotesize $^{*}$Project-Leads. Ordering was decided with a coin toss.}%
  \fancyfoot[R]{\footnotesize \textcopyright\ 2026 NVIDIA. All rights reserved.}%
}

\pretitle{\centering\fontsize{19.5}{23}\selectfont\bfseries\color{titlecolor}}

\posttitle{\par\vspace{0.5em}}
\preauthor{\raggedright\large}
\postauthor{\par\vspace{0.5em}}
\predate{\raggedright\normalsize}
\postdate{}

\renewenvironment{abstract}
  {\noindent\raggedright\textbf{Abstract.}}
  {}

\usepackage{titlesec}
\titleformat{\section}[hang]
  {\normalfont\Large\bfseries} 
  {\thesection.}               
  {1em}                       
  {}                          

\title{%
\raisebox{-7pt}{\includegraphics[height=28pt]{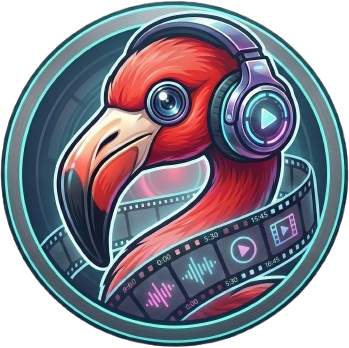}}\hspace{8pt}%
  Nemotron-Labs-Audio-Visual Flamingo: Open Audio-Visual
  Intelligence for Long and Complex Videos%
}
\author{
    \makebox[\textwidth][c]{
        \vspace{20pt}
        \begin{tabular}{c}
            Sreyan Ghosh$^{1,2,*}$ \quad Arushi Goel$^{1,*}$ \quad Kaousheik Jayakumar$^{2}$ \quad Lasha Koroshinadze$^{2}$ \\
            Nishit Anand$^{2}$ \quad Siddharth Gururani$^{1}$ \quad Hanrong Ye$^{1}$ \quad Pritam Biswas$^{1}$ \quad Yuanhang Su$^{1}$ \\
            Ehsan Hosseini-Asl$^{1}$ \quad Sang-gil Lee$^{1}$ \quad Zhifeng Kong$^{1}$ \quad Jaehyeon Kim$^{1}$ \quad Sungwon Kim$^{1}$  \\ Karan Sapra$^{1}$ \quad S Sakshi$^{2}$ \quad Ramani Duraiswami$^{2}$ \quad Dinesh Manocha$^{2}$  \quad Andrew Tao$^{1}$ \\
            Mohammad Shoeybi$^{1}$ \quad Bryan Catanzaro$^{1}$ \quad Ming-Yu Liu$^{1}$ \quad Wei Ping$^{1}$ \\
            [3mm]
            {\large $^{1}$NVIDIA, USA \quad $^{2}$University of Maryland, USA } \\ [3mm]
            {\small \href{https://github.com/NVIDIA/audio-flamingo}{Code} \quad \href{http://huggingface.co/nvidia/audio-visual-flamingo-hf}{Model} \quad \href{https://avflamingo.pages.dev/}{Project Page} \quad \href{https://huggingface.co/datasets/nvidia/AV-Skills}{Dataset} \quad 
            \href{https://huggingface.co/spaces/nvidia/audio-visual-flamingo}{Demo}}
        \end{tabular}
    }
}
\date{} 

\begin{document}

\maketitle
\thispagestyle{firstpage} 

\vspace{-4mm}
\begin{abstract}
 We present \textbf{Nemotron-Labs-Audio-Visual Flamingo} (AV-Flamingo), a fully open state-of-the-art audio-visual large language model (AV-LLM) for joint understanding and reasoning over audio, images, and long-form videos. Unlike prior AV-LLMs that primarily focus on short clips, AV-Flamingo is designed for understanding and reasoning over long and complex real-world (audio-visual) videos. To support this, we make three key contributions: \textbf{(i)} Audio-Visual-Skills, a large-scale collection of real-world videos with $\approx$7M caption and question-answer training instances designed to emphasize temporal, compositional, and cross-modal audio-visual reasoning; \textbf{(ii)} a novel three-stage curriculum that progressively trains the model from short-range perception to long-horizon multi-event reasoning; and \textbf{(iii)} Temporal Audio-Visual Interleaved Chain-of-Thought, a reasoning framework that explicitly grounds intermediate reasoning steps to timestamps in long audio-visual streams, improving temporal alignment and interpretability. Extensive experiments across 15+ AV, omni-modal, audio, and vision benchmarks show that AV-Flamingo outperforms similarly sized open models by clear margins and remains highly competitive with, and in some cases surpasses, much larger open-weight and closed models, particularly on long and complex real-world audio-visual understanding and reasoning tasks. Beyond benchmark performance, AV-Flamingo exhibits strong real-world utility and transfers well to unseen tasks, highlighting its robustness and generalization ability.
\end{abstract}
\vspace{-10pt}

\section{Introduction}
\begin{wrapfigure}[14]{r}{0.5\textwidth}
\vspace{-10pt}
     \centering
     \begin{subfigure}[b]{0.5\textwidth}
         \centering
         \includegraphics[trim= 0.4cm 0.5cm 0cm 1.5cm, width=0.8\textwidth]{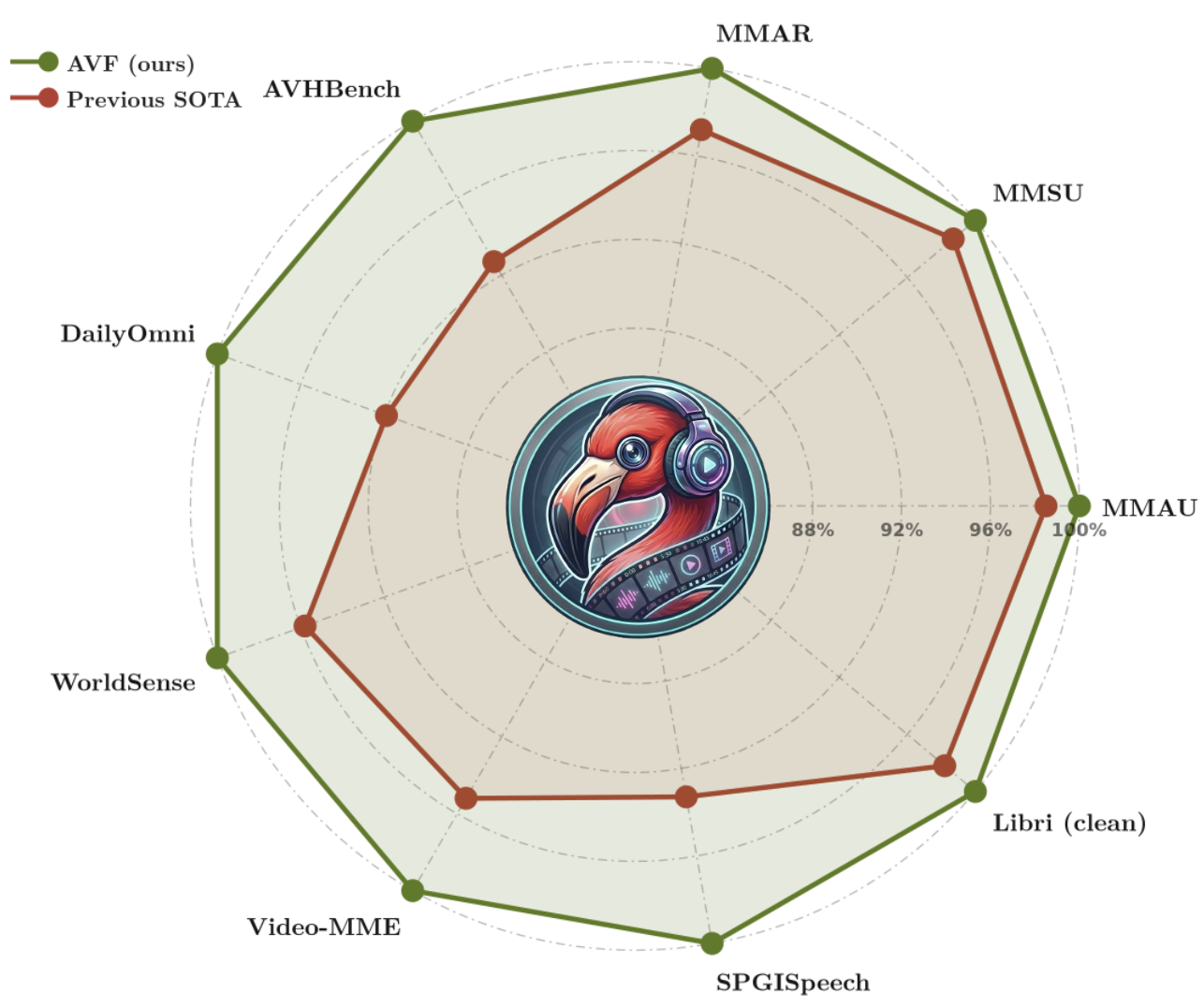}
     \end{subfigure}
        \caption{\small Comparison of AVF with current SOTA model on various benchmarks.}
        \label{fig:avf_radial}
\end{wrapfigure}
\vspace{-10pt}
Video constitutes 82\% of global internet traffic, and users spend over 2.5 hours per day watching online videos on average~\citep{kumar2026videomarketing}. These videos are often long, diverse, and information-rich, combining complex visuals, graphics, speech, sounds, and music in a continuous stream. This closely mirrors how humans perceive the world: as an ongoing audio-visual sequence. Yet, despite the remarkable progress of Large Language Models (LLMs) in many real-world tasks, their ability to understand and reason over videos in a human-like manner remains relatively underexplored. Hereafter, unless stated otherwise, we use \textit{video} to refer to audio-visual sequences. 

Most video understanding models still do not process audio jointly with visuals~\citep{zhang2023video, zhang2025videollama, zhu2025internvl3exploringadvancedtraining}. However, audio--including speech, sounds, and music--is a fundamental modality and has even been described as ``50\% of the movie-going experience''~\citep{lucas1992sound}. 

Recent audio-visual and omni-modal LLMs have shown promise~\citep{xu2025qwen25omnitechnicalreport, qwen35omniblog}, but most focus primarily on short-video understanding ~\citep{chowdhury2024meerkat}. In contrast, much of the video content humans consume is medium- to long-form, such as movies, television, lectures, and documentaries. 

A central challenge for long-form audio-visual understanding is the lack of suitable data: publicly available resources are often audio-only, video-only, or concentrated on short-form videos with captions and QA annotations. 
Moreover, current models still exhibit substantial weaknesses in joint AV perception~\citep{gong2024avodysseybenchmultimodalllms, zhou2026dailyomniaudiovisualreasoningtemporal, goel2026mmou, Chowdhury_2025_ICCV, fu2025video}, which we attribute to both architectural limitations and the scarcity of large-scale, high-quality paired audio-visual data~\citep{seth-etal-2025-egoillusion, selvakumar-etal-2025-multivox}. Recent mechanistic analysis attributes this partly to a visual bias in which the deeper layers privilege vision and suppress latent audio representations~\citep{selvakumar2026avllm}. In practice, many existing omni models are trained separately for audio and visual understanding, and are expected to acquire cross-modal reasoning only implicitly. Finally, the strongest models for audio-visual understanding are either closed-source~\citep{team2024gemini} or only open-weight~\citep{xu2025qwen3, qwen35omniblog}, with their training data, code, and methodology undisclosed. This limits both open progress and understanding of how such systems are built.

\noindent \textbf{Main Contributions.} In this work, we present \textbf{Nemotron-Labs-Audio-Visual Flamingo (AV-Flamingo or AVF)}, a fully open multimodal large language model for joint audio-visual understanding of long and complex real-world videos. AV-Flamingo is a first step toward scaling open audio-visual intelligence beyond academic benchmarks by leveraging internet-scale audio-visual data and targeted post-training for reasoning. To enable this, AV-Flamingo centers on three technical components. First, we introduce \textbf{Audio-Visual-Skills (AV-Skills)}, a large-scale dataset for joint audio-visual understanding and reasoning. Unlike prior omni models that rely heavily on single-modality data and hope to learn cross-modal reasoning implicitly, AV-Skills is explicitly curated for cross-modal learning. It contains videos collected from diverse sources, paired with captions and QA annotations designed to test and train audio-visual reasoning, totaling $\approx$7M caption and QA training instances, including $\approx$4.8M QA pairs, across both short and long videos. Second, we propose \textbf{Temporal Audio-Visual Interleaved Chain-of-Thought (TAVIT)}, a reasoning framework that explicitly grounds intermediate reasoning steps to timestamps in long audio-visual streams. Third, we develop a novel three-stage training curriculum consisting of pre-training, mid-training, and post-training with different data mixtures. AV-Flamingo outperforms similarly sized or larger models across more than 15 audio-visual, omni-modal, audio, and vision benchmarks. In summary, our main contributions are:

\begin{enumerate}
\vspace{-2mm}
\setlength{\parskip}{0pt}

\item We introduce \textbf{Nemotron-Labs-Audio-Visual Flamingo}, a fully open frontier AVLLM for joint audio-visual understanding and reasoning over long and complex real-world videos.

\item We present a scalable recipe for next-generation AV-LLMs, including internet-scale audio-visual data curation, targeted capability expansion, and temporally grounded reasoning for long videos.

\item We open-source the model, training, and inference code, and associated techniques to support future research in open audio-visual large language models.

\item AV-Flamingo outperforms similar-sized AV and omni-modal LLMs on 15+ benchmarks, and remains highly competitive with, and in some cases surpasses, much larger open-weight and closed models. Additionally, it shows substantially stronger robustness on long and complex real-world videos.

\end{enumerate}

\vspace{-2mm}
\section{Related Works}
\vspace{-2mm}

\textbf{Audio-Visual Large Language Models}. Recent progress in multimodal LLMs has produced increasingly capable vision-language models, from foundational work like Flamingo~\citep{alayrac2022flamingo}, BLIP-2~\citep{li2023blip}, and LLaVA~\citep{liu2023visual} to recent systems such as InternVL 2.5~\citep{chen2024expanding}, Qwen3-VL~\citep{bai2025qwen3}, VideoLLaMA 3~\citep{zhang2025videollama}, and Qwen3.5~\citep{qwen35blog}. In parallel, audio-language models such as LTU~\citep{gong2023listen}, SALMONN~\citep{tang2023salmonn}, Qwen2-Audio~\citep{chu2024qwen2}, and Audio Flamingo 3~\citep{goel2025audio} have advanced speech, sound, and music understanding. However, integrating audio and visual streams into a single model remains qualitatively harder, requiring cross-modal alignment, temporal synchronization, and handling of diverse input combinations. Early AV-LLMs such as Video-LLaMA~\citep{zhang2023video} used separate Q-Formers for each modality; Video-LLaMA 2~\citep{cheng2024videollama} improved on this with a spatial-temporal convolution connector; video-SALMONN~\citep{sun2024video, tang2025video} proposed multi-resolution causal Q-Formers; Dolphin~\citep{guo2025alignedbetterlistenbetter} introduced multi-scale adapters; and TriSense~\citep{li2025watch} proposed query-based adaptive modality reweighting. On the omni-modal frontier, proprietary systems like GPT-4o~\citep{hurst2024gpt} and Gemini 1.5 Pro~\citep{team2024gemini} set high benchmarks, while open models have rapidly closed the gap: Qwen2.5-Omni~\citep{xu2025qwen25omnitechnicalreport} introduced the thinker-talker architecture, Qwen3-Omni~\citep{xu2025qwen3} and Qwen3.5-Omni~\citep{qwen35omniblog} scaled this further, Phi-4-Multimodal~\citep{abouelenin2025phi} demonstrated Mixture-of-LoRAs for compact omni-modal models, and OmniVinci~\citep{ye2025omnivinci} introduced shared omni-modal alignment with a 24M-sample synthetic pipeline. However, most of these systems focus on short-to-medium clips rather than long-form, temporally grounded audio-visual reasoning. Furthermore, many remain closed-source or open-weights-only, hindering reproducibility -- a gap AV-Flamingo addresses. 

\textbf{Long-Form Video Understanding and Temporal Reasoning}. Understanding long videos remains one of the central open challenges in multimodal AI, as most current video LLMs are still limited to clips shorter than one minute. Recent progress in long-form video understanding has focused largely on vision-only settings, often without audio~\citep{li2024llama, cheng2024videollama}, and many approaches rely on context reduction to handle long videos, which can hurt performance in real-world scenarios~\citep{zhang2024long,qian2024streaming,li2025improvingllmvideounderstanding}. Recent omni models are a notable exception, with some beginning to support long audio-visual sequences~\cite{qwen35omniblog, deshmukh2026nemotron}. However, benchmarks like Video-MME~\citep{fu2025video} and MMOU~\citep{goel2026mmou} reveal systematic performance degradation as video duration increases -- MMOU shows the best proprietary models achieve only 64.2\% accuracy on long audio-visual reasoning, while open-source models plateau at 46.8\%. Separately, temporal chain-of-thought reasoning has emerged as a promising direction: Temporal CoT~\citep{arnab2025temporal} narrows attention progressively for long-video QA, Open-o3-Video~\citep{meng2025open} integrates spatiotemporal evidence into reasoning chains, UniTime~\citep{liu2024unitime} interleaves timestamp tokens with video tokens, and Audio Flamingo 3~\citep{goel2025audio} introduced on-demand thinking for audio events. In the audio domain, R1-AQA~\citep{li2025reinforcement} and Step-Audio-R1~\citep{tian2025step} have shown CoT gains, though primarily on short audio. However, no prior work jointly grounds intermediate reasoning steps to timestamps in both audio and visual streams, and we propose to bridge this gap with AV-Think.

\textbf{Data Curation}. High-quality data remains a major bottleneck in building multimodal foundation models, including for audio~\citep{panayotov2015librispeech, gemmeke2017audioset, galvez2021peoplesspeechlargescalediverse}, video~\citep{miech2019howto100mlearningtextvideoembedding, bain2022frozentimejointvideo, chen2024panda70mcaptioning70mvideos}, and especially long-form understanding. This challenge is even more pronounced for audio-visual models~\citep{liu2025ola}, where high-quality joint audio-visual supervision is rarely available in the open~\citep{geng2025longvalevisionaudiolanguageeventbenchmarktimeaware}. Most existing resources are either unimodal or, when audio-visual, are largely limited to short videos and recognition-oriented tasks. For example, many current audio-visual LLMs are trained primarily on short-form audio-visual data~\citep{sun2024video, cheng2024videollama, xu2025qwen25omnitechnicalreport, fu2025vitaopensourceinteractiveomni}, while models such as OmniVinci rely heavily on separate unimodal audio and video corpora. To bridge this gap, we introduce Audio-Visual-Skills, a large-scale dataset designed to support temporal, compositional, and cross-modal reasoning across 13 skill categories in long, real-world videos.

\vspace{-4mm}
\vspace{-2mm}
\section{Methodology}
\vspace{-2mm}                                                              

\begin{figure*}[t]
  \centering
  \includegraphics[width=\textwidth]{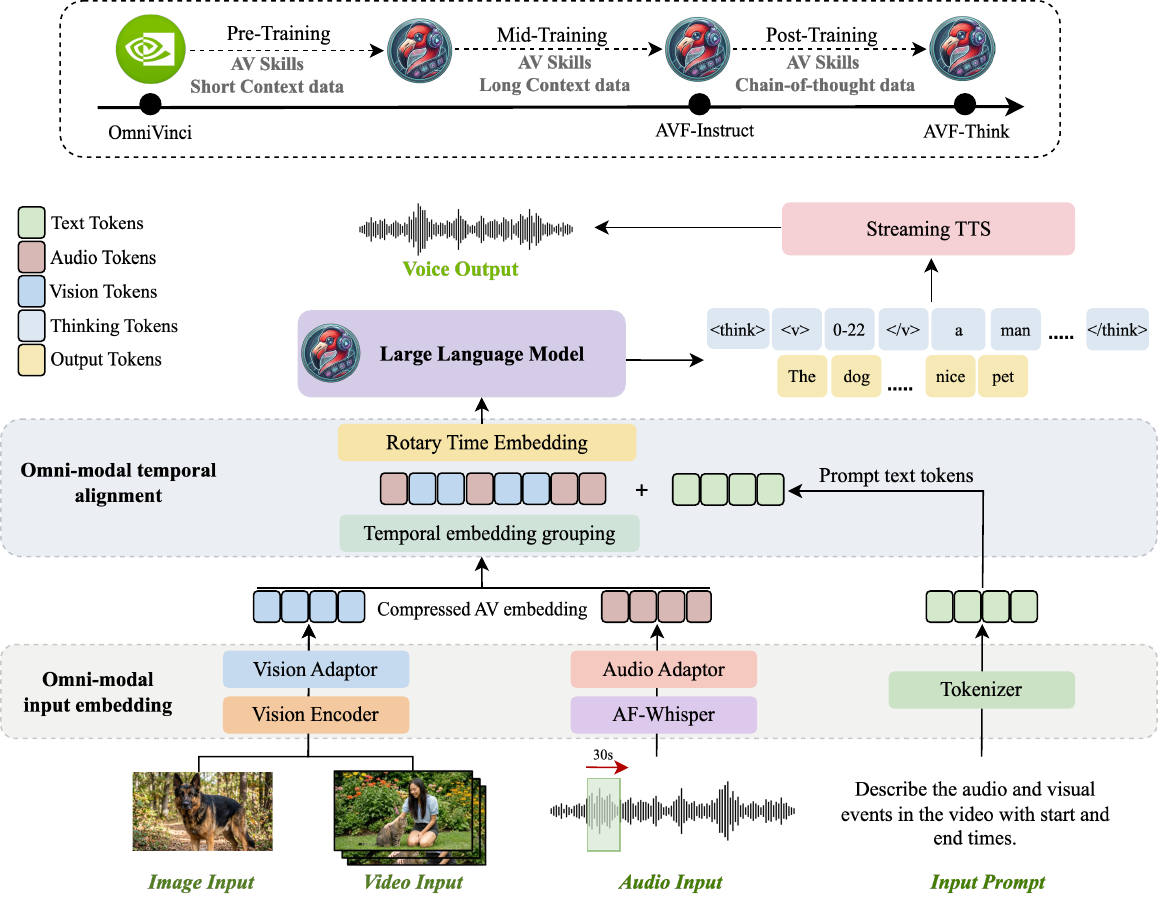}  
  \caption{\small \textbf{AV-Flamingo training and architecture.} \textbf{Top:} Starting from OmniVinci, AV-Flamingo is trained in three stages: pre-training on AV-Skills short-context data, mid-training on AV-Skills long-context data, and post-training on chain-of-thought data, producing AVF-Instruct and AVF-Think. \textbf{Bottom:} AV-Flamingo accepts image, video, audio, and text inputs. Visual and audio streams are encoded through separate encoders and adaptors, compressed into audio-visual embeddings, temporally grouped, and aligned with prompt text tokens using Rotary Time Embedding before being processed by the LLM.}
  \label{fig:arch}
\vspace{-3mm}
\end{figure*}

\subsection{Architecture}
\label{sec:architecture}
\vspace{-2mm}
                      
In this section, we discuss the architecture for Audio-Visual Flamingo as shown in~\Cref{fig:arch}. AVF is an AVLM designed to jointly reason over visual and auditory inputs (both or either) while supporting voice-based interaction through speech synthesis. The architecture is similar to OmniVinci and consists of five key components: i) a SigLip vision encoder that extracts rich spatial features from visual inputs (\textit{e.g.}, images, video frames), ii) AF-Whisper (borrowed from the Audio Flamingo series), an audio encoder with a sliding-window feature extraction mechanism for processing long-form audio, iii) a cross-modal interleaving and temporal alignment module that fuses video and audio representations along the time axis, iv) a text-only LLM that serves as the central reasoning backbone, and v) a streaming text-to-speech (TTS) module that enables real-time voice output. We discuss each component in detail below.

\noindent \textbf{SigLip Vision Encoder.} To encode visual inputs, we employ the SigLip~\citep{zhai2023sigmoidlosslanguageimage} vision encoder, which produces dense feature representations from images or video frames, $V$. Given a raw input, we first extract feature maps from the pre-trained SigLip model. These feature maps are then passed through a ``Spatial-Scale-then-Compress'' Dynamic S2 module, following the design principles of~\citet{liu2025nvilaefficientfrontiervisual} and~\citet{ye2025omnivinci}. The Dynamic S2 module operates by first encoding the input image at multiple spatial scales, then compressing the multi-scale representations into a compact token sequence. This multi-scale strategy enables AVF to encode higher-resolution images and longer videos with more frames while preserving fine-grained spatial and temporal details -- critically, without a proportional increase in the total number of tokens consumed by the LLM. As a result, we obtain a feature representation for a single image (or each frame of a video) denoted as $h_v = f_v(V)$, where $h_v \in \mathbb{R}^{HW \times d_v}$, $H$ and $W$ are the spatial height and width of the compressed visual feature map, and $d_v$ is the hidden dimension of the vision encoder.

\noindent \textbf{AF-Whisper Audio Encoder.}  Following Audio Flamingo 3, Next, and Music Flamingo, we adopt the Whisper-based AF-Whisper audio encoder from AF-Next~\citep{goel2025audio,ghosh2025music}. For each audio input, $A$, we first resample it to 16\,kHz mono to standardize the sampling rate across diverse audio sources. The raw resampled waveform is then transformed into a 128-channel log-mel spectrogram using a window size of 25\,ms and a hop size of 10\,ms, producing a time-frequency representation suitable for the downstream encoder. This mel-spectrogram is processed by the encoder~\citep{goel2025audio, ye2025omnivinci}, which applies a sliding-window mechanism to handle audio of arbitrary length: the spectrogram is segmented into non-overlapping 30-second chunks, each of which is independently encoded and then concatenated along the temporal axis. This yields audio features denoted as $h_a = f_a(A)$, where $h_a \in \mathbb{R}^{N \times d_a}$, $N$ is the total number of 30-second audio chunks, and $d_a$ is the hidden dimension of the Whisper encoder. The sliding-window design allows AVF to process long-form audio (e.g., full-length podcast episodes or movie soundtracks) without truncation or excessive memory consumption.

\noindent \textbf{Vision and Audio Adaptors. }For each modality, we have projection layers, specifically, a 2-layer MLP for audio, denoted $\mathcal{A}(\cdot)$, and a 2-layer MLP for vision, denoted $\mathcal{V}(\cdot)$. These projections map the encoder outputs into the LLM's embedding space: the projected audio embeddings are $a = \mathcal{A}(h_a)$ and the projected image/video embeddings are $v = \mathcal{V}(h_v)$.

\noindent \textbf{Cross-Modal Alignment of Vision and Audio.} A central challenge in multimodal architectures is the fusion of visual and audio tokens before feeding into the LLM. Following OmniVinci~\citep{ye2025omnivinci}, we enforce a cross-modal alignment between the two modalities through temporal interleaving. The projected video and audio token sequences are first divided along the time dimension into multiple synchronized chunks and then interleaved according to their timestamps, so that visual tokens from a given temporal window are placed adjacent to the audio tokens from the same window.  This interleaved arrangement allows the LLM's self-attention mechanism to naturally attend across co-occurring visual and auditory events. Finally, before injecting these interleaved token embeddings into the LLM, we augment them with periodic temporal information via Constrained Rotary Time Embeddings (CRTE)~\citep{goel2024omcat, ye2025omnivinci}. This encodes absolute temporal position using sinusoidal rotary transformations, enabling the model to distinguish the ordering and relative timing of multimodal events even after interleaving.                                                                                                         

\noindent \textbf{Large Language Model (LLM).} The core reasoning component of AVF is a large language model that processes the interleaved multimodal token sequence and generates textual responses. 
Following OmniVinci~\citep{ye2025omnivinci}, we employ Qwen2.5-7B~\citep{qwen2025qwen25technicalreport} as our base LLM, which comprises 7B parameters, 36 hidden layers, and 16 attention heads per layer. The LLM receives the temporally aligned, interleaved vision-audio embeddings as input prefix tokens, followed by any text-based instructions or queries, and autoregressively generates a textual response. To handle long-context training for long videos of upto 15 minutes long as discussed in \Cref{sec.curriculum}, we use hybrid sequence parallelism~\citep{fang2024usp} along with fully-sharded data parallel~\citep{rasley2020deepspeed}. This approach reduces memory usage by splitting the multi-modal sequence dimension across GPUs, using Ulysses~\citep{jacobs2023deepspeed} for intra-node communication and Ring-Attention~\citep{liu2023ring} for inter-node communication. Optionally, to support voice-to-voice interaction, similar to Audio Flamingo 3~\citep{goel2025audio} and OmniVinci~\citep{ye2025omnivinci}, Audio-Visual Flamingo incorporates a streaming TTS module. For more details, we refer our readers to ~\citet{goel2025audio}.

\noindent \textbf{Streaming TTS.} To support voice-to-voice interaction, similar to Audio Flamingo 3~\citep{goel2025audio} and OmniVinci~\citep{ye2025omnivinci}, Audio-Visual Flamingo incorporates a streaming TTS module. The module is implemented as a decoder-only transformer that predicts the next audio token conditioned on incoming subword text tokens from the LLM and previously generated audio tokens. For more details, we refer our readers to ~\citet{goel2025audio}.

\vspace{-2mm}
\subsection{Training Data}
\vspace{-2mm}

Data remains one of the most critical yet least openly discussed components in building open foundational omni-modal and audio-visual models~\citep{ye2025omnivinci}. While recent efforts have improved the availability of large-scale audio-only~\citep{galvez2021peoplesspeechlargescalediverse, ghosh2026audioflamingonextnextgeneration} and image/video-only datasets~\citep{schuhmann2022laion, zhang2024llava}, comparable resources for joint audio-visual learning remain limited, particularly for tasks requiring integrated understanding of sound and vision and cross-modal reasoning. Existing audio-visual QA datasets, often derived from foundational benchmarks centered on recognition tasks such as AVSR, event classification, or AV localization~\citep{Chen20, tian2018audiovisualeventlocalizationunconstrained}, are insufficient for training models with strong expert-level reasoning abilities~\citep{goel2026mmou}.

In AVF, we place particular emphasis on developing reasoning and problem-solving capabilities through large-scale, high-quality audio-visual QA data. Inspired by the Audio Flamingo series, in addition to existing unimodal image/audio/video datasets (largely borrowed from OmniVinci, with some additions, detailed stats in Table~\ref{tab:dataset-details}), we introduce \textbf{Audio-Visual-Skills (AV-Skills)}, a large-scale dataset of real-world videos with diverse durations, sourced from both existing datasets and the open internet, paired with high-quality captions and QA annotations. Unlike prior data mixtures that rely heavily on unimodal supervision, AV-Skills is curated specifically for joint cross-modal audio-visual understanding, with a focus on teaching models complex skills beyond recognition, including reasoning over temporally extended, compositional, and real-world long-form videos.

As a first step in data curation, we identify the key skills required for strong real-world audio-visual understanding. To do so, we evaluate frontier omni-modal and audio-visual models, including the Qwen-Omni series, Gemini, and VideoLLaMA, on challenging benchmarks such as WorldSense and MMOU, and analyze their outputs to identify systematic capability gaps. In particular, we convert benchmark QA pairs into open-ended form, since multiple-choice settings can allow models to exploit option biases or elimination heuristics. Our analysis reveals both gaps in core skills, such as counting and temporal understanding, and distributional weaknesses caused by limited training exposure to diverse video types. For instance, much of the data used in existing open models is dominated by vlog-style talking-head videos, with relatively limited coverage of environmental sounds, background music, and complex real-world audio-visual scenes.

To address these limitations, we curate two large-scale training sets: \textbf{AV-Skills-Short}, consisting of videos of up to 60 seconds, and \textbf{AV-Skills-Long}, consisting of videos between 60 seconds and 15 minutes. These datasets are built from two sources: publicly available datasets and raw videos collected from the open internet, which we subsequently annotate synthetically. For public datasets, we primarily use YouTube-8M~\citep{abu2016youtube}, HD-VILA~\citep{xue2022advancinghighresolutionvideolanguagerepresentation}, InternVid~\citep{wang2024internvidlargescalevideotextdataset}, and VidChapters~\citep{yang2023vidchapters7mvideochaptersscale} for short-form data, and HarmonySet~\citep{zhou2025harmonyset}, LSMDC~\citep{rohrbach2016moviedescription}, MMTrail~\citep{chi2024mmtrailmultimodaltrailervideo}, MovieClips~\citep{bose2022movieclipvisualscenerecognition}, and MiraData~\citep{ju2024miradatalargescalevideodataset} for longer-form data. The remaining data is collected from the open internet with an emphasis on categorical diversity~\citep{goel2026mmou}, spanning domains such as podcasts, city tours, interviews, sports, and more (Fig.~\ref{fig:data_categories}). Our final AV-Skills-Short dataset comprises 100K hours of video and 3.8M training instances, including 1M captions and 2.8M QA pairs (detailed stats in Table~\ref{tab:dataset-details}). The long-video training data comprises approximately 140K hours of video and 3.2M training instances, including 1.2M captions and 2.0M QA pairs. For \textbf{AV-Skills-Short}, we focus on the following skills (examples shown in ~\Cref{sec.dataset_examples}): (i) \textbf{Relation Reasoning} involves understanding the relationships, interactions, or distinctions between entities in a video, including people, objects, roles, and their relative configurations. (ii) \textbf{Emotion Change} focuses on tracking how a person’s or character’s emotional state evolves over time, including the initial emotion, the final emotion, and the transition between them. (iii) \textbf{Temporal Reasoning} captures the chronological structure of a video, including event order, what happens before or after a target event, and how actions unfold over time. (iv) \textbf{Spatial Perception} requires perceiving spatial properties in the visual scene, such as direction, orientation, relative position, alignment, and placement of people or objects. (v) \textbf{Causal Reasoning} aims to infer why an event, action, behavior, or outcome occurs by connecting observed effects to their underlying causes. (vi) \textbf{Hallucination Detection} evaluates whether a question is grounded in information actually present in the video, requiring the model to reject false premises or unsupported claims. (vii) \textbf{Audio Counting} tests the ability to count the number of times a particular sound, spoken word, instrument, or other audio event occurs in the video. (viii) \textbf{Video Counting} measures the ability to count the number of times a particular object, entity, action, or visual event appears in the video.

\begin{figure*}[!t]
  \centering
  \includegraphics[width=\textwidth]{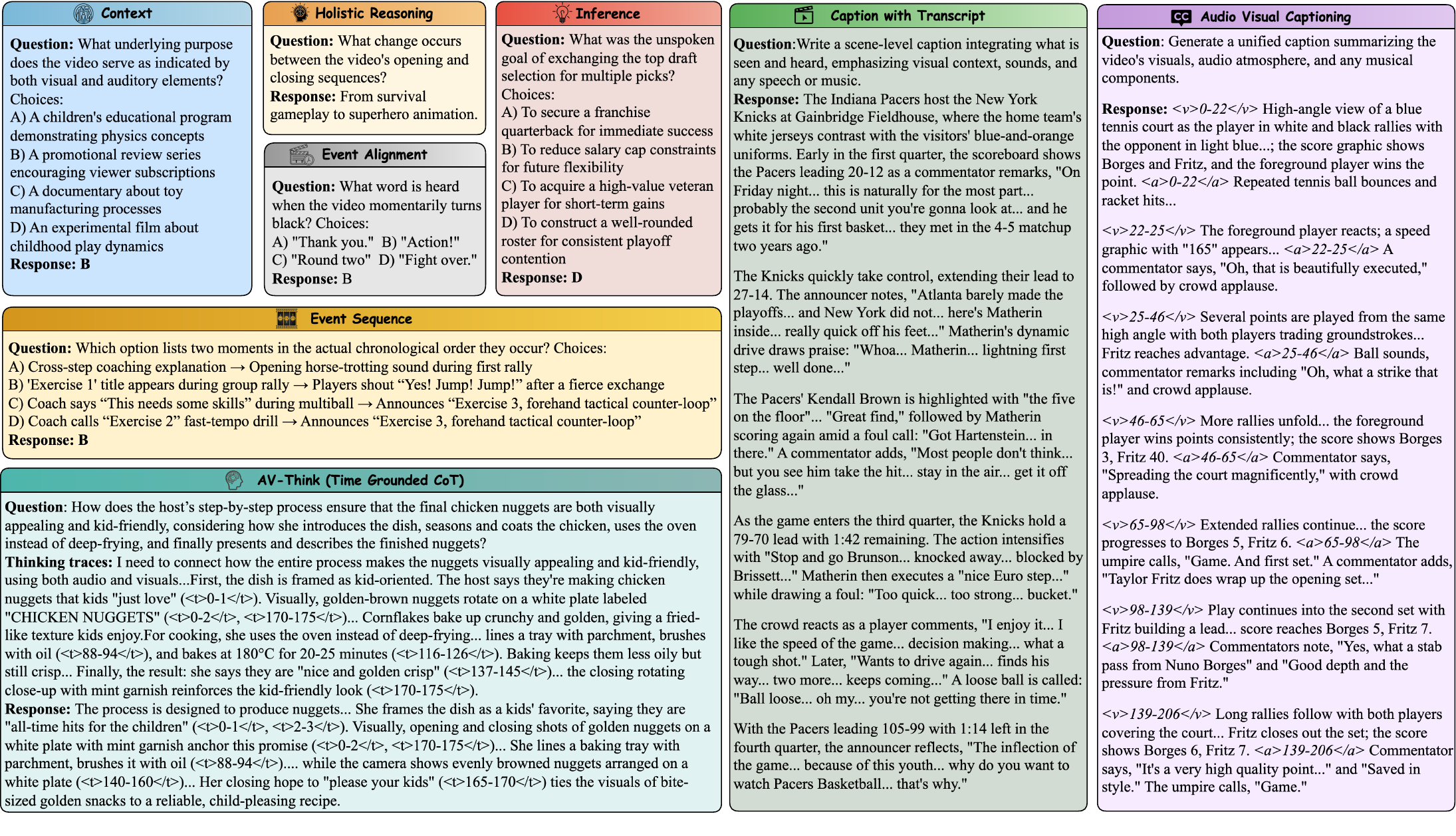}
  \caption{\small Examples of types of questions in Audio-Visual-Skills. More examples are shown in Fig~\ref{fig:examples_appendix} and Table \ref{tab:avf-short-dataset-examples}, and details are provided in Appendix \ref{app:dataset_examples}.}
  \label{fig:av-skills-long}
\vspace{-3mm}
\end{figure*}

For \textbf{AV-Skills-Long}, we focus on the following skills (examples shown in Fig.~\ref{fig:av-skills-long} and Table~\ref{tab:dataset-details}): (i) \textbf{Needle-in-the-Haystack Reasoning:} Answering questions about a specific but important moment in a long video, or a localized event that still requires broader audio-visual context to identify correctly. (ii) \textbf{Temporal Reasoning:} \textbf{\textit{a) Temporal Referring:}} Identify what happens immediately before or after a target event. \textbf{\textit{b) Temporal Order:}} Determine the correct order of multiple audio-visual events. \textbf{\textit{c) Temporal Attribute:}} Reason about how scene dynamics such as intensity, pace, or atmosphere evolve over time. (iii) \textbf{Sub-scene Understanding:} Understanding and describing a meaningful intermediate portion of a long video based on the events that precede and follow it. (iv) \textbf{Holistic Reasoning:} Inferring high-level goals, motivations, or outcomes by integrating audio-visual evidence distributed across a substantial part of the video. (v) \textbf{Counting:} Counting the number of times a particular audio-visual event, interaction, or recurring pattern appears throughout the video. (vi) \textbf{Audio-Visual Referring:} Linking an event in one modality to its counterpart in the other, such as identifying the visual scene associated with a sound or the sound associated with a visual event. (vii) \textbf{Topic-level Reasoning:} Identifying the main activity, objective, or overall theme of the video by synthesizing evidence across the full timeline. (viii) \textbf{Detailed Captioning:} Generating a comprehensive audio-visual description of the full video, capturing key actions, scene changes, sounds, and spoken content without redundancy. (ix) \textbf{Event Sequence Reasoning:} Determining the relative ordering of key audio-visual events across the video timeline. (x) \textbf{Audio-Visual Event Alignment:} Identifying the exact sound that coincides with a visual moment, or the exact visual event synchronized with a target audio cue. (xi) \textbf{Inference:} Drawing conclusions about implicit intentions, causes, or outcomes by combining multiple audio-visual clues that are not stated directly. (xii) \textbf{Comparative Reasoning:} Identifying important differences or similarities between two audio-visual events, scenes, or presentations in the same video. (xiii) \textbf{Context Understanding:} Inferring the broader situational setting, background context, or scene conditions by jointly reasoning over visual details, sounds, and speech.

\noindent \textbf{AV-Think.} Finally, we introduce \textbf{Temporal Audio-Visual Interleaved Chain-of-Thought (TAVIT)}, a reasoning framework that teaches the model to ground intermediate reasoning steps to timestamps in interdependent audio and visual streams (example in Fig.~\ref{fig:av-skills-long}). Prior work on CoT for video understanding remains limited, largely focusing on video-only settings or shallow reasoning traces in recent omni models. Existing approaches often yield only modest gains, and sometimes even degrade performance~\citep{qwen35omniblog}, partly because current audio or audio-visual CoT data is mostly restricted to short clips and simple QA pairs with reasoning appended post hoc.

We argue that explicit reasoning is most useful for long, complex real-world videos, where evidence is distributed across multiple, overlapping, and temporally dispersed events. To address this gap, we construct \textbf{AV-Think}, a dataset of question-answer-reasoning triplets for temporally grounded audio-visual reasoning. Unlike prior work, AV-Think explicitly interleaves audio and visual evidence within timestamped reasoning steps, bridging reasoning accuracy with temporal grounding.

AV-Think is curated from challenging long-form videos, including trailers, movie recaps, mystery stories, and multi-party conversations, paired with questions requiring extended temporal reasoning. We ground reasoning to time because \textbf{(i)} timestamped thoughts help the model navigate and reason over long audio-visual streams, and \textbf{(ii)} grounding intermediate reasoning in temporal events can improve recognition performance~\citep{kumar2026tactimestampedaudiocaptioning}. To build the dataset, we first generate time-stamped captions for each video using a TAC-style pipeline, and then prompt an LLM over these captions to synthesize reasoning triplets (see Prompt in Fig~\ref{fig:prompt_avthink}). AV-Think contains approximately 24K training samples, with reasoning chains averaging 635.7 words.

\vspace{-2mm}
\subsection{Training Curriculum}
\label{sec.curriculum}
\vspace{-2mm}

We design a three-stage curriculum to train AVF.  In each stage, we employ a distinct data mixture designed to gradually increase context length and complexity of tasks. We hypothesize that different capabilities emerge at different stages of training: foundational auditory and visual skills are best acquired early on unimodal and short-context data, during which the model also acquires weak audio-visual alignment. On the other hand, strong audio-visual alignment, long-context understanding, and more complex skills like temporal understanding require later-stage specialization with long, real-world, complex data. We provide the full data mixing ratios in Table~\ref{tab:dataset-details} and describe the training technique, including training hyperparameters, in Section~\ref{sec:experiments}.    

\noindent \textbf{Pre-training.} This stage consists of two phases designed to establish foundational capabilities. In \textbf{Stage 1: Initialization stage}, we initialize AVF from the pre-trained OmniVinci~\citep{ye2025omnivinci} checkpoint, which provides a strong starting point with already-aligned vision, audio, and language representations. In \textbf{Stage 2: Short-Context Training}, we perform full fine-tuning of the model on a mixture of unimodal datasets--spanning audio-only and vision-only tasks such as classification, captioning, ASR, and image/video question answering--alongside our newly curated AV-Skills-Short data. The unimodal data ensures that modality-specific capabilities inherited from OmniVinci are preserved and strengthened, while the AV-Skills-Short data introduces cross-modal reasoning abilities unique to Audio-Visual Flamingo. The maximum a-v duration in this stage is capped at 5 minutes, keeping the focus on high-quality short- and medium-length examples where skill-specific supervision is most reliable and scalable. The total context length in this stage is capped at 16K tokens.

\vspace{1mm}

\begin{table}[!t]
\centering
\caption{\small Performance comparison of AVF model variants on Omni, Audio, Video, and Speech Understanding benchmarks (WER ↓ and ACC ↑). We highlight \textcolor{closedGray}{closed source}, \textcolor{qwenPurple}{open weights}, and \textcolor{nvidiaGreen}{open source} models.}
\vspace{2mm}
\resizebox{\linewidth}{!}{%
\begin{tabular}{llccc}
\toprule
\textbf{Task} & \textbf{Dataset} & \textbf{Model} & \textbf{Metrics} & \textbf{Results} \\
\midrule
\multirow{13}{*}{\shortstack[l]{\textbf{Omni} \\ \textbf{Understanding}}}

& \multirow{4}{*}{\textbf{WorldSense}} 
& \textcolor{qwenPurple}{Qwen2.5-O} & \multirow{4}{*}{ACC ↑} & 45.4 \\
& & \textcolor{qwenPurple}{OmniVinci} &  & 48.2 \\
& & \textcolor{nvidiaGreen}{AVF-Instruct} & & \textbf{50.3} \\
& & \textcolor{nvidiaGreen}{AVF-Think} & & \textcolor{black!50}{51.6} \\
\cmidrule{2-5}

& \multirow{3}{*}{\textbf{DailyOmni}} 
& \textcolor{qwenPurple}{OmniVinci} & \multirow{3}{*}{ACC ↑} & 66.5 \\
& & \textcolor{nvidiaGreen}{AVF-Instruct} & & \textbf{72.4}\\
& & \textcolor{nvidiaGreen}{AVF-Think} & & \textcolor{black!50}{73.9} \\
\cmidrule{2-5}
& \multirow{3}{*}{\textbf{OmniBench}} 
& \textcolor{closedGray}{Gemini-1.5 Pro} & \multirow{3}{*}{ACC ↑} & 47.6 \\
& & \textcolor{nvidiaGreen}{AVF-Instruct} & & \textbf{48.5} \\
& & \textcolor{nvidiaGreen}{AVF-Think} & & \textcolor{black!50}{50.6}\\
\cmidrule{2-5}
& \multirow{4}{*}{\textbf{MMOU}} 
& \textcolor{closedGray}{Gemini-2.5 Pro} & \multirow{4}{*}{ACC ↑} & \textbf{64.2} \\
& & \textcolor{qwenPurple}{Minicpm-o 4.5} & & 46.8 \\
& & \textcolor{nvidiaGreen}{AVF-Instruct} & & 56.9 \\
& & \textcolor{nvidiaGreen}{AVF-Think} & & \textcolor{black!50}{60.2} \\
\cmidrule{2-5}
& \multirow{3}{*}{\shortstack[l]{\textbf{AVHBench} \\ \textit{A->V Hall. | V->A Hall.} }}
& \textcolor{closedGray}{Gemini-2.0 Flash} & \multirow{3}{*}{ACC ↑} & \textbf{83.3} | 63.3 \\
& & \textcolor{nvidiaGreen}{AVF-Instruct} & & 77.0 | \textbf{81.1} \\
& & \textcolor{nvidiaGreen}{AVF-Think} & & \textcolor{black!50}{79.0 | 85.9} \\
\midrule
\multirow{9}{*}{\shortstack[l]{\textbf{Audio} \\ \textbf{Understanding}}}

& \multirow{2}{*}{\textbf{MMAR}} 
& \textcolor{qwenPurple}{OmniVinci} & \multirow{3}{*}{ACC ↑} & 58.4 \\
& & \textcolor{nvidiaGreen}{AVF-Instruct} & & \textbf{60.1} \\
\cmidrule{2-5}

& \multirow{2}{*}{\textbf{MMSU}} 
& \textcolor{closedGray}{Gemini 1.5 Pro} & \multirow{2}{*}{ACC ↑} & 60.7 \\
& & \textcolor{nvidiaGreen}{AVF-Instruct} & & \textbf{61.5} \\
\cmidrule{2-5}

& \multirow{3}{*}{\shortstack[l]{\textbf{MMAU-v05.15.25 (test)} \\ \textit{Sound | Music | Speech | Avg} }}
& \textcolor{nvidiaGreen}{Audio Flamingo 3} & \multirow{3}{*}{ACC ↑} & 75.83 | 74.47 | 66.97 | 72.42\\
& & \textcolor{qwenPurple}{OmniVinci} & & 73.57 |73.07| 68.17 | 71.60\\
& & \textcolor{nvidiaGreen}{AVF-Instruct} & & \textbf{77.97} |\textbf{73.17}| \textbf{69.33} | \textbf{73.49} \\
\cmidrule{2-5}

& \multirow{2}{*}{\textbf{CMM Hallucination}} 
& \textcolor{closedGray}{Gemini 2.5 Pro} & \multirow{2}{*}{ACC ↑} & 82.0 \\
& & \textcolor{nvidiaGreen}{AVF-Instruct} & & \textbf{86.7} \\
\midrule

\multirow{8}{*}{\shortstack[l]{\textbf{Video} \\ \textbf{Understanding}}}

& \multirow{3}{*}{\shortstack[l]{\textbf{Video-MME} \\ \textit{w/o subtitles | w/ subtitles} }}
& \textcolor{nvidiaGreen}{NVILA} & \multirow{3}{*}{ACC ↑} & 64.2 | - \\
& & \textcolor{qwenPurple}{OmniVinci} &  & 67.3 | 68.6 \\
& & \textcolor{nvidiaGreen}{AVF-Instruct} & & \textbf{70.7} | \textbf{71.2} \\
\cmidrule{2-5}

& \multirow{3}{*}{\textbf{LongVideoBench}} 
& \textcolor{qwenPurple}{OmniVinci} & \multirow{3}{*}{ACC ↑} & \textbf{62.0} \\
& & \textcolor{nvidiaGreen}{NVILA} &  & 58.7 \\
& & \textcolor{nvidiaGreen}{AVF-Instruct} & & 60.1 \\

\cmidrule{2-5}

& \multirow{2}{*}{\textbf{MVHBench}} 
& \textcolor{qwenPurple}{OmniVinci} & \multirow{2}{*}{ACC ↑} & 70.6 \\
& & \textcolor{nvidiaGreen}{AVF-Instruct} & & \textbf{71.7} \\
\midrule

\multirow{10}{*}{\shortstack[l]{\textbf{Automatic Speech} \\ \textbf{Recognition}}}

& \multirow{2}{*}{\shortstack[l]{\textbf{LibriSpeech (en)} \\ \textit{test-clean | test-other}}}
& \textcolor{qwenPurple}{Phi-4-mm} | \textcolor{qwenPurple}{Qwen2.5-O} 
& \multirow{2}{*}{WER ↓} 
& 1.67 | \textbf{3.4} \\
& & \textcolor{nvidiaGreen}{AVF-Instruct} & & \textbf{1.64} | 3.5 \\
\cmidrule{2-5}

& \multirow{2}{*}{\textbf{SPGISpeech (en)}} 
& \textcolor{qwenPurple}{Qwen2-A-Inst} & \multirow{2}{*}{WER ↓} & 3.0 \\
& & \textcolor{nvidiaGreen}{AVF-Instruct} & & \textbf{2.8} \\
\cmidrule{2-5}

& \multirow{2}{*}{\textbf{TEDLIUM (en)}} 
& \textcolor{qwenPurple}{Phi-4-mm} & \multirow{2}{*}{WER ↓} & \textbf{2.9} \\
& & \textcolor{nvidiaGreen}{AVF-Instruct} & & 3.0 \\
\cmidrule{2-5}

& \multirow{2}{*}{\textbf{GigaSpeech (en)}} 
& \textcolor{qwenPurple}{Phi-4-mm} & \multirow{2}{*}{WER ↓} & \textbf{9.78} \\
& & \textcolor{nvidiaGreen}{AVF-Instruct} & & 10.2 \\
\cmidrule{2-5}

& \multirow{2}{*}{\textbf{VoxPopuli (en)}} 
& \textcolor{qwenPurple}{Phi-4-mm} & \multirow{2}{*}{WER ↓} & 5.9 \\
& & \textcolor{nvidiaGreen}{AVF-Instruct} & & \textbf{5.8} \\

\bottomrule
\end{tabular}%
}
\vspace{-2mm}
\label{tab:simplified_results}
\end{table}

{\noindent \textbf{Mid-training.}} Building on the capabilities acquired after stage 2 of pre-training, this stage focuses on extending AVF's understanding to longer audio-visual inputs. We expand the data mixture with our AV-Skills-Long data, focused on long-form captioning, time-aware annotations that require precise temporal grounding, and long-context question answering datasets. To maintain a balanced training distribution, we additionally include a down-sampled subset of the AV-Skills-Short data from the pre-training stage, ensuring that previously acquired skills are retained while the model learns to reason over extended temporal contexts. The maximum audio/video duration in this stage is increased to 15 minutes, and the total context length is expanded to 32K tokens. The fully trained model resulting from this process is referred to as \textbf{AVF-Instruct}.
\vspace{1mm}

{\noindent \textbf{Post-training.}} Finally, we train AVF for temporally grounded CoT on the AV-Think dataset. Starting from AVF-Instruct, we first perform SFT on AV-Think that encourages the model to produce structured, step-by-step rationales anchored to specific instances in the audio-visual input. We then train the model using GRPO-based RL~\citep{shao2024deepseekmath} (more details in Section~\ref{sec:grpo-rl}) to further strengthen the model's reasoning quality. The final model obtained from this stage is referred to as \textbf{AVF-Think}.

\vspace{-4mm}
\section{Experiments}
\label{sec:experiments}
\vspace{-2mm}

{\noindent \textbf{Experimental Setup.}} We perform pre-training, long-context training, and post-training of Audio Visual Flamingo on 512 NVIDIA H100 GPUs. Further details on batch size, learning rates, and optimizers for each stage of training are in Appendix~\ref{sec.avf_training_details}. 
\vspace{1mm}

\noindent \textbf{Baselines. }We evaluate our proposed model against recent SOTA open-source and proprietary MLLMs (Large Multimodal Language Models, including Qwen2.5-O(mni)~\citep{xu2025qwen25omnitechnicalreport}, Qwen3-O(mni)~\citep{xu2025qwen3}, OmniVinci~\citep{ye2025omnivinci}, Gemini (2.0 Flash, 1.5 Pro, 2.5 Flash and 2.5 Pro)~\citep{team2024gemini, comanici2025gemini25pushingfrontier} as well as GPT-4o~\citep{hurst2024gpt}. Furthermore, for audio and speech understanding, we also compare against specialized Large Audio Language Models (LALMs) such as Audio Flamingo~\citep{kong2024audioflamingonovelaudio}, Audio Flamingo 2~\citep{ghosh2025audioflamingo2audiolanguage}, Audio Flamingo 3~\citep{goel2025audio}, Qwen-A(udio)~\citep{chu2023qwenaudioadvancinguniversalaudio}, Qwen2-A(udio)~\citep{chu2024qwen2}, Qwen2-A(udio)-(Inst)ruct, R1-AQA~\cite{li2025reinforcement}, Pengi~\citep{deshmukh2024pengiaudiolanguagemodel}, Phi-4-mm~\citep{abouelenin2025phi}, Baichun Audio~\citep{li2025baichuanaudiounifiedframeworkendtoend}, Step-Audio-Chat~\citep{tian2025step}, LTU~\citep{gong2023listen}, LTU-AS~\citep{gong2023jointaudiospeechunderstanding}, SALMONN~\citep{tang2023salmonn} and AudioGPT~\citep{huang2023audiogptunderstandinggeneratingspeech}. Similarly, for video understanding, we compare against specialized vision language models (VLMs) such as VILA~\citep{lin2024vilapretrainingvisuallanguage}, LongVILA~\citep{chen2024longvilascalinglongcontextvisual}, NVILA~\citep{liu2025nvilaefficientfrontiervisual}, InternVL~\citep{chen2024internvlscalingvisionfoundation}, Qwen2-VL~\citep{wang2024qwen2vlenhancingvisionlanguagemodels}, LLaVA-OneVision~\citep{li2024llavaonevisioneasyvisualtask}, and Florence-VL~\citep{chen2024florencevlenhancingvisionlanguagemodels}.
\vspace{1mm}

\noindent \textbf{Evaluation Datasets.} We evaluate our model on a diverse set of benchmarks spanning \textit{audio understanding} (MMAR~\citep{ma2025mmarchallengingbenchmarkdeep}, MMSU~\citep{wang2026mmsumassivemultitaskspoken}, MMAU (v05.15.25)~\citep{sakshi2024mmaumassivemultitaskaudio}), which covers sound, music, and speech reasoning; \textit{video understanding} (Video-MME~\citep{fu2025video}, ActivityNetQA~\citep{yu2019activitynetqadatasetunderstandingcomplex}, LongVideoBench~\citep{wu2024longvideobenchbenchmarklongcontextinterleaved}), capturing visual comprehension as well as temporal reasoning; \textit{omni/multimodal understanding} (WorldSense~\citep{hong2026worldsenseevaluatingrealworldomnimodal} and DailyOmni~\citep{zhou2026dailyomniaudiovisualreasoningtemporal}), evaluating joint reasoning across audio and video signals; and ASR (LibriSpeech (clean and other)~\citep{panayotov2015librispeech}, SPGISpeech~\citep{oneill2021spgispeech5000hourstranscribed}, TEDLIUM~\citep{Hernandez_2018}, GigaSpeech~\citep{Chen_2021}, Common Voice 15), which span clean, noisy, and diverse real-world speech conditions. 

\vspace{-4mm}
\section{Results}
\vspace{-2mm}

\noindent \textbf{Audio Understanding and Reasoning.}
AVF-Instruct shows strong transfer to audio-only understanding and reasoning despite being trained as a joint audio-visual model. It improves over OmniVinci on MMAR (60.1 vs.\ 58.4) and exceeds Gemini 1.5 Pro on MMSU (61.5 vs.\ 60.7). On MMAU, AVF-Instruct achieves the best overall average score (73.49), outperforming both AF3 and OmniVinci, with especially strong gains on sound and speech understanding. For ASR, AVF achieves the best result on LibriSpeech test-clean (1.64 WER), SPGISpeech (2.8), and VoxPopuli (5.8), while remaining close to the strongest baselines on others.

\noindent \textbf{Video and Omni Understanding.}
AVF-Instruct also demonstrates consistent gains on both video-only and omni-modal evaluation. On Video-MME, it achieves the best performance both without subtitles (70.7) and with subtitles (71.2), outperforming both NVILA and OmniVinci. On WorldSense and DailyOmni, AVF-Instruct again sets the strongest results, indicating better joint perception and reasoning across modalities. On LongVideoBench, while OmniVinci remains slightly stronger, AVF-Instruct still outperforms NVILA and remains competitive on long-form video understanding, suggesting that it scales effectively to challenging real-world video settings. Results on MMOU emphasize AVF is best-in-class for long and complex audio-visual understanding. Further qualitative examples are provided in Table~\ref{tab:av-cot-examples} and our project page. \textbf{Ablations:} Table~\ref{tab:app_training_recipe} summarizes the compact AVF training recipe. Table~\ref{tab:ablation-avskills} provides the appendix ablation on our AV-Skills curriculum.

\section{Conclusion, Limitations and Future Work}
\label{sec:conclusion}
\vspace{-2mm}

We presented AV-Flamingo, a fully open AV-LLM for joint understanding and reasoning over long and complex real-world videos. AV-Flamingo combines AV-Skills, a large-scale dataset for joint cross-modal learning, a three-stage curriculum for scaling from short-context perception to long-horizon reasoning, and TAVIT, a reasoning framework that grounds intermediate thoughts to timestamps in audio and visual streams. Experiments show strong performance across diverse benchmarks, with especially notable gains on long and complex real-world videos.

AV-Flamingo still has several limitations. AV-Skills is built from public datasets and open-internet videos, which may introduce source bias and potential overlap with prior training data. In addition, reasoning over very long and highly dense videos remains challenging, especially when evidence is sparse or temporally dispersed. Finally, current benchmarks do not fully capture open-ended real-world deployment. In future work, we plan to expand AV-Skills to broader domains and more challenging long-form settings, improve long-context reasoning, and develop more realistic evaluation protocols for open audio-visual systems.



\bibliographystyle{plainnat}
\bibliography{references} 

@article{alayrac2022flamingo,
  title={Flamingo: a visual language model for few-shot learning},
  author={Alayrac, Jean-Baptiste and Donahue, Jeff and Luc, Pauline and Miech, Antoine and Barr, Iain and Hasson, Yana and Lenc, Karel and Mensch, Arthur and Millican, Katherine and Reynolds, Malcolm and others},
  journal={Advances in neural information processing systems},
  volume={35},
  pages={23716--23736},
  year={2022}
}

@inproceedings{li2023blip,
  title={Blip-2: Bootstrapping language-image pre-training with frozen image encoders and large language models},
  author={Li, Junnan and Li, Dongxu and Savarese, Silvio and Hoi, Steven},
  booktitle={International conference on machine learning},
  pages={19730--19742},
  year={2023},
  organization={PMLR}
}

@article{liu2023visual,
  title={Visual instruction tuning},
  author={Liu, Haotian and Li, Chunyuan and Wu, Qingyang and Lee, Yong Jae},
  journal={Advances in neural information processing systems},
  volume={36},
  pages={34892--34916},
  year={2023}
}

@article{goel2024omcat,
  title={Omcat: Omni context aware transformer},
  author={Goel, Arushi and Sapra, Karan and Le, Matthieu and Valle, Rafael and Tao, Andrew and Catanzaro, Bryan},
  journal={arXiv preprint arXiv:2410.12109},
  year={2024}
}

@article{chen2024expanding,
  title={Expanding performance boundaries of open-source multimodal models with model, data, and test-time scaling},
  author={Chen, Zhe and Wang, Weiyun and Cao, Yue and Liu, Yangzhou and Gao, Zhangwei and Cui, Erfei and Zhu, Jinguo and Ye, Shenglong and Tian, Hao and Liu, Zhaoyang and others},
  journal={arXiv preprint arXiv:2412.05271},
  year={2024}
}

@article{bai2025qwen3,
  title={Qwen3-vl technical report},
  author={Bai, Shuai and Cai, Yuxuan and Chen, Ruizhe and Chen, Keqin and Chen, Xionghui and Cheng, Zesen and Deng, Lianghao and Ding, Wei and Gao, Chang and Ge, Chunjiang and others},
  journal={arXiv preprint arXiv:2511.21631},
  year={2025}
}

@article{zhang2025videollama,
  title={Videollama 3: Frontier multimodal foundation models for image and video understanding},
  author={Zhang, Boqiang and Li, Kehan and Cheng, Zesen and Hu, Zhiqiang and Yuan, Yuqian and Chen, Guanzheng and Leng, Sicong and Jiang, Yuming and Zhang, Hang and Li, Xin and others},
  journal={arXiv preprint arXiv:2501.13106},
  year={2025}
}

@misc{qwen35blog,
    title = {Qwen3.5: Accelerating Productivity with Native Multimodal Agents},
    url = {https://qwen.ai/blog?id=qwen3.5},
    author = {Qwen Team},
    month = {February},
    year = {2026}
}

@article{gong2023listen,
  title={Listen, think, and understand},
  author={Gong, Yuan and Luo, Hongyin and Liu, Alexander H and Karlinsky, Leonid and Glass, James},
  journal={arXiv preprint arXiv:2305.10790},
  year={2023}
}

@article{tang2023salmonn,
  title={Salmonn: Towards generic hearing abilities for large language models},
  author={Tang, Changli and Yu, Wenyi and Sun, Guangzhi and Chen, Xianzhao and Tan, Tian and Li, Wei and Lu, Lu and Ma, Zejun and Zhang, Chao},
  journal={arXiv preprint arXiv:2310.13289},
  year={2023}
}

@article{chu2024qwen2,
  title={Qwen2-audio technical report},
  author={Chu, Yunfei and Xu, Jin and Yang, Qian and Wei, Haojie and Wei, Xipin and Guo, Zhifang and Leng, Yichong and Lv, Yuanjun and He, Jinzheng and Lin, Junyang and others},
  journal={arXiv preprint arXiv:2407.10759},
  year={2024}
}

@article{goel2025audio,
  title={Audio flamingo 3: Advancing audio intelligence with fully open large audio language models},
  author={Goel, Arushi and Ghosh, Sreyan and Kim, Jaehyeon and Kumar, Sonal and Kong, Zhifeng and Lee, Sang-gil and Yang, Chao-Han Huck and Duraiswami, Ramani and Manocha, Dinesh and Valle, Rafael and others},
  journal={arXiv preprint arXiv:2507.08128},
  year={2025}
}

@inproceedings{zhang2023video,
  title={Video-llama: An instruction-tuned audio-visual language model for video understanding},
  author={Zhang, Hang and Li, Xin and Bing, Lidong},
  booktitle={Proceedings of the 2023 conference on empirical methods in natural language processing: system demonstrations},
  pages={543--553},
  year={2023}
}

@article{cheng2024videollama,
  title={Videollama 2: Advancing spatial-temporal modeling and audio understanding in video-llms},
  author={Cheng, Zesen and Leng, Sicong and Zhang, Hang and Xin, Yifei and Li, Xin and Chen, Guanzheng and Zhu, Yongxin and Zhang, Wenqi and Luo, Ziyang and Zhao, Deli and others},
  journal={arXiv preprint arXiv:2406.07476},
  year={2024}
}

@article{sun2024video,
  title={video-salmonn: Speech-enhanced audio-visual large language models},
  author={Sun, Guangzhi and Yu, Wenyi and Tang, Changli and Chen, Xianzhao and Tan, Tian and Li, Wei and Lu, Lu and Ma, Zejun and Wang, Yuxuan and Zhang, Chao},
  journal={arXiv preprint arXiv:2406.15704},
  year={2024}
}

@article{tang2025video,
  title={video-SALMONN 2: Caption-Enhanced Audio-Visual Large Language Models},
  author={Tang, Changli and Li, Yixuan and Yang, Yudong and Zhuang, Jimin and Sun, Guangzhi and Li, Wei and Ma, Zejun and Zhang, Chao},
  journal={arXiv preprint arXiv:2506.15220},
  year={2025}
}

@article{li2025watch,
  title={Watch and listen: Understanding audio-visual-speech moments with multimodal llm},
  author={Li, Zinuo and Zhang, Xian and Guo, Yongxin and Bennamoun, Mohammed and Boussaid, Farid and Dwivedi, Girish and Gong, Luqi and Ke, Qiuhong},
  journal={arXiv preprint arXiv:2505.18110},
  year={2025}
}

@article{team2024gemini,
  title={Gemini 1.5: Unlocking multimodal understanding across millions of tokens of context},
  author={Team, Gemini and Georgiev, Petko and Lei, Ving Ian and Burnell, Ryan and Bai, Libin and Gulati, Anmol and Tanzer, Garrett and Vincent, Damien and Pan, Zhufeng and Wang, Shibo and others},
  journal={arXiv preprint arXiv:2403.05530},
  year={2024}
}

@misc{xu2025qwen25omnitechnicalreport,
      title={Qwen2.5-Omni Technical Report}, 
      author={Jin Xu and Zhifang Guo and Jinzheng He and Hangrui Hu and Ting He and Shuai Bai and Keqin Chen and Jialin Wang and Yang Fan and Kai Dang and Bin Zhang and Xiong Wang and Yunfei Chu and Junyang Lin},
      year={2025},
      eprint={2503.20215},
      archivePrefix={arXiv},
      primaryClass={cs.CL},
      url={https://arxiv.org/abs/2503.20215}, 
}

@article{xu2025qwen3,
  title={Qwen3-omni technical report},
  author={Xu, Jin and Guo, Zhifang and Hu, Hangrui and Chu, Yunfei and Wang, Xiong and He, Jinzheng and Wang, Yuxuan and Shi, Xian and He, Ting and Zhu, Xinfa and others},
  journal={arXiv preprint arXiv:2509.17765},
  year={2025}
}

@article{abouelenin2025phi,
  title={Phi-4-mini technical report: Compact yet powerful multimodal language models via mixture-of-loras},
  author={Abouelenin, Abdelrahman and Ashfaq, Atabak and Atkinson, Adam and Awadalla, Hany and Bach, Nguyen and Bao, Jianmin and Benhaim, Alon and Cai, Martin and Chaudhary, Vishrav and Chen, Congcong and others},
  journal={arXiv preprint arXiv:2503.01743},
  year={2025}
}

@article{ye2025omnivinci,
  title={OmniVinci: Enhancing Architecture and Data for Omni-Modal Understanding LLM},
  author={Ye, Hanrong and Yang, Chao-Han Huck and Goel, Arushi and Huang, Wei and Zhu, Ligeng and Su, Yuanhang and Lin, Sean and Cheng, An-Chieh and Wan, Zhen and Tian, Jinchuan and others},
  journal={arXiv preprint arXiv:2510.15870},
  year={2025}
}

@inproceedings{li2024llama,
  title={Llama-vid: An image is worth 2 tokens in large language models},
  author={Li, Yanwei and Wang, Chengyao and Jia, Jiaya},
  booktitle={European Conference on Computer Vision},
  pages={323--340},
  year={2024},
  organization={Springer}
}

@inproceedings{fu2025video,
  title={Video-mme: The first-ever comprehensive evaluation benchmark of multi-modal llms in video analysis},
  author={Fu, Chaoyou and Dai, Yuhan and Luo, Yongdong and Li, Lei and Ren, Shuhuai and Zhang, Renrui and Wang, Zihan and Zhou, Chenyu and Shen, Yunhang and Zhang, Mengdan and others},
  booktitle={Proceedings of the IEEE/CVF conference on computer vision and pattern recognition},
  pages={24108--24118},
  year={2025}
}

@article{goel2026mmou,
  title={MMOU: A Massive Multi-Task Omni Understanding and Reasoning Benchmark for Long and Complex Real-World Videos},
  author={Goel, Arushi and Ghosh, Sreyan and Agarwal, Vatsal and Anand, Nishit and Jayakumar, Kaousheik and Koroshinadze, Lasha and Xu, Yao and Lyons, Katie and Case, James and Sapra, Karan and others},
  journal={arXiv preprint arXiv:2603.14145},
  year={2026}
}

@article{meng2025open,
  title={Open-o3 Video: Grounded Video Reasoning with Explicit Spatio-Temporal Evidence},
  author={Meng, Jiahao and Li, Xiangtai and Wang, Haochen and Tan, Yue and Zhang, Tao and Kong, Lingdong and Tong, Yunhai and Wang, Anran and Teng, Zhiyang and Wang, Yujing and others},
  journal={arXiv preprint arXiv:2510.20579},
  year={2025}
}

@misc{zhu2025internvl3exploringadvancedtraining,
      title={InternVL3: Exploring Advanced Training and Test-Time Recipes for Open-Source Multimodal Models}, 
      author={Jinguo Zhu and Weiyun Wang and Zhe Chen and Zhaoyang Liu and Shenglong Ye and Lixin Gu and Hao Tian and Yuchen Duan and Weijie Su and Jie Shao and Zhangwei Gao and Erfei Cui and Xuehui Wang and Yue Cao and Yangzhou Liu and Xingguang Wei and Hongjie Zhang and Haomin Wang and Weiye Xu and Hao Li and Jiahao Wang and Nianchen Deng and Songze Li and Yinan He and Tan Jiang and Jiapeng Luo and Yi Wang and Conghui He and Botian Shi and Xingcheng Zhang and Wenqi Shao and Junjun He and Yingtong Xiong and Wenwen Qu and Peng Sun and Penglong Jiao and Han Lv and Lijun Wu and Kaipeng Zhang and Huipeng Deng and Jiaye Ge and Kai Chen and Limin Wang and Min Dou and Lewei Lu and Xizhou Zhu and Tong Lu and Dahua Lin and Yu Qiao and Jifeng Dai and Wenhai Wang},
      year={2025},
      eprint={2504.10479},
      archivePrefix={arXiv},
      primaryClass={cs.CV},
      url={https://arxiv.org/abs/2504.10479}, 
}

@inproceedings{liu2024unitime,
  title={Unitime: A language-empowered unified model for cross-domain time series forecasting},
  author={Liu, Xu and Hu, Junfeng and Li, Yuan and Diao, Shizhe and Liang, Yuxuan and Hooi, Bryan and Zimmermann, Roger},
  booktitle={Proceedings of the ACM Web Conference 2024},
  pages={4095--4106},
  year={2024}
}

@article{arnab2025temporal,
  title={Temporal chain of thought: Long-video understanding by thinking in frames},
  author={Arnab, Anurag and Iscen, Ahmet and Caron, Mathilde and Fathi, Alireza and Schmid, Cordelia},
  journal={arXiv preprint arXiv:2507.02001},
  year={2025}
}

@article{li2025reinforcement,
  title={Reinforcement learning outperforms supervised fine-tuning: A case study on audio question answering},
  author={Li, Gang and Liu, Jizhong and Dinkel, Heinrich and Niu, Yadong and Zhang, Junbo and Luan, Jian},
  journal={arXiv preprint arXiv:2503.11197},
  year={2025}
}

@article{tian2025step,
  title={Step-Audio-R1 Technical Report},
  author={Tian, Fei and Zhang, Xiangyu Tony and Zhang, Yuxin and Zhang, Haoyang and Li, Yuxin and Liu, Daijiao and Deng, Yayue and Wu, Donghang and Chen, Jun and Zhao, Liang and others},
  journal={arXiv preprint arXiv:2511.15848},
  year={2025}
}

@misc{qwen35omniblog,

    title = {Qwen3.5-Omni: Scaling Up, Toward Native Omni-Modal AGI},

    url = {https://qwen.ai/blog?id=qwen3.5-omni},

    author = {Qwen Team},

    month = {March},

    year = {2026}

}

@article{deshmukh2026nemotron,
  title={Nemotron 3 Nano Omni: Efficient and Open Multimodal Intelligence},
  author={Deshmukh, Amala Sanjay and Chumachenko, Kateryna and Rintamaki, Tuomas and Le, Matthieu and Poon, Tyler and Taheri, Danial Mohseni and Karmanov, Ilia and Liu, Guilin and Seppanen, Jarno and Goel, Arushi and others},
  journal={arXiv preprint arXiv:2604.24954},
  year={2026}
}

@article{hurst2024gpt,
  title={Gpt-4o system card},
  author={Hurst, Aaron and Lerer, Adam and Goucher, Adam P and Perelman, Adam and Ramesh, Aditya and Clark, Aidan and Ostrow, AJ and Welihinda, Akila and Hayes, Alan and Radford, Alec and others},
  journal={arXiv preprint arXiv:2410.21276},
  year={2024}
}

@misc{li2025improvingllmvideounderstanding,
      title={Improving LLM Video Understanding with 16 Frames Per Second}, 
      author={Yixuan Li and Changli Tang and Jimin Zhuang and Yudong Yang and Guangzhi Sun and Wei Li and Zejun Ma and Chao Zhang},
      year={2025},
      eprint={2503.13956},
      archivePrefix={arXiv},
      primaryClass={cs.CV},
      url={https://arxiv.org/abs/2503.13956}, 
}

@article{zhang2024long,
  title={Long context transfer from language to vision},
  author={Zhang, Peiyuan and Zhang, Kaichen and Li, Bo and Zeng, Guangtao and Yang, Jingkang and Zhang, Yuanhan and Wang, Ziyue and Tan, Haoran and Li, Chunyuan and Liu, Ziwei},
  journal={arXiv preprint arXiv:2406.16852},
  year={2024}
}

@article{qian2024streaming,
  title={Streaming long video understanding with large language models},
  author={Qian, Rui and Dong, Xiaoyi and Zhang, Pan and Zang, Yuhang and Ding, Shuangrui and Lin, Dahua and Wang, Jiaqi},
  journal={Advances in Neural Information Processing Systems},
  volume={37},
  pages={119336--119360},
  year={2024}
}

@misc{guo2025alignedbetterlistenbetter,
      title={Aligned Better, Listen Better for Audio-Visual Large Language Models}, 
      author={Yuxin Guo and Shuailei Ma and Shijie Ma and Xiaoyi Bao and Chen-Wei Xie and Kecheng Zheng and Tingyu Weng and Siyang Sun and Yun Zheng and Wei Zou},
      year={2025},
      eprint={2504.02061},
      archivePrefix={arXiv},
      primaryClass={cs.CV},
      url={https://arxiv.org/abs/2504.02061}, 
}

@article{ghosh2025music,
  title={Music flamingo: Scaling music understanding in audio language models},
  author={Ghosh, Sreyan and Goel, Arushi and Koroshinadze, Lasha and Lee, Sang-gil and Kong, Zhifeng and Santos, Joao Felipe and Duraiswami, Ramani and Manocha, Dinesh and Ping, Wei and Shoeybi, Mohammad and others},
  journal={arXiv preprint arXiv:2511.10289},
  year={2025}
}

@inproceedings{Yu2022M2MeT,
  title={M2{M}e{T}: The {ICASSP} 2022 Multi-Channel Multi-Party Meeting Transcription Challenge},
  author={Yu, Fan and Zhang, Shiliang and Fu, Yihui and Xie, Lei and Zheng, Siqi and Du, Zhihao and Huang, Weilong and Guo, Pengcheng and Yan, Zhijie and Ma, Bin and Xu, Xin and Bu, Hui},
  booktitle={Proc. ICASSP},
  year={2022},
  organization={IEEE}
}

@inproceedings{vincent2013second,
  title={The Second {CHiME} Speech Separation and Recognition Challenge: Datasets, Tasks and Baselines},
  author={Vincent, Emmanuel and Barker, Jon and Watanabe, Shinji and Le Roux, Jonathan and Nesta, Francesco and Matassoni, Marco},
  booktitle={Proc. IEEE International Conference on Acoustics, Speech and Signal Processing (ICASSP)},
  year={2013},
  address={Vancouver, Canada},
  organization={IEEE}
}

@inproceedings{he2024emilia,
  title={Emilia: An extensive, multilingual, and diverse speech dataset for large-scale speech generation},
  author={He, Haorui and Shang, Zengqiang and Wang, Chaoren and Li, Xuyuan and Gu, Yicheng and Hua, Hua and Liu, Liwei and Yang, Chen and Li, Jiaqi and Shi, Peiyang and others},
  booktitle={2024 IEEE Spoken Language Technology Workshop (SLT)},
  pages={885--890},
  year={2024},
  organization={IEEE}
}

@article{cattoni2021must,
  title={MuST-C: A multilingual corpus for end-to-end speech translation},
  author={Cattoni, Roldano and Di Gangi, Mattia Antonino and Bentivogli, Luisa and Negri, Matteo and Turchi, Marco},
  journal={Computer speech \& language},
  volume={66},
  pages={101155},
  year={2021},
  publisher={Elsevier}
}

@article{wang2020covost,
  title={Covost 2 and massively multilingual speech-to-text translation},
  author={Wang, Changhan and Wu, Anne and Pino, Juan},
  journal={arXiv preprint arXiv:2007.10310},
  year={2020}
}

@inproceedings{aishell_2017,
  title={AIShell-1: An Open-Source Mandarin Speech Corpus and A Speech Recognition Baseline},
  author={Hui Bu and Jiayu Du and Xingyu Na and Bengu Wu and Hao Zheng},
  booktitle={Oriental COCOSDA 2017},
  year={2017}
}

@misc{aidatatang_200zh,
  title={aidatatang\_200zh: A Free {Chinese} {Mandarin} Speech Corpus},
  author={{Beijing DataTang Technology Co., Ltd.}},
  year={2018},
  howpublished={\url{https://www.openslr.org/62/}},
  note={OpenSLR resource 62}
}

@article{shao2024deepseekmath,
  title={Deepseekmath: Pushing the limits of mathematical reasoning in open language models},
  author={Shao, Zhihong and Wang, Peiyi and Zhu, Qihao and Xu, Runxin and Song, Junxiao and Bi, Xiao and Zhang, Haowei and Zhang, Mingchuan and Li, YK and Wu, Yang and others},
  journal={arXiv preprint arXiv:2402.03300},
  year={2024}
}

@inproceedings{zhou2025harmonyset,
  title={Harmonyset: A comprehensive dataset for understanding video-music semantic alignment and temporal synchronization},
  author={Zhou, Zitang and Mei, Ke and Lu, Yu and Wang, Tianyi and Rao, Fengyun},
  booktitle={Proceedings of the Computer Vision and Pattern Recognition Conference},
  pages={3152--3162},
  year={2025}
}

@misc{rohrbach2016moviedescription,
      title={Movie Description}, 
      author={Anna Rohrbach and Atousa Torabi and Marcus Rohrbach and Niket Tandon and Christopher Pal and Hugo Larochelle and Aaron Courville and Bernt Schiele},
      year={2016},
      eprint={1605.03705},
      archivePrefix={arXiv},
      primaryClass={cs.CV},
      url={https://arxiv.org/abs/1605.03705}, 
}

@misc{chi2024mmtrailmultimodaltrailervideo,
      title={MMTrail: A Multimodal Trailer Video Dataset with Language and Music Descriptions}, 
      author={Xiaowei Chi and Yatian Wang and Aosong Cheng and Pengjun Fang and Zeyue Tian and Yingqing He and Zhaoyang Liu and Xingqun Qi and Jiahao Pan and Rongyu Zhang and Mengfei Li and Ruibin Yuan and Yanbing Jiang and Wei Xue and Wenhan Luo and Qifeng Chen and Shanghang Zhang and Qifeng Liu and Yike Guo},
      year={2024},
      eprint={2407.20962},
      archivePrefix={arXiv},
      primaryClass={cs.CV},
      url={https://arxiv.org/abs/2407.20962}, 
}

@misc{bose2022movieclipvisualscenerecognition,
      title={MovieCLIP: Visual Scene Recognition in Movies}, 
      author={Digbalay Bose and Rajat Hebbar and Krishna Somandepalli and Haoyang Zhang and Yin Cui and Kree Cole-McLaughlin and Huisheng Wang and Shrikanth Narayanan},
      year={2022},
      eprint={2210.11065},
      archivePrefix={arXiv},
      primaryClass={cs.CV},
      url={https://arxiv.org/abs/2210.11065}, 
}

@misc{ju2024miradatalargescalevideodataset,
      title={MiraData: A Large-Scale Video Dataset with Long Durations and Structured Captions}, 
      author={Xuan Ju and Yiming Gao and Zhaoyang Zhang and Ziyang Yuan and Xintao Wang and Ailing Zeng and Yu Xiong and Qiang Xu and Ying Shan},
      year={2024},
      eprint={2407.06358},
      archivePrefix={arXiv},
      primaryClass={cs.CV},
      url={https://arxiv.org/abs/2407.06358}, 
}

@misc{xue2022advancinghighresolutionvideolanguagerepresentation,
      title={Advancing High-Resolution Video-Language Representation with Large-Scale Video Transcriptions}, 
      author={Hongwei Xue and Tiankai Hang and Yanhong Zeng and Yuchong Sun and Bei Liu and Huan Yang and Jianlong Fu and Baining Guo},
      year={2022},
      eprint={2111.10337},
      archivePrefix={arXiv},
      primaryClass={cs.CV},
      url={https://arxiv.org/abs/2111.10337}, 
}

@misc{wang2024internvidlargescalevideotextdataset,
      title={InternVid: A Large-scale Video-Text Dataset for Multimodal Understanding and Generation}, 
      author={Yi Wang and Yinan He and Yizhuo Li and Kunchang Li and Jiashuo Yu and Xin Ma and Xinhao Li and Guo Chen and Xinyuan Chen and Yaohui Wang and Conghui He and Ping Luo and Ziwei Liu and Yali Wang and Limin Wang and Yu Qiao},
      year={2024},
      eprint={2307.06942},
      archivePrefix={arXiv},
      primaryClass={cs.CV},
      url={https://arxiv.org/abs/2307.06942}, 
}

@misc{yang2023vidchapters7mvideochaptersscale,
      title={VidChapters-7M: Video Chapters at Scale}, 
      author={Antoine Yang and Arsha Nagrani and Ivan Laptev and Josef Sivic and Cordelia Schmid},
      year={2023},
      eprint={2309.13952},
      archivePrefix={arXiv},
      primaryClass={cs.CV},
      url={https://arxiv.org/abs/2309.13952}, 
}

@article{abu2016youtube,
  title={{YouTube-8M}: A Large-Scale Video Classification Benchmark},
  author={Abu-El-Haija, Sami and Kothari, Nisarg and Lee, Joonseok and Natsev, Paul and Toderici, George and Varadarajan, Balakrishnan and Vijayanarasimhan, Sudheendra},
  journal={arXiv preprint arXiv:1609.08675},
  year={2016}
}

@inproceedings{gemmeke2017audioset,
  title={Audio {Set}: An Ontology and Human-Labeled Dataset for Audio Events},
  author={Gemmeke, Jort F. and Ellis, Daniel P. W. and Freedman, Dylan and Jansen, Aren and Lawrence, Wade and Moore, R. Channing and Plakal, Manoj and Ritter, Marvin},
  booktitle={Proc. IEEE International Conference on Acoustics, Speech and Signal Processing (ICASSP)},
  year={2017},
  address={New Orleans, LA, USA},
  organization={IEEE}
}

@misc{zhai2023sigmoidlosslanguageimage,
      title={Sigmoid Loss for Language Image Pre-Training}, 
      author={Xiaohua Zhai and Basil Mustafa and Alexander Kolesnikov and Lucas Beyer},
      year={2023},
      eprint={2303.15343},
      archivePrefix={arXiv},
      primaryClass={cs.CV},
      url={https://arxiv.org/abs/2303.15343}, 
}

@misc{qwen2025qwen25technicalreport,
      title={Qwen2.5 Technical Report}, 
      author={Qwen Team.},
      year={2025},
      eprint={2412.15115},
      archivePrefix={arXiv},
      primaryClass={cs.CL},
      url={https://arxiv.org/abs/2412.15115}, 
}

@misc{godfrey1993switchboard,
  title        = {Switchboard-1 Release 2 (LDC97S62)},
  author       = {Godfrey, John J. and Holliman, Edward},
  year         = {1993},
  howpublished = {Linguistic Data Consortium, Philadelphia},
  note         = {Web Download}
}

@misc{gong2023jointaudiospeechunderstanding,
      title={Joint Audio and Speech Understanding}, 
      author={Yuan Gong and Alexander H. Liu and Hongyin Luo and Leonid Karlinsky and James Glass},
      year={2023},
      eprint={2309.14405},
      archivePrefix={arXiv},
      primaryClass={cs.SD},
      url={https://arxiv.org/abs/2309.14405}, 
}

@misc{sakshi2024mmaumassivemultitaskaudio,
      title={MMAU: A Massive Multi-Task Audio Understanding and Reasoning Benchmark}, 
      author={S Sakshi and Utkarsh Tyagi and Sonal Kumar and Ashish Seth and Ramaneswaran Selvakumar and Oriol Nieto and Ramani Duraiswami and Sreyan Ghosh and Dinesh Manocha},
      year={2024},
      eprint={2410.19168},
      archivePrefix={arXiv},
      primaryClass={eess.AS},
      url={https://arxiv.org/abs/2410.19168}, 
}

@misc{ma2025mmarchallengingbenchmarkdeep,
      title={MMAR: A Challenging Benchmark for Deep Reasoning in Speech, Audio, Music, and Their Mix}, 
      author={Ziyang Ma and Yinghao Ma and Yanqiao Zhu and Chen Yang and Yi-Wen Chao and Ruiyang Xu and Wenxi Chen and Yuanzhe Chen and Zhuo Chen and Jian Cong and Kai Li and Keliang Li and Siyou Li and Xinfeng Li and Xiquan Li and Zheng Lian and Yuzhe Liang and Minghao Liu and Zhikang Niu and Tianrui Wang and Yuping Wang and Yuxuan Wang and Yihao Wu and Guanrou Yang and Jianwei Yu and Ruibin Yuan and Zhisheng Zheng and Ziya Zhou and Haina Zhu and Wei Xue and Emmanouil Benetos and Kai Yu and Eng-Siong Chng and Xie Chen},
      year={2025},
      eprint={2505.13032},
      archivePrefix={arXiv},
      primaryClass={cs.SD},
      url={https://arxiv.org/abs/2505.13032}, 
}

@misc{wang2026mmsumassivemultitaskspoken,
      title={MMSU: A Massive Multi-task Spoken Language Understanding and Reasoning Benchmark}, 
      author={Dingdong Wang and Junan Li and Jincenzi Wu and Dongchao Yang and Xueyuan Chen and Tianhua Zhang and Helen Meng},
      year={2026},
      eprint={2506.04779},
      archivePrefix={arXiv},
      primaryClass={cs.CL},
      url={https://arxiv.org/abs/2506.04779}, 
}

@misc{yu2019activitynetqadatasetunderstandingcomplex,
      title={ActivityNet-QA: A Dataset for Understanding Complex Web Videos via Question Answering}, 
      author={Zhou Yu and Dejing Xu and Jun Yu and Ting Yu and Zhou Zhao and Yueting Zhuang and Dacheng Tao},
      year={2019},
      eprint={1906.02467},
      archivePrefix={arXiv},
      primaryClass={cs.CV},
      url={https://arxiv.org/abs/1906.02467}, 
}

@misc{wu2024longvideobenchbenchmarklongcontextinterleaved,
      title={LongVideoBench: A Benchmark for Long-context Interleaved Video-Language Understanding}, 
      author={Haoning Wu and Dongxu Li and Bei Chen and Junnan Li},
      year={2024},
      eprint={2407.15754},
      archivePrefix={arXiv},
      primaryClass={cs.CV},
      url={https://arxiv.org/abs/2407.15754}, 
}

@misc{hong2026worldsenseevaluatingrealworldomnimodal,
      title={WorldSense: Evaluating Real-world Omnimodal Understanding for Multimodal LLMs}, 
      author={Jack Hong and Shilin Yan and Jiayin Cai and Xiaolong Jiang and Yao Hu and Weidi Xie},
      year={2026},
      eprint={2502.04326},
      archivePrefix={arXiv},
      primaryClass={cs.CV},
      url={https://arxiv.org/abs/2502.04326}, 
}

@misc{zhou2026dailyomniaudiovisualreasoningtemporal,
      title={Daily-Omni: Towards Audio-Visual Reasoning with Temporal Alignment across Modalities}, 
      author={Ziwei Zhou and Rui Wang and Zuxuan Wu and Yu-Gang Jiang},
      year={2026},
      eprint={2505.17862},
      archivePrefix={arXiv},
      primaryClass={cs.AI},
      url={https://arxiv.org/abs/2505.17862}, 
}

@misc{wang2024qwen2vlenhancingvisionlanguagemodels,
      title={Qwen2-VL: Enhancing Vision-Language Model's Perception of the World at Any Resolution}, 
      author={Peng Wang and Shuai Bai and Sinan Tan and Shijie Wang and Zhihao Fan and Jinze Bai and Keqin Chen and Xuejing Liu and Jialin Wang and Wenbin Ge and Yang Fan and Kai Dang and Mengfei Du and Xuancheng Ren and Rui Men and Dayiheng Liu and Chang Zhou and Jingren Zhou and Junyang Lin},
      year={2024},
      eprint={2409.12191},
      archivePrefix={arXiv},
      primaryClass={cs.CV},
      url={https://arxiv.org/abs/2409.12191}, 
}

@misc{chen2024florencevlenhancingvisionlanguagemodels,
      title={Florence-VL: Enhancing Vision-Language Models with Generative Vision Encoder and Depth-Breadth Fusion}, 
      author={Jiuhai Chen and Jianwei Yang and Haiping Wu and Dianqi Li and Jianfeng Gao and Tianyi Zhou and Bin Xiao},
      year={2024},
      eprint={2412.04424},
      archivePrefix={arXiv},
      primaryClass={cs.CV},
      url={https://arxiv.org/abs/2412.04424}, 
}

@misc{chen2024internvlscalingvisionfoundation,
      title={InternVL: Scaling up Vision Foundation Models and Aligning for Generic Visual-Linguistic Tasks}, 
      author={Zhe Chen and Jiannan Wu and Wenhai Wang and Weijie Su and Guo Chen and Sen Xing and Muyan Zhong and Qinglong Zhang and Xizhou Zhu and Lewei Lu and Bin Li and Ping Luo and Tong Lu and Yu Qiao and Jifeng Dai},
      year={2024},
      eprint={2312.14238},
      archivePrefix={arXiv},
      primaryClass={cs.CV},
      url={https://arxiv.org/abs/2312.14238}, 
}

@misc{chen2024longvilascalinglongcontextvisual,
      title={LongVILA: Scaling Long-Context Visual Language Models for Long Videos}, 
      author={Yukang Chen and Fuzhao Xue and Dacheng Li and Qinghao Hu and Ligeng Zhu and Xiuyu Li and Yunhao Fang and Haotian Tang and Shang Yang and Zhijian Liu and Ethan He and Hongxu Yin and Pavlo Molchanov and Jan Kautz and Linxi Fan and Yuke Zhu and Yao Lu and Song Han},
      year={2024},
      eprint={2408.10188},
      archivePrefix={arXiv},
      primaryClass={cs.CV},
      url={https://arxiv.org/abs/2408.10188}, 
}

@misc{liu2025nvilaefficientfrontiervisual,
      title={NVILA: Efficient Frontier Visual Language Models}, 
      author={Zhijian Liu and Ligeng Zhu and Baifeng Shi and Zhuoyang Zhang and Yuming Lou and Shang Yang and Haocheng Xi and Shiyi Cao and Yuxian Gu and Dacheng Li and Xiuyu Li and Yunhao Fang and Yukang Chen and Cheng-Yu Hsieh and De-An Huang and An-Chieh Cheng and Vishwesh Nath and Jinyi Hu and Sifei Liu and Ranjay Krishna and Daguang Xu and Xiaolong Wang and Pavlo Molchanov and Jan Kautz and Hongxu Yin and Song Han and Yao Lu},
      year={2025},
      eprint={2412.04468},
      archivePrefix={arXiv},
      primaryClass={cs.CV},
      url={https://arxiv.org/abs/2412.04468}, 
}

@misc{lin2024vilapretrainingvisuallanguage,
      title={VILA: On Pre-training for Visual Language Models}, 
      author={Ji Lin and Hongxu Yin and Wei Ping and Yao Lu and Pavlo Molchanov and Andrew Tao and Huizi Mao and Jan Kautz and Mohammad Shoeybi and Song Han},
      year={2024},
      eprint={2312.07533},
      archivePrefix={arXiv},
      primaryClass={cs.CV},
      url={https://arxiv.org/abs/2312.07533}, 
}

@misc{li2025baichuanaudiounifiedframeworkendtoend,
      title={Baichuan-Audio: A Unified Framework for End-to-End Speech Interaction}, 
      author={Tianpeng Li and Jun Liu and Tao Zhang and Yuanbo Fang and Da Pan and Mingrui Wang and Zheng Liang and Zehuan Li and Mingan Lin and Guosheng Dong and Jianhua Xu and Haoze Sun and Zenan Zhou and Weipeng Chen},
      year={2025},
      eprint={2502.17239},
      archivePrefix={arXiv},
      primaryClass={cs.CL},
      url={https://arxiv.org/abs/2502.17239}, 
}

@misc{deshmukh2024pengiaudiolanguagemodel,
      title={Pengi: An Audio Language Model for Audio Tasks}, 
      author={Soham Deshmukh and Benjamin Elizalde and Rita Singh and Huaming Wang},
      year={2024},
      eprint={2305.11834},
      archivePrefix={arXiv},
      primaryClass={eess.AS},
      url={https://arxiv.org/abs/2305.11834}, 
}

@misc{chu2023qwenaudioadvancinguniversalaudio,
      title={Qwen-Audio: Advancing Universal Audio Understanding via Unified Large-Scale Audio-Language Models}, 
      author={Yunfei Chu and Jin Xu and Xiaohuan Zhou and Qian Yang and Shiliang Zhang and Zhijie Yan and Chang Zhou and Jingren Zhou},
      year={2023},
      eprint={2311.07919},
      archivePrefix={arXiv},
      primaryClass={eess.AS},
      url={https://arxiv.org/abs/2311.07919}, 
}

@misc{kong2024audioflamingonovelaudio,
      title={Audio Flamingo: A Novel Audio Language Model with Few-Shot Learning and Dialogue Abilities}, 
      author={Zhifeng Kong and Arushi Goel and Rohan Badlani and Wei Ping and Rafael Valle and Bryan Catanzaro},
      year={2024},
      eprint={2402.01831},
      archivePrefix={arXiv},
      primaryClass={cs.SD},
      url={https://arxiv.org/abs/2402.01831}, 
}

@misc{ghosh2025audioflamingo2audiolanguage,
      title={Audio Flamingo 2: An Audio-Language Model with Long-Audio Understanding and Expert Reasoning Abilities}, 
      author={Sreyan Ghosh and Zhifeng Kong and Sonal Kumar and S Sakshi and Jaehyeon Kim and Wei Ping and Rafael Valle and Dinesh Manocha and Bryan Catanzaro},
      year={2025},
      eprint={2503.03983},
      archivePrefix={arXiv},
      primaryClass={cs.SD},
      url={https://arxiv.org/abs/2503.03983}, 
}

@inproceedings{panayotov2015librispeech,
  title     = {Librispeech: An ASR corpus based on public domain audio books},
  author    = {Panayotov, Vassil and Chen, Guoguo and Povey, Daniel and Khudanpur, Sanjeev},
  booktitle = {2015 IEEE International Conference on Acoustics, Speech and Signal Processing (ICASSP)},
  pages     = {5206--5210},
  year      = {2015},
  organization = {IEEE},
  doi       = {10.1109/ICASSP.2015.7178964}
}

@misc{oneill2021spgispeech5000hourstranscribed,
      title={SPGISpeech: 5,000 hours of transcribed financial audio for fully formatted end-to-end speech recognition}, 
      author={Patrick K. O'Neill and Vitaly Lavrukhin and Somshubra Majumdar and Vahid Noroozi and Yuekai Zhang and Oleksii Kuchaiev and Jagadeesh Balam and Yuliya Dovzhenko and Keenan Freyberg and Michael D. Shulman and Boris Ginsburg and Shinji Watanabe and Georg Kucsko},
      year={2021},
      eprint={2104.02014},
      archivePrefix={arXiv},
      primaryClass={cs.CL},
      url={https://arxiv.org/abs/2104.02014}, 
}

@inbook{Hernandez_2018,
   title={TED-LIUM 3: Twice as Much Data and Corpus Repartition for Experiments on Speaker Adaptation},
   ISBN={9783319995793},
   ISSN={1611-3349},
   url={http://dx.doi.org/10.1007/978-3-319-99579-3_21},
   DOI={10.1007/978-3-319-99579-3_21},
   booktitle={Speech and Computer},
   publisher={Springer International Publishing},
   author={Hernandez, François and Nguyen, Vincent and Ghannay, Sahar and Tomashenko, Natalia and Estève, Yannick},
   year={2018},
   pages={198–208} }

@inproceedings{Chen_2021,
   title={GigaSpeech: An Evolving, Multi-Domain ASR Corpus with 10,000 Hours of Transcribed Audio},
   url={http://dx.doi.org/10.21437/Interspeech.2021-1965},
   DOI={10.21437/interspeech.2021-1965},
   booktitle={Interspeech 2021},
   publisher={ISCA},
   author={Chen, Guoguo and Chai, Shuzhou and Wang, Guan-Bo and Du, Jiayu and Zhang, Wei-Qiang and Weng, Chao and Su, Dan and Povey, Daniel and Trmal, Jan and Zhang, Junbo and Jin, Mingjie and Khudanpur, Sanjeev and Watanabe, Shinji and Zhao, Shuaijiang and Zou, Wei and Li, Xiangang and Yao, Xuchen and Wang, Yongqing and You, Zhao and Yan, Zhiyong},
   year={2021},
   month=aug,
   pages={3670--3674}
}

@misc{ghosh2026audioflamingonextnextgeneration,
      title={Audio Flamingo Next: Next-Generation Open Audio-Language Models for Speech, Sound, and Music}, 
      author={Sreyan Ghosh and Arushi Goel and Kaousheik Jayakumar and Lasha Koroshinadze and Nishit Anand and Zhifeng Kong and Siddharth Gururani and Sang-gil Lee and Jaehyeon Kim and Aya Aljafari and Chao-Han Huck Yang and Sungwon Kim and Ramani Duraiswami and Dinesh Manocha and Mohammad Shoeybi and Bryan Catanzaro and Ming-Yu Liu and Wei Ping},
      year={2026},
      eprint={2604.10905},
      archivePrefix={arXiv},
      primaryClass={cs.SD},
      url={https://arxiv.org/abs/2604.10905}, 
}

@misc{kumar2026videomarketing,
  author       = {Kumar, Naveen},
  title        = {93 Video Marketing Statistics 2026 [Latest Data \& Trends]},
  howpublished = {\url{https://www.demandsage.com/video-marketing-statistics/}},
  year         = {2026},
  month        = apr,
  note         = {DemandSage. Published April 9, 2026. Accessed April 14, 2026}
}

@article{lucas1992sound,
  author  = {Lucas, George},
  title   = {{Sound is half the experience of seeing a film}},
  journal = {The New York Times},
  year    = {1992},
  month   = may,
  day     = {3},
  note    = {Quoted in C. Hodenfield, ``In the Action With `Star Wars' Sound,'' Section 2, p. 24}
}

@misc{gong2024avodysseybenchmultimodalllms,
      title={AV-Odyssey Bench: Can Your Multimodal LLMs Really Understand Audio-Visual Information?}, 
      author={Kaixiong Gong and Kaituo Feng and Bohao Li and Yibing Wang and Mofan Cheng and Shijia Yang and Jiaming Han and Benyou Wang and Yutong Bai and Zhuoran Yang and Xiangyu Yue},
      year={2024},
      eprint={2412.02611},
      archivePrefix={arXiv},
      primaryClass={cs.CV},
      url={https://arxiv.org/abs/2412.02611}, 
}

@misc{comanici2025gemini25pushingfrontier,
      title={Gemini 2.5: Pushing the Frontier with Advanced Reasoning, Multimodality, Long Context, and Next Generation Agentic Capabilities}, 
      author={Gemini Team},
      year={2025},
      eprint={2507.06261},
      archivePrefix={arXiv},
      primaryClass={cs.CL},
      url={https://arxiv.org/abs/2507.06261}, 
}

@misc{galvez2021peoplesspeechlargescalediverse,
      title={The People's Speech: A Large-Scale Diverse English Speech Recognition Dataset for Commercial Usage}, 
      author={Daniel Galvez and Greg Diamos and Juan Ciro and Juan Felipe Cerón and Keith Achorn and Anjali Gopi and David Kanter and Maximilian Lam and Mark Mazumder and Vijay Janapa Reddi},
      year={2021},
      eprint={2111.09344},
      archivePrefix={arXiv},
      primaryClass={cs.LG},
      url={https://arxiv.org/abs/2111.09344}, 
}

@misc{miech2019howto100mlearningtextvideoembedding,
      title={HowTo100M: Learning a Text-Video Embedding by Watching Hundred Million Narrated Video Clips}, 
      author={Antoine Miech and Dimitri Zhukov and Jean-Baptiste Alayrac and Makarand Tapaswi and Ivan Laptev and Josef Sivic},
      year={2019},
      eprint={1906.03327},
      archivePrefix={arXiv},
      primaryClass={cs.CV},
      url={https://arxiv.org/abs/1906.03327}, 
}

@misc{bain2022frozentimejointvideo,
      title={Frozen in Time: A Joint Video and Image Encoder for End-to-End Retrieval}, 
      author={Max Bain and Arsha Nagrani and Gül Varol and Andrew Zisserman},
      year={2022},
      eprint={2104.00650},
      archivePrefix={arXiv},
      primaryClass={cs.CV},
      url={https://arxiv.org/abs/2104.00650}, 
}

@misc{chen2024panda70mcaptioning70mvideos,
      title={Panda-70M: Captioning 70M Videos with Multiple Cross-Modality Teachers}, 
      author={Tsai-Shien Chen and Aliaksandr Siarohin and Willi Menapace and Ekaterina Deyneka and Hsiang-wei Chao and Byung Eun Jeon and Yuwei Fang and Hsin-Ying Lee and Jian Ren and Ming-Hsuan Yang and Sergey Tulyakov},
      year={2024},
      eprint={2402.19479},
      archivePrefix={arXiv},
      primaryClass={cs.CV},
      url={https://arxiv.org/abs/2402.19479}, 
}

@misc{fu2025vitaopensourceinteractiveomni,
      title={VITA: Towards Open-Source Interactive Omni Multimodal LLM}, 
      author={Chaoyou Fu and Haojia Lin and Zuwei Long and Yunhang Shen and Yuhang Dai and Meng Zhao and Yi-Fan Zhang and Shaoqi Dong and Yangze Li and Xiong Wang and Haoyu Cao and Di Yin and Long Ma and Xiawu Zheng and Rongrong Ji and Yunsheng Wu and Ran He and Caifeng Shan and Xing Sun},
      year={2025},
      eprint={2408.05211},
      archivePrefix={arXiv},
      primaryClass={cs.CV},
      url={https://arxiv.org/abs/2408.05211}, 
}

@article{liu2025ola,
  title={Ola: Pushing the frontiers of omni-modal language model},
  author={Liu, Zuyan and Dong, Yuhao and Wang, Jiahui and Liu, Ziwei and Hu, Winston and Lu, Jiwen and Rao, Yongming},
  journal={arXiv preprint arXiv:2502.04328},
  year={2025}
}

@misc{geng2025longvalevisionaudiolanguageeventbenchmarktimeaware,
      title={LongVALE: Vision-Audio-Language-Event Benchmark Towards Time-Aware Omni-Modal Perception of Long Videos}, 
      author={Tiantian Geng and Jinrui Zhang and Qingni Wang and Teng Wang and Jinming Duan and Feng Zheng},
      year={2025},
      eprint={2411.19772},
      archivePrefix={arXiv},
      primaryClass={cs.CV},
      url={https://arxiv.org/abs/2411.19772}, 
}

@article{schuhmann2022laion,
  title={Laion-5b: An open large-scale dataset for training next generation image-text models},
  author={Schuhmann, Christoph and Beaumont, Romain and Vencu, Richard and Gordon, Cade and Wightman, Ross and Cherti, Mehdi and Coombes, Theo and Katta, Aarush and Mullis, Clayton and Wortsman, Mitchell and others},
  journal={Advances in neural information processing systems},
  volume={35},
  pages={25278--25294},
  year={2022}
}

@article{zhang2024llava,
  title={Llava-video: Video instruction tuning with synthetic data},
  author={Zhang, Yuanhan and Wu, Jinming and Li, Wei and Li, Bo and Ma, Zejun and Liu, Ziwei and Li, Chunyuan},
  journal={arXiv preprint arXiv:2410.02713},
  year={2024}
}

@InProceedings{Chen20,
  author       = "Honglie Chen and Weidi Xie and Andrea Vedaldi and Andrew Zisserman",
  title        = "VGGSound: A Large-scale Audio-Visual Dataset",
  booktitle    = "International Conference on Acoustics, Speech, and Signal Processing (ICASSP)",
  year         = "2020",
}

@misc{huang2023audiogptunderstandinggeneratingspeech,
      title={AudioGPT: Understanding and Generating Speech, Music, Sound, and Talking Head}, 
      author={Rongjie Huang and Mingze Li and Dongchao Yang and Jiatong Shi and Xuankai Chang and Zhenhui Ye and Yuning Wu and Zhiqing Hong and Jiawei Huang and Jinglin Liu and Yi Ren and Zhou Zhao and Shinji Watanabe},
      year={2023},
      eprint={2304.12995},
      archivePrefix={arXiv},
      primaryClass={cs.CL},
      url={https://arxiv.org/abs/2304.12995}, 
}

@article{fang2024usp,
  title={Usp: A unified sequence parallelism approach for long context generative ai},
  author={Fang, Jiarui and Zhao, Shangchun},
  journal={arXiv preprint arXiv:2405.07719},
  year={2024}
}

@article{jacobs2023deepspeed,
  title={Deepspeed ulysses: System optimizations for enabling training of extreme long sequence transformer models},
  author={Jacobs, Sam Ade and Tanaka, Masahiro and Zhang, Chengming and Zhang, Minjia and Song, Shuaiwen Leon and Rajbhandari, Samyam and He, Yuxiong},
  journal={arXiv preprint arXiv:2309.14509},
  year={2023}
}

@article{liu2023ring,
  title={Ring attention with blockwise transformers for near-infinite context},
  author={Liu, Hao and Zaharia, Matei and Abbeel, Pieter},
  journal={arXiv preprint arXiv:2310.01889},
  year={2023}
}

@misc{li2024llavaonevisioneasyvisualtask,
      title={LLaVA-OneVision: Easy Visual Task Transfer}, 
      author={Bo Li and Yuanhan Zhang and Dong Guo and Renrui Zhang and Feng Li and Hao Zhang and Kaichen Zhang and Peiyuan Zhang and Yanwei Li and Ziwei Liu and Chunyuan Li},
      year={2024},
      eprint={2408.03326},
      archivePrefix={arXiv},
      primaryClass={cs.CV},
      url={https://arxiv.org/abs/2408.03326}, 
}

@misc{kumar2026tactimestampedaudiocaptioning,
      title={TAC: Timestamped Audio Captioning}, 
      author={Sonal Kumar and Prem Seetharaman and Ke Chen and Oriol Nieto and Jiaqi Su and Zhepei Wang and Rithesh Kumar and Dinesh Manocha and Nicholas J. Bryan and Zeyu Jin and Justin Salamon},
      year={2026},
      eprint={2602.15766},
      archivePrefix={arXiv},
      primaryClass={cs.SD},
      url={https://arxiv.org/abs/2602.15766}, 
}

@misc{tian2018audiovisualeventlocalizationunconstrained,
      title={Audio-Visual Event Localization in Unconstrained Videos}, 
      author={Yapeng Tian and Jing Shi and Bochen Li and Zhiyao Duan and Chenliang Xu},
      year={2018},
      eprint={1803.08842},
      archivePrefix={arXiv},
      primaryClass={cs.CV},
      url={https://arxiv.org/abs/1803.08842}, 
}

@inproceedings{rasley2020deepspeed,
  title={Deepspeed: System optimizations enable training deep learning models with over 100 billion parameters},
  author={Rasley, Jeff and Rajbhandari, Samyam and Ruwase, Olatunji and He, Yuxiong},
  booktitle={Proceedings of the 26th ACM SIGKDD international conference on knowledge discovery \& data mining},
  pages={3505--3506},
  year={2020}
}

@inproceedings{chowdhury2024meerkat,
  title={Meerkat: Audio-visual large language model for grounding in space and time},
  author={Chowdhury, Sanjoy and Nag, Sayan and Dasgupta, Subhrajyoti and Chen, Jun and Elhoseiny, Mohamed and Gao, Ruohan and Manocha, Dinesh},
  booktitle={European Conference on Computer Vision},
  pages={52--70},
  year={2024},
  organization={Springer}
}

@InProceedings{selvakumar2026avllm,
    author    = {Selvakumar, Ramaneswaran and Jayakumar, Kaousheik and Sakshi, S and Ghosh, Sreyan and Gao, Ruohan and Manocha, Dinesh},
    title     = {Do Audio-Visual Large Language Models Really See and Hear?},
    booktitle = {Proceedings of the IEEE/CVF Conference on Computer Vision and Pattern Recognition (CVPR) Findings},
    month     = {June},
    year      = {2026},
    pages     = {5892--5902}
}

@InProceedings{Chowdhury_2025_ICCV,
    author    = {Chowdhury, Sanjoy and Gani, Hanan and Anand, Nishit and Nag, Sayan and Gao, Ruohan and Elhoseiny, Mohamed and Khan, Salman and Manocha, Dinesh},
    title     = {AURELIA: Test-time Reasoning Distillation in Audio-Visual LLMs},
    booktitle = {Proceedings of the IEEE/CVF International Conference on Computer Vision (ICCV)},
    month     = {October},
    year      = {2025},
    pages     = {22899-22910}
}

@inproceedings{seth-etal-2025-egoillusion,
    title = "{EGOILLUSION}: Benchmarking Hallucinations in Egocentric Video Understanding",
    author = "Seth, Ashish  and
      Tyagi, Utkarsh  and
      Selvakumar, Ramaneswaran  and
      Anand, Nishit  and
      Kumar, Sonal  and
      Ghosh, Sreyan  and
      Duraiswami, Ramani  and
      Agarwal, Chirag  and
      Manocha, Dinesh",
    editor = "Christodoulopoulos, Christos  and
      Chakraborty, Tanmoy  and
      Rose, Carolyn  and
      Peng, Violet",
    booktitle = "Proceedings of the 2025 Conference on Empirical Methods in Natural Language Processing",
    month = nov,
    year = "2025",
    address = "Suzhou, China",
    publisher = "Association for Computational Linguistics",
    url = "https://aclanthology.org/2025.emnlp-main.1446/",
    doi = "10.18653/v1/2025.emnlp-main.1446",
    pages = "28461--28480",
    ISBN = "979-8-89176-332-6"
}

@inproceedings{selvakumar-etal-2025-multivox,
    title = "{MULTIVOX}: A Benchmark for Evaluating Voice Assistants for Multimodal Interactions",
    author = "Selvakumar, Ramaneswaran  and
      Seth, Ashish  and
      Anand, Nishit  and
      Tyagi, Utkarsh  and
      Kumar, Sonal  and
      Ghosh, Sreyan  and
      Manocha, Dinesh",
    editor = "Christodoulopoulos, Christos  and
      Chakraborty, Tanmoy  and
      Rose, Carolyn  and
      Peng, Violet",
    booktitle = "Proceedings of the 2025 Conference on Empirical Methods in Natural Language Processing",
    month = nov,
    year = "2025",
    address = "Suzhou, China",
    publisher = "Association for Computational Linguistics",
    url = "https://aclanthology.org/2025.emnlp-main.1447/",
    doi = "10.18653/v1/2025.emnlp-main.1447",
    pages = "28481--28493",
    ISBN = "979-8-89176-332-6"
}

\appendix

\section{Audio-Visual Flamingo Training Datasets}
\label{sec.avf_training_data}

Table~\ref{tab:dataset-details} summarizes all datasets used to train Audio-Visual Flamingo, including total hours, number of training pairs, and the number of epochs (passes over the dataset) used at each training stage.
Similar to ~\citep{ye2025omnivinci, goel2025audio}, we convert all foundational datasets (captioning, classification, etc.) into QA formats, using the same set of prompts for each task.

\noindent
\textbf{AV-Safety QA.} In addition to the skill-focused subsets, Table~\ref{tab:dataset-details} includes AV-Safety QA, a set of 92K question--answer pairs over 536 hours of video used during long-context SFT. This subset pairs audio-visual inputs with questions that probe unsafe, harmful, or policy-violating requests (e.g., requests to identify private individuals, generate harassing descriptions, or extract sensitive information from the audio-visual content), together with target responses that safely decline or redirect while remaining helpful on the benign portion of the request. Its purpose is to preserve safety alignment while the model acquires long-context audio-visual capabilities.

\begin{table*}[h]
    \centering
    \caption{\small List of pre-training and fine-tuning datasets together with their training composition.}
    \resizebox{\textwidth}{!}{
    \begin{tabular}{lccccc}
    \toprule
Dataset & Hours  & Num. Instances & AVF-Short-SFT & AVF-Long-SFT & CoT-Training \\ \midrule
        AV-Skills-Short (Ours) & 100k hrs & 3.8M   & 1.0 & 0.5 & -   \\
        Long AV Captioning (Ours) & 51k hrs & 1.2M   & - & 2.0 & -  \\
        Long Comparative AVQA (Ours) &  8.3K hrs& 169K      & - & 2.0 & -  \\
      Long Context AVQA (Ours)& 8.3K hrs  & 175K  &  - & 2.0 & -    \\
    Long Subscene AVQA (Ours)& 6.8K hrs  & 157K  & - & 2.0 & -  \\
      Long Counting AVQA (Ours)&6.2K hrs   & 136K  & - & 2.0 & -  \\
      Long Needle AVQA (Ours)& 8K hrs  & 169K  &  - & 2.0 & -  \\
      Long Referring AVQA (Ours)& 7.2K hrs  & 160K  &  - & 2.0 & -   \\
      Long Topic AVQA (Ours)& 7.3K hrs   & 172K  &  - & 2.0 & - \\
      Long Holistic Reas. AVQA (Ours)& 8.9K hrs  & 180K  & - & 2.0 & -  \\
      Long Event Alignment AVQA (Ours)& 7.9K hrs  &160K  &  - & 2.0 & -   \\
      Long Event Sequence AVQA (Ours)& 7.4K hrs  &153K  &  - & 2.0 & -    \\
      Long Inference AVQA (Ours)&4.5K  hrs &  63K  & - & 2.0 & -   \\
      Long Temporal AVQA (Ours)& 9K  hrs &  200K  & - & 2.0 & -  \\
      AV-Safety QA (Ours) & 536 hrs & 92K   & - & 2.0 & -  \\
        AF3-training mix~\citep{goel2025audio} & 308k hrs & 48.6M & 1.0 & - & -  \\
        AudioSkills-XL ~\citep{goel2025audio} & - & 9700K & 1.0 & 0.5 & - \\
        MF-Skills~\citep{ghosh2025music} & - & 3M & 1.0 & 0.5 & -  \\
        Image-text~\citep{ye2025omnivinci} & - & 3M & 0.5 & - & -  \\
        Video-text~\citep{ye2025omnivinci} & - & 3M & 0.5 & - & - \\


        CHiME~\citep{vincent2013second} & 342 hrs & 30k & $1.0$ & 1.0 &-\\
        AliMeeting~\citep{Yu2022M2MeT} & 118.75 hrs & 387k & $1.0$ & 1.0 &- \\
        EMILIA~\citep{he2024emilia} & 5000 hrs & 1.7M & $1.0$ & 1.0 &- \\
        MuST-C~\citep{cattoni2021must} & 500 hrs & 245k & $1.0$ & 1.0 &-  \\
        CoVoST~\citep{wang2020covost} & 2880 hrs & 5M & $1.0$ & 1.0 &-\\
        Aidatatang~\citep{aidatatang_200zh} & 139.39 hrs  & 165k & $1.0$ & 1.0 &- \\
        Aishell-1~\citep{aishell_2017} & 150.85 hrs & 120k & $1.0$ & 1.0 &- \\
        Multi-talker Switchboard~\citep{godfrey1993switchboard} & 109.9 hrs & 76.6K & $1.0$ & 1.0 & -\\
        Audio-Think-Time (Ours) & 3954 hrs & 43K & - &  2.0 & -  \\
        Video-Think-Time (Ours) & 1238 hrs & 22K & - &  2.0&  -\\
        AV-Think (Ours) & 1500 hrs & 24K  & - &  - &  2.0 \\

    \bottomrule
    \end{tabular}}

    \label{tab:dataset-details}
\end{table*}

\section{Dataset Examples and Category Distribution}
\label{app:dataset_examples}
\label{sec.dataset_examples}
\begin{figure*}[t]
  \centering
  \includegraphics[width=\textwidth]{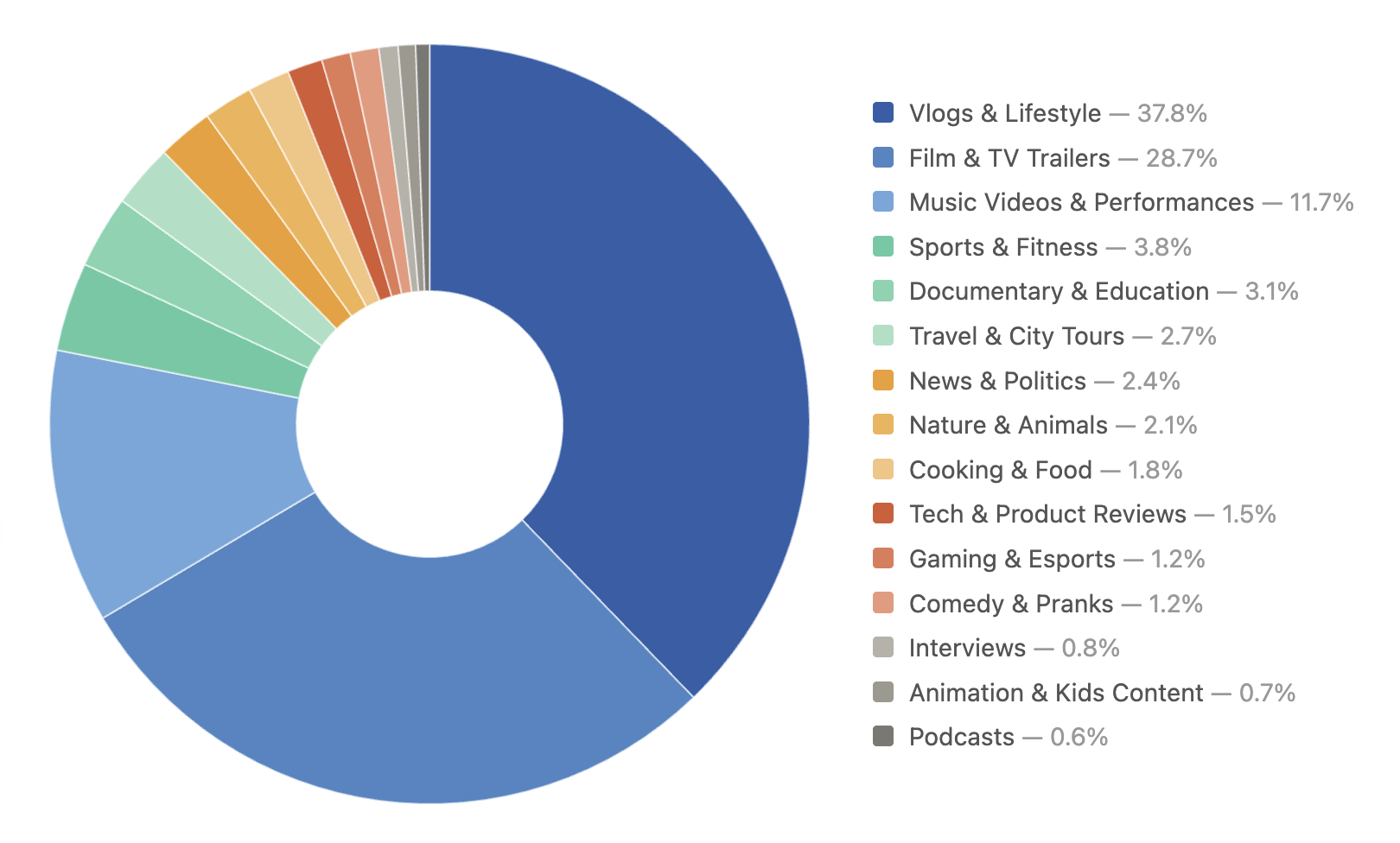}
  \caption{\small Category distribution of the videos used in the training datasets.}
  \label{fig:data_categories}
\vspace{-2mm}
\end{figure*}

\begin{figure*}[t]
  \centering
  \includegraphics[width=\textwidth]{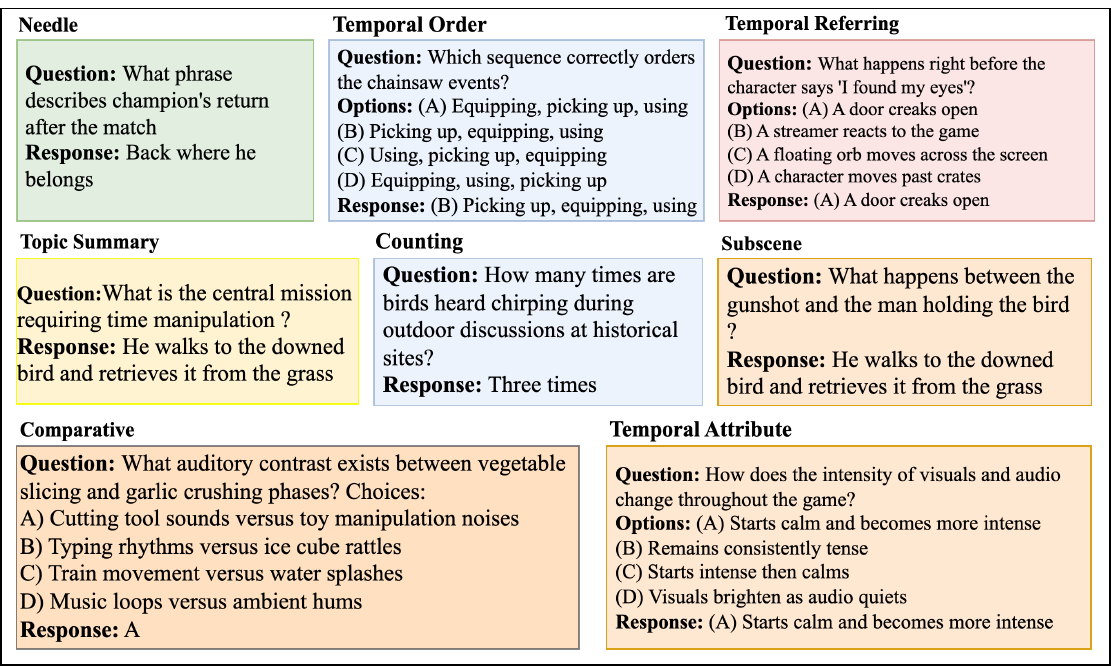}
\caption{\small \textbf{Additional QA pair examples from AV-Skills-Long.} Representative samples spanning eight long-form skill categories used during mid-training: Needle-in-the-Haystack, Temporal Order, Temporal Referring, Topic Summary, Counting, Sub-scene, Comparative, and Temporal Attribute. Examples include both open-ended and multiple-choice formats.}

  \label{fig:examples_appendix}
\vspace{-2mm}
\end{figure*}

Figure~\ref{fig:data_categories} shows the category distribution of the 15 major video categories used for training. Figure~\ref{fig:examples_appendix} and Table~\ref{tab:avf-short-dataset-examples} provide representative AV-Skills-Long and AV-Skills-Short examples.

\begin{table*}[h]
    \centering
    \caption{\small Examples of AV-Skills-Short training data}
    \resizebox{\textwidth}{!}{
    \begin{tabular}{p{2.8cm} | p{11cm} | p{5cm}}
    \toprule
    \textbf{Category} & \textbf{Prompt} & \textbf{Response} \\
    \midrule

    Causal Reasoning & Why did the person cut the cake roll after attempting to roll it? & Because when rolling the cake, the cream overflowed and since others disagreed with eating the overflowed portion, the person decided to cut the cake instead of continuing to roll it. \\
    \midrule

    Spatial Perception & During the scene when room with wooden floor, large TV, person on floor with dog, where is the small fluffy dog named Lala located relative to the person wearing star-patterned pajamas playing with dog? \newline A) Lala is in front of the person \newline B) Lala is under the person's legs \newline C) Lala is on the person's lap \newline D) Lala is standing beside the person & C) Lala is on the person's lap \\
    \midrule

    Emotion Change & What is emotion / mood change of the woman with long dark hair, glasses, and a sweater who sits at the right side of the panel during the panel discussion on child-abuse reporting when the question appears and a comic-style bubble with `wae?' (why?) pops up? \newline A) surprised \newline B) confused \newline C) amused \newline D) curious & A) surprised \\
    \midrule

    Hallucination & Does Mickey Mouse wear a red hat in the video? \newline A. I am not sure. \newline B. No, Mickey does not wear a red hat. \newline C. None of the above. \newline D. Yes, Mickey is wearing a red hat. & B. No, Mickey does not wear a red hat. \\
    \midrule

    Audio Counting & How many times does a sniffle occur in the audio? \newline A. 0 \newline B. 2 \newline C. 3 \newline D. 1 & D. 1 \\
    \midrule

    Video Counting & How many times does the hand adjust a pressure switch? \newline A. Five \newline B. Six \newline C. Four \newline D. Three & C. Four \\
    \midrule

    Temporal Reasoning & Arrange the below list of audio-visual events in the order of their occurrence in the video. Use arrow `$\rightarrow$' to indicate the order of the events. \newline A) Speaker discussing surface preparation while gloved hands open a white tube of `CHASER', placed next to an open can of Bondo filler. \newline B) Speaker mentions applying filler like Bondo while a person applies purple filler to a hairdryer. \newline C) Speaker mentions `400 grit sandpaper' as text appears on screen while a person holds a sanding block. \newline D) A person wearing gloves mixes Bondo and hardener as on-screen text explains the process. & A $\rightarrow$ D $\rightarrow$ B $\rightarrow$ C \\
    \midrule

    Relation Reasoning & During the scene when room with aquarium and instrumental electronic music, where is the man near the aquarium positioned relative to the man in the hoodie? \newline A) The man in the hoodie is behind the man in the t-shirt \newline B) The man in the hoodie is in front of the man in the t-shirt \newline C) The man in the hoodie is to the right of the man in the t-shirt \newline D) The man in the hoodie is to the left of the man in the t-shirt & C) The man in the hoodie is to the right of the man in the t-shirt \\

    \bottomrule
    \end{tabular}}
    \label{tab:avf-short-dataset-examples}
\end{table*}

\section{Audio-Visual Flamingo Training Details}
\label{sec.avf_training_details}
In this section, we present the training settings of our model across all stages, each with specific configurations. Details are in \cref{tab:blend_weights}; a compact stage-level recipe is summarized in \cref{tab:app_training_recipe}.

\begin{table}[t]
\centering
\caption{\small Training settings across stages.}
\vspace{2pt}
\resizebox{0.8\linewidth}{!}{
\begin{tabular}{lccc}
\toprule
\textbf{Settings} & \textbf{Pre-training} & \textbf{Mid-Training} & \textbf{Post-Training} \\
\midrule
global batch size & 128 & 128 & 64 \\
learning rate & 1e-5 & 1e-5 & 2e-5\\
learning schedule & \multicolumn{3}{c}{Cosine decay} \\
warm up ratio & \multicolumn{3}{c}{0.03}\\
weight decay & \multicolumn{3}{c}{0.0}\\
epoch & 1 & 1 & 2\\
bf16 & \checkmark & \checkmark & \checkmark \\
grad accumulate & \multicolumn{3}{c}{8} \\
Parallelism  & Zero-3 & Zero-3 + SP & Zero-3 + SP \\
GPUs & \multicolumn{3}{c}{512$\times$H100} \\
\bottomrule
\end{tabular}}

\label{tab:blend_weights}

\end{table}

\begin{table}[t]
\centering
\small
\caption{\small Compact training recipe for AV-Flamingo. Dataset-level mixing ratios are provided in Table~\ref{tab:dataset-details}.}
\vspace{2pt}
\resizebox{\linewidth}{!}{
\begin{tabular}{lllllll}
\toprule
\textbf{Stage} & \textbf{Main data mixture} & \textbf{Max input / context} & \textbf{Epochs} & \textbf{Batch} & \textbf{LR} & \textbf{Compute / parallelism} \\
\midrule
Short-context SFT & AV-Skills-Short + AF3/AudioSkills/MF + image/video/speech & 5 min / 16K & 1 & 128 & 1e-5 & 512$\times$H100; ZeRO-3 \\
Long-context SFT & AV-Skills-Long + downsampled short/audio/music/speech + think-time data & 15 min / 32K & 1 & 128 & 1e-5 & 512$\times$H100; ZeRO-3 + SP \\
CoT post-training & AV-Think reasoning triplets with SFT + GRPO & 15 min / 32K & 2 & 64 & 2e-5 & 512$\times$H100; ZeRO-3 + SP \\
\bottomrule
\end{tabular}}
\label{tab:app_training_recipe}
\end{table}

\section{Additional Ablation Study}

\noindent
\textbf{Effect of AV-Skills-Short and AV-Skills-Long.} This section isolates the contribution of the curated AV-Skills data across the training curriculum. We compare three models: (i) \textbf{OmniVinci}, the base checkpoint from which AVF is initialized; (ii) \textbf{AVF-Stage1}, our model after pre-training (Stage 1 + Stage 2) on AV-Skills-Short alongside unimodal data; and (iii) \textbf{AVF-Instruct}, the final model obtained after mid-training on AV-Skills-Long.

As shown in Table~\ref{tab:ablation-avskills}, training on AV-Skills-Short (AVF-Stage1) yields consistent improvements over OmniVinci across both audio-visual and video-only benchmarks, confirming that curated AV-Skills data injects cross-modal reasoning that cannot be captured by unimodal supervision alone. Adding AV-Skills-Long during mid-training (AVF-Instruct) further improves reasoning over extended temporal contexts and tighter audio-visual grounding.

\begin{table}[!ht]
\centering
\small
\caption{Ablation study on the effect of AV-Skills-Short and AV-Skills-Long. AVF-Stage1 isolates the contribution of AV-Skills-Short during pre-training, while AVF-Instruct adds AV-Skills-Long during mid-training. Higher is better.}
\vspace{2pt}
\begin{tabular}{lccc}
\toprule

\textbf{Model} & \textbf{DailyOmni} & \textbf{WorldSense} & \textbf{VideoMME} \\
\midrule
OmniVinci      & 66.5 & 48.2 & 67.3 \\
AVF-Stage1     & 69.5 & 48.5 & 68.5 \\
AVF-Instruct   & \textbf{72.4} & \textbf{50.3} & \textbf{70.7} \\
\bottomrule
\end{tabular}

\label{tab:ablation-avskills}
\end{table}

\section{GRPO-based RL training Details}
\label{sec:grpo-rl}
\textbf{GRPO for Audio-Visual Understanding and Reasoning.}
Building on recent advances in Group Relative Policy Optimization (GRPO), we adopt the standard GRPO framework to train AV-Flamingo. GRPO removes the need for an explicit value function and instead estimates the advantage using the average reward of multiple sampled outputs for the same input. For each input question or caption prompt $q$, the policy model generates a group of candidate outputs $\{o_1, o_2, \ldots, o_G\}$ from the old policy $\pi_{\theta_{\text{old}}}$, along with their corresponding rewards $\{r_1, r_2, \ldots, r_G\}$ computed using rule-based reward functions such as format, answer accuracy, and structured reward matching. The policy $\pi_{\theta}$ is then optimized using the GRPO objective:
\begin{align}
\mathcal{J}(\theta)\!\! = \!\!\mathbb{E}_{q,\{o_i\}} \!\!\left[\! \tfrac{1}{G} \!\sum_{i=1}^{G} \!\Big(\!\! \min \!\Big(\! \tfrac{\pi_\theta(o_i|q)}{\pi_{\theta_{\text{old}}}(o_i|q)} A_i, \, \text{clip}\!\big(\tfrac{\pi_\theta(o_i|q)}{\pi_{\theta_{\text{old}}}(o_i|q)},\! 1-\epsilon, \!1+\epsilon \big) A_i \Big)\! -\! \beta D_{\text{KL}}(\pi_\theta \,\|\, \pi_{\text{ref}}) \!\Big)\! \right]\!,
\label{eq:grpo_objective_avf}
\end{align}
where $\epsilon$ is the clipping range for the importance sampling ratio, $\beta$ is the coefficient for the KL regularization term that keeps the learned policy close to the reference policy, and $G$ denotes the group size, i.e., the number of sampled candidate outputs per input, which we set to $5$ in our experiments. To stabilize optimization, rewards are normalized within each group to compute the advantage:
\[
A_i=\frac{r_i-\mathrm{mean}(\{r_1,\ldots,r_G\})}{\mathrm{std}(\{r_1,\ldots,r_G\})}.
\]
We next describe the reward functions used for AV-Flamingo.

\noindent
\textbf{Format Reward.}
To encourage well-structured outputs, we use a regex-based format reward similar to prior reasoning work. The model is required to produce intermediate reasoning within \texttt{<think>} \texttt{</think>} tags, followed by the final response within \texttt{<answer>} \texttt{</answer>} tags. Outputs that strictly follow this format receive a reward of 1, and 0 otherwise. This binary reward encourages the model to consistently generate structured reasoning traces and final answers.

\noindent
\textbf{Accuracy Reward.}
For question-answering instances, we use an \textit{accuracy reward} based on the final answer extracted from the \texttt{<answer>} \texttt{</answer>} tags. Given a question and its ground-truth answer, the reward is computed by matching the normalized predicted answer with the normalized reference answer. This reward directly encourages correct final responses for audio-visual QA.

\noindent
\textbf{Structured Reward for Open-Ended Captions and Responses.}
For open-ended captioning and free-form response generation, strict string matching is not suitable due to the long-form and semantically diverse nature of valid outputs. To address this, we design a structured reward that first parses both the reference output and the generated output into structured JSON using an LLM, and then compares the resulting fields. Specifically, given a ground-truth caption or response, we prompt an LLM to extract salient structured metadata such as scene type, participants, environment, topic, events, emotional tone, and key conclusions. For example, for a podcast video discussing artificial intelligence, the parsed JSON may contain fields such as:
\begin{mdframed}[linewidth=1pt, linecolor=black, leftmargin=1pt, rightmargin=1pt, innerleftmargin=10pt, innerrightmargin=10pt, innertopmargin=4pt, innerbottommargin=2pt, backgroundcolor=gray!20, roundcorner=5pt]
\{``\textbf{Scene}'': podcast studio, ``\textbf{Participants}'': two individuals, speaker in light blue shirt; listener in black blazer, ``\textbf{Environment}'': professional studio, beige walls, microphones, ``\textbf{Topic}'': artificial intelligence, ``\textbf{Main Discussion}'': limitations of neural networks for human-level intelligence, ``\textbf{Speaker 1 View}'': skeptical of scaling neural networks alone, ``\textbf{Speaker 2 View}'': emphasizes physical-world understanding for reasoning and planning, ``\textbf{Tone}'': analytical, technical, collaborative, ``\textbf{Key Themes}'': human-level AI, neural network limitations, reasoning, planning, collaboration in research community \}
\end{mdframed}
We then parse the model-generated caption or response into the same schema and compute a structured matching reward by measuring overlap between the predicted and reference fields. The final reward is obtained by averaging field-level matches across categories, allowing semantically rich open-ended outputs to be evaluated in a structured and consistent way.

The overall reward function in GRPO combines the format reward with the accuracy reward for question-answering data, and the format reward with the structured reward for open-ended captioning and response-generation data. This enables AV-Flamingo to improve both answer correctness and the quality of long-form grounded audio-visual generations.

\section{AV-Think and TAVIT Details}
This section documents the AV-Think generation constraints in \cref{sec:app_avthink_generation_constraints} and representative Temporal Audio-Visual Interleaved Chain-of-Thought (TAVIT) examples in \cref{app:tavit}.

\subsection{AV-Think Generation Constraints}
\label{sec:app_avthink_generation_constraints}
For each long-form video, the generator receives timestamped audio segments, including speech transcriptions and non-speech event descriptions, together with timestamped visual captions. It produces open-ended QA-reasoning triplets, multiple-choice variants with plausible distractors, and an integrated audio-visual caption. We explicitly require questions to depend on both modalities, demand multi-hop temporal reasoning rather than surface-level transcription, and ground each intermediate reasoning step with \texttt{<t>...</t>} timestamp tags; questions answerable from audio alone or vision alone are excluded.

\subsection{TAVIT Examples}
\label{app:tavit}

Table~\ref{tab:av-cot-examples} shows representative TAVIT examples with the question, grounded thinking chain, and ground-truth response. Table~\ref{tab:tavit-samples} provides five additional TAVIT samples pairing the merged audio-visual caption (with source YouTube link) with the corresponding chain-of-thought question and ground-truth response.

\begin{table*}[h]
    \centering
    \caption{\small Examples of Temporal Audio-Visual Interleaved Chain-of-Thought (TAVIT) data}
    \resizebox{\textwidth}{!}{
    \begin{tabular}{p{5.5cm} | p{7cm} | p{6cm}}
    \toprule
    \textbf{Question [YouTube URL]} & \textbf{Thinking Chain} & \textbf{Ground Truth Response} \\
    \midrule

    How does the video use both the upbeat music, the brief spoken subscription message, and the evolving cooking visuals to guide the viewer from learning how to make mushroom tarts to being nudged toward engaging further with the channel, and what specific moments show this shift in purpose? [\href{https://www.youtube.com/watch?v=-2tPqC6quZQ}{\textcolor{blue}{https://www.youtube.com/watch?v=-2tPqC6quZQ}}]
    &
    I need to trace how the video's purpose evolves and how both sound and visuals participate in that change. The clip opens with a close-up of a finished mushroom tart and a graphic that labels it ``MUSHROOM TART'' (0--3s), then immediately shows raw puff pastry with a text overlay specifying ``200 grams puff pastry dough'' (3--5s). That tells me the initial goal is instructional. The visuals proceed step by step through the cooking process: rolling dough (5--9s), cutting rounds (9--15s), brushing with milk and baking (17--22s), saut\'eing garlic and mushrooms with seasonings (22--37s), filling baked pastry and topping with mozzarella (42--49s), and garnishing with thyme (49--51s). Upbeat funk/disco music plays throughout with no speech during cooking steps. The turning point is when ``enjoy!'' appears (51--53s). Immediately after, visuals change to an outro screen with video thumbnails, a YouTube Subscribe button, and social media links for Sanjeev Kapoor (53--63s). The male speaker's line---``If you like the videos that we share on YouTube, do subscribe to us on our YouTube channel''---occurs at (54--59s), explicitly promotional. The first phase combines cooking visuals with music to teach the recipe (0--51s); the second phase shifts to channel branding and a spoken call to subscribe (53--63s).
    &
    The video first focuses on teaching the viewer how to make mushroom tarts, then gradually pivots to encouraging ongoing engagement with the channel. During the cooking portion, there is no speech---only a steady, upbeat funk/disco track (0--40s). Visually, this phase walks step-by-step through the recipe: the finished tart and title overlay (0--3s), rolling and cutting puff pastry (3--15s), baking (17--22s), saut\'eing garlic and mushrooms (24--37s), and filling, topping, rebaking, and garnishing (42--51s). The purpose shifts once ``enjoy!'' appears (51--53s). The visuals change to an outro screen with thumbnails, a Subscribe button, and social media links for Sanjeev Kapoor (53--63s). A friendly male voice says, ``If you like the videos that we share on YouTube, do subscribe to us on our YouTube channel'' (54--59s). The cooking visuals plus music build trust, and once that value is delivered, the audio shifts to a call to action while the visuals shift to subscribe prompts and social links. \\
    \midrule

    Across the entire highlight, what is the main reason the sequence of play is presented as a showcase of rugby sevens excellence, rather than just a routine attacking move? \newline (A) Commentators focus on crowd chants over players' actions \newline (B) Clip shows scoreboard and time remaining \newline (C) Sequence combines ecstatic live call and crowd with slow-motion replays and detailed analysis of quick taps, dummy passes, and footwork \newline (D) Visuals center on referee decisions \newline (E) Player shown arguing with teammates \newline (F) Focus on scrums and mauls \newline (G) Clip shows pre-match warm-ups \newline (H) Video emphasizes sponsor logos \newline (I) Commentators describe a defensive masterclass

    [\href{https://www.youtube.com/watch?v=-2hY4Ap7xkA}{\textcolor{blue}{https://www.youtube.com/watch?v=-2hY4Ap7xkA}}]
    &
    I need the option that best captures why the whole clip frames this as a showcase of sevens excellence. From the audio: loud crowd cheering (4--24s), an excited commentator shouting about ``some footwork'', ``out to Watton, he throws the dummy'', and ``That's why we love the game of seven so much'' (10--24s). Then analytical commentary: ``Quick tap from Lindsay, off one foot'', ``He's electric and then the one-handed pass'', ``The show and go, the English footwork paying off hours at the Lensbury Club'' (25--46s). From the visuals: live aerial/close-up shots of a white-shirted player breaking through red defenders (4--17s), then slow-motion replays showing number 9 tapping quickly (28--30s), number 11 making a one-handed pass (32--35s), and number 1 repeatedly stepping and dodging (30--47s). This combination---live emotional high plus slow-motion technical breakdown---is exactly option (C). Other options eliminated: (A) crowd isn't the focus; (B) no scoreboard shown; (D) no refereeing debate; (E) hands-on-head is not anger; (F) this is open play; (G) Lensbury Club only mentioned verbally; (H) branding is just framing; (I) attack succeeds.
    &
    (C) Because the sequence combines an ecstatic live call and roaring crowd with slow-motion replays and detailed analysis of quick taps, dummy passes, and practiced footwork, highlighting both the excitement and the underlying skill. \\

    \bottomrule
    \end{tabular}}
    \label{tab:av-cot-examples}
\end{table*}

{\small
\setlength{\tabcolsep}{4pt}
\renewcommand{\arraystretch}{1.2}
\begin{longtable}{| p{0.30\textwidth} | p{0.24\textwidth} | p{0.40\textwidth} |}

\caption{Five overlapping samples from TAVIT: merged audio-visual caption (with YouTube link), chain-of-thought question, and ground-truth (GT) response. All entries are truncated; full text available in the dataset.}
\label{tab:tavit-samples} \\
\hline
\textbf{[YouTubeID] Caption} & \textbf{Question} & \textbf{GT Response} \\
\hline
\endfirsthead

\hline
\multicolumn{3}{|c|}{\textit{Table~\thetable\ -- continued from previous page}} \\
\hline
\textbf{[YouTubeID] Caption} & \textbf{Question} & \textbf{GT Response} \\
\hline
\endhead

\multicolumn{3}{|r|}{\textit{Continued on next page}} \\
\hline
\endfoot

\endlastfoot

\href{https://www.youtube.com/watch?v=-0EraoJmdhk}{[youtube.com/watch?v=-0EraoJmdhk]}

A bartender shakes a cocktail shaker containing a pink liquid.
The liquid is poured through a fine mesh strainer into a martini glass.
A close-up shows the finished pink cocktail garnished with an orange peel;
text overlays display ``Cosmopolitan'', its Hindi transliteration, and ``DRINK IT EASY 2.0''.
The bartender sets down an empty shaker, removes its lid, then measures and pours
vodka, lemon juice, cranberry juice, and triple sec with bilingual text callouts
for each ingredient.
Ice cubes are added, the shaker sealed and shaken vigorously.
The cocktail is strained into a frosted martini glass and finished with a twisted
orange peel garnish before the final beauty shot.
&
How do the changes in the lo-fi hip-hop soundtrack support the visual progression
from ingredient preparation to the final presentation and branding of the
Cosmopolitan cocktail, and what does this reveal about the intended mood and
viewing experience of the video as a whole?
&
The soundtrack and visuals are tightly coordinated so that the music's evolving
texture mirrors the cocktail's journey from raw ingredients to a polished, branded
product, creating a relaxed, tutorial-meets-aesthetic experience rather than a purely
instructional one.
At the very beginning, the ambient pad without drums (0--2\,s) plays over the first
glimpse of shaking and pouring and the hero close-up of the finished Cosmopolitan
with multilingual overlays (0--8\,s), framing the drink as aspirational, not just a recipe.
When the main synth melody and full beat enter (2--18\,s), the visuals shift into
step-by-step recipe work: measuring and pouring each spirit with bilingual callouts,
adding ice, sealing, and shaking. The steady, looping groove makes these repetitive
actions feel smooth and unhurried.
A slightly thinned bridge (27--35\,s) coincides with the refinement phase: the frosted
glass, straining, and the orange-peel garnish.
Finally, during the gentle fade-out (35--44\,s), the camera lingers on the completed
cocktail before cutting to a branded outro screen with social-media handles and a
website. [\ldots]
\\
\hline

\href{https://www.youtube.com/watch?v=-0ItZ8fGUNc}{[youtube.com/watch?v=-0ItZ8fGUNc]}

A close-up of a dark bowl of rich, reddish-brown stew garnished with coriander;
text reads ``Pork Sorpotel''.
A dark pan shows whole cloves, black peppercorns (``4--5''), green cardamoms (``3''),
half a cinnamon stick, cumin seeds, and 3--4 dried red chillies added one by one.
The spices are shaken on an induction cooktop, transferred into warm water to soak
for 30--40 minutes, then blended into a liquid paste in a Wonderchef blender.
Oil is poured into a black pot; boiled, diced pork and pork liver are saut\'{e}ed
until golden brown and set aside.
Ginger, garlic, green chillies, onions, and salt are saut\'{e}ed until the onions
brown; the red spice paste and pork stock are added and stirred.
Sugar, tamarind pulp, vinegar, and salt are added; the pot is covered and simmered
for 15--20 minutes before the finished dish is ladled into a serving bowl and
garnished with fresh coriander.
&
How does the combination of the steadily building EDM soundtrack and the evolving
visuals of the Pork Sorpotel recipe work together to guide the viewer from raw
spices and meat through to the final plated dish, and what does this reveal about
how the video balances methodical instruction with an energetic, appetizing mood?
&
The video uses the EDM track as a continuous emotional backbone that parallels the
visual progression from raw ingredients to finished Pork Sorpotel, turning a
methodical recipe into something that feels energetic and appetizing rather than slow
or tedious.
At the start, the music is in a steady, moderate groove but with a light arrangement
(0--2\,s), matching the clean title close-up of the garnished dish.
As whole spices are toasted and added one by one (3--16\,s), the beat maintains
energy, making the repetitive individual shots feel purposeful.
When the pork and liver are saut\'{e}ed to golden brown (38--49\,s), the EDM
transitions to a more energetic section, lifting the energy as the recipe moves from
preparation into active cooking.
A brief breakdown (75--91\,s) accompanies the final seasoning steps---sugar, tamarind,
vinegar---signalling an imminent climax.
The full groove returns as the pot simmers (91--104\,s), then fades as the dish is
plated and garnished, completing the parallel arcs of music and recipe. [\ldots]
\\
\hline

\href{https://www.youtube.com/watch?v=--ccsQMua_s}{[youtube.com/watch?v=--ccsQMua\_s]}

The \textit{Fox \& Friends} logo appears on an orange-and-blue graphic.
Biden is shown at the Resolute Desk in the Oval Office; on-screen text reads
``BIDEN DELIVERS FAREWELL ADDRESS FROM OVAL OFFICE''.
Peter Doocy reports live from a snowy White House lawn.
Biden warns of ``a dangerous concentration of power in the hands of a very few
ultra-wealthy people'' and ``an oligarchy taking shape in America''.
Biden invokes Eisenhower's military-industrial complex, coining ``tech-industrial
complex''.
A ghost image of Biden is overlaid on a shot of his black-clad family seated
alongside Kamala Harris and Doug Emhoff.
CNN poll graphics appear: Biden job approval 36\%, with 61\% calling his presidency
a failure.
Studio hosts discuss Biden's legacy while a split screen shows him alongside the
\textit{Fox \& Friends} panel.
&
How does the segment use both President Biden's own farewell warnings and the
studio's visual and conversational framing---especially the ghostly overlay of him
on his black-clad family, the polling graphics, and the contrast with Trump and
tech leaders---to recast his stated concerns about oligarchy and a tech-industrial
complex into a narrative of a failed, bitter presidency out of step with the
country's future?
&
The segment systematically reframes Biden's farewell warnings from principled
concerns into evidence of a failed, embittered presidency by weaving together his
own words with highly curated visuals and studio commentary.
Biden is visually anchored at the Resolute Desk as a formal, traditional president
(1--15\,s, 42--60\,s), while the audio has him warning about concentrated wealth
and a tech-industrial complex (48--60\,s, 128--145\,s).
Rather than engaging with the substance, the hosts pivot immediately to mockery:
Steve Doocy jokes about ``teleprompter-archy'' (130--147\,s), and the ghostly
overlay of Biden on his black-clad family (152--157\,s) visually associates his
warnings with funereal gloom rather than civic urgency.
The CNN poll graphic showing 36\% approval and 61\% ``failure'' (374--388\,s) is
then presented as the electorate's verdict on those same warnings.
The contrast is completed by hosts praising tech CEOs for the economic growth
Biden's presidency benefited from, inverting his oligarchy framing into ingratitude.
[\ldots]
\\
\hline

\href{https://www.youtube.com/watch?v=-0YyxReol5E}{[youtube.com/watch?v=-0YyxReol5E]}

A close-up of a South African rugby player in a green, black, and yellow jersey
with a white headband, eyes closed, as rain falls.
The camera moves to his back showing the ``SOUTH AFRICA RUGBY'' crest and
``RUGBY WORLD CUP FRANCE 2023'' text.
A second player sings with eyes open; the camera pans a line of players singing
with serious expressions in a packed stadium.
Spectators in a box seat watch and photograph the ceremony.
Close-ups of a light-skinned and a bald dark-skinned player, both sweating and
singing with eyes closed.
Fans in green and yellow jerseys sing looking upward; a player with long blonde
hair sings with eyes closed for an extended shot.
All 15 players stand arms-around-shoulders, joined by a young boy in a
black-and-white kit, before turning and walking away as the crowd erupts.
&
How do the lyrics and musical build-up of the choral anthem combine with the
behavior and arrangement of the South African players, officials, and supporters
to show that this moment is not just a routine pre-match ritual but a deeply
spiritual and nationally unifying ceremony?
&
The moment becomes clearly more than a routine pre-match ritual because the
religious, sacrificial language and swelling structure of the anthem are mirrored
by the intensity, posture, and inclusiveness of everyone shown in the stadium.
The choir sings of a people beloved by God---lines such as ``Para que seas su
pueblo, y su naci\'{o}n'' frame the singers as a chosen community under divine
care (6--16\,s). Repeated invocations of ``\textit{Oh, Se\~{n}or}'' and
references to ``Su sangre'' (37--57\,s) give the anthem a liturgical character,
while the music grows into a full orchestral and choral crescendo (33--84\,s).
Visually, the players' closed eyes, sweat, and upward tilted heads mirror the
posture of private prayer (0--84\,s).
The inclusion of officials and fans singing in the stands (17--33\,s) extends
the ceremony into a national act.
Most powerfully, the young boy joining the players' line (77--83\,s) signals
intergenerational transmission of identity, and the crowd's eruption the instant
the anthem ends (84--86\,s) confirms that the entire stadium participated in a
shared spiritual moment. [\ldots]
\\
\hline

\end{longtable}
}

\section{Prompt Templates}
\label{sec:prompt_templates}
\begin{figure*}[t]
  \centering
  \includegraphics[width=\textwidth]{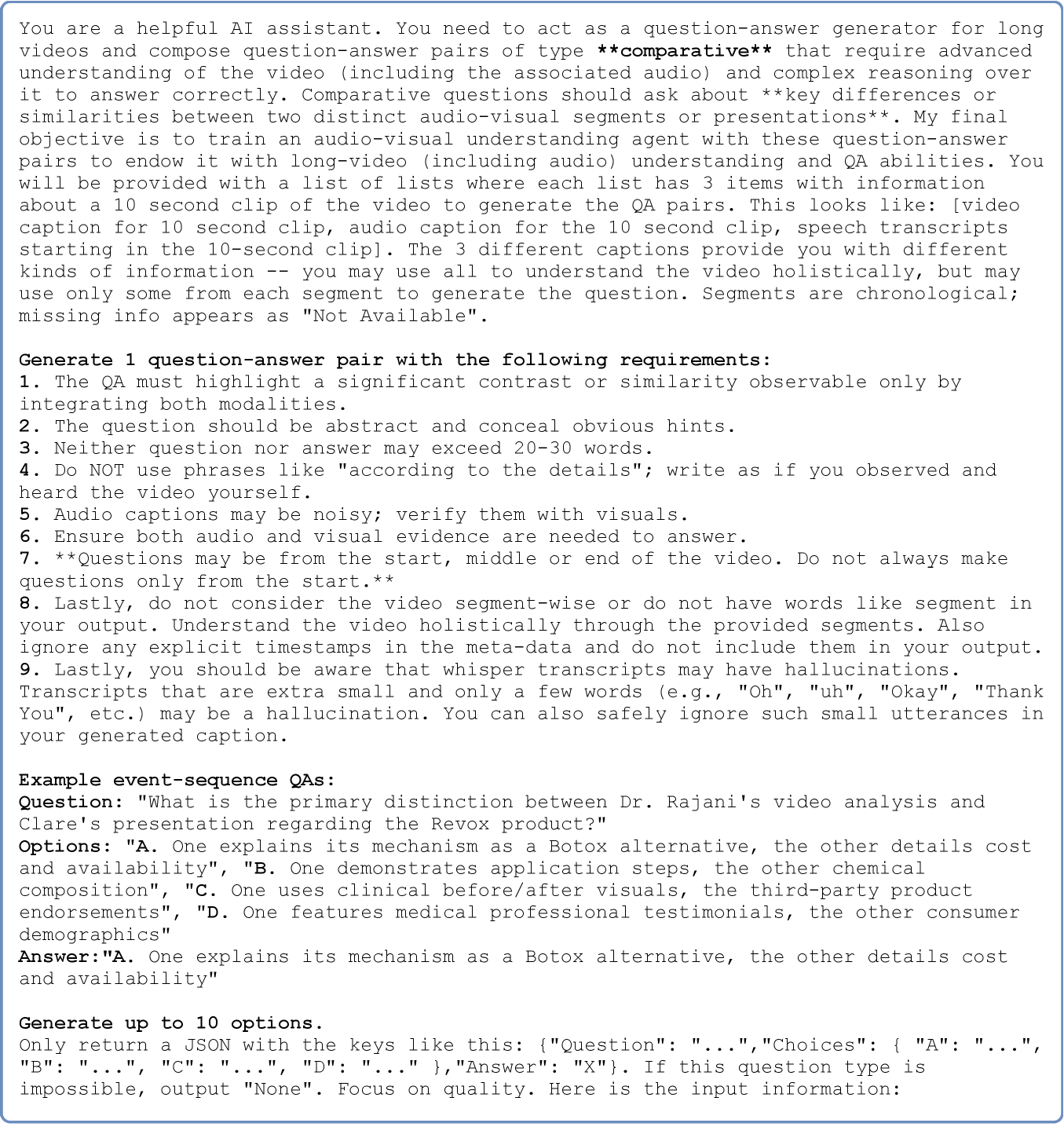}
  \caption{\small \textbf{Prompt for Comparative Reasoning QA generation.} Instructs the LLM to produce a multiple-choice question (with up to 10 options) that hinges on a key contrast or similarity between two distinct audio-visual segments in a long video, requiring both modalities to answer.}

  \label{fig:comparative}
\vspace{-2mm}
\end{figure*}

\begin{figure*}[t]
  \centering
  \includegraphics[width=\textwidth]{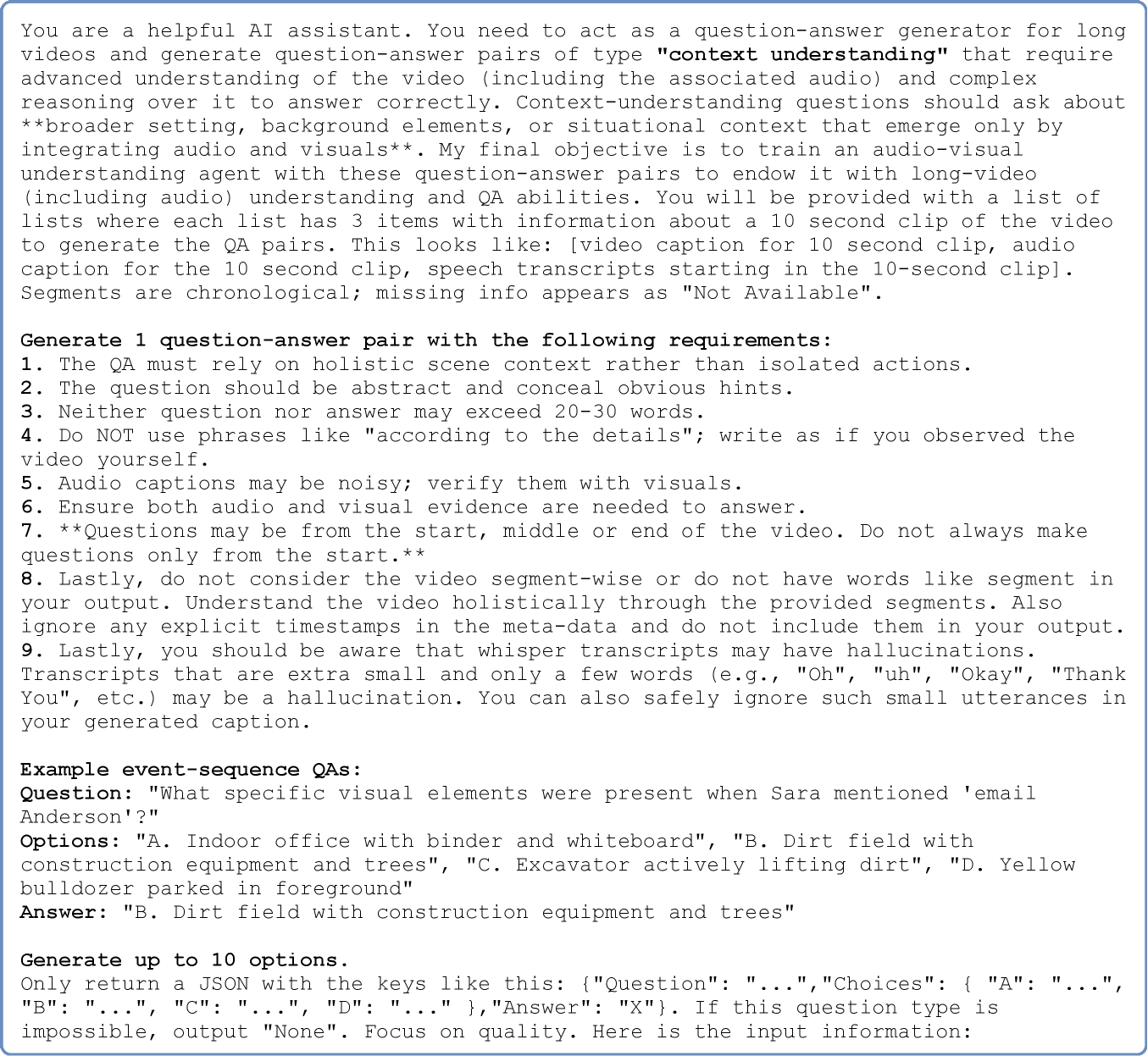}
\caption{\small \textbf{Prompt for Context Understanding QA generation.} Instructs the LLM to produce a multiple-choice question that depends on the broader situational setting, background, or scene conditions inferable only by jointly reasoning over visuals, sounds, and speech.}
  \label{fig:context}
\vspace{-2mm}
\end{figure*}

\begin{figure*}[t]
  \centering
  \includegraphics[width=\textwidth]{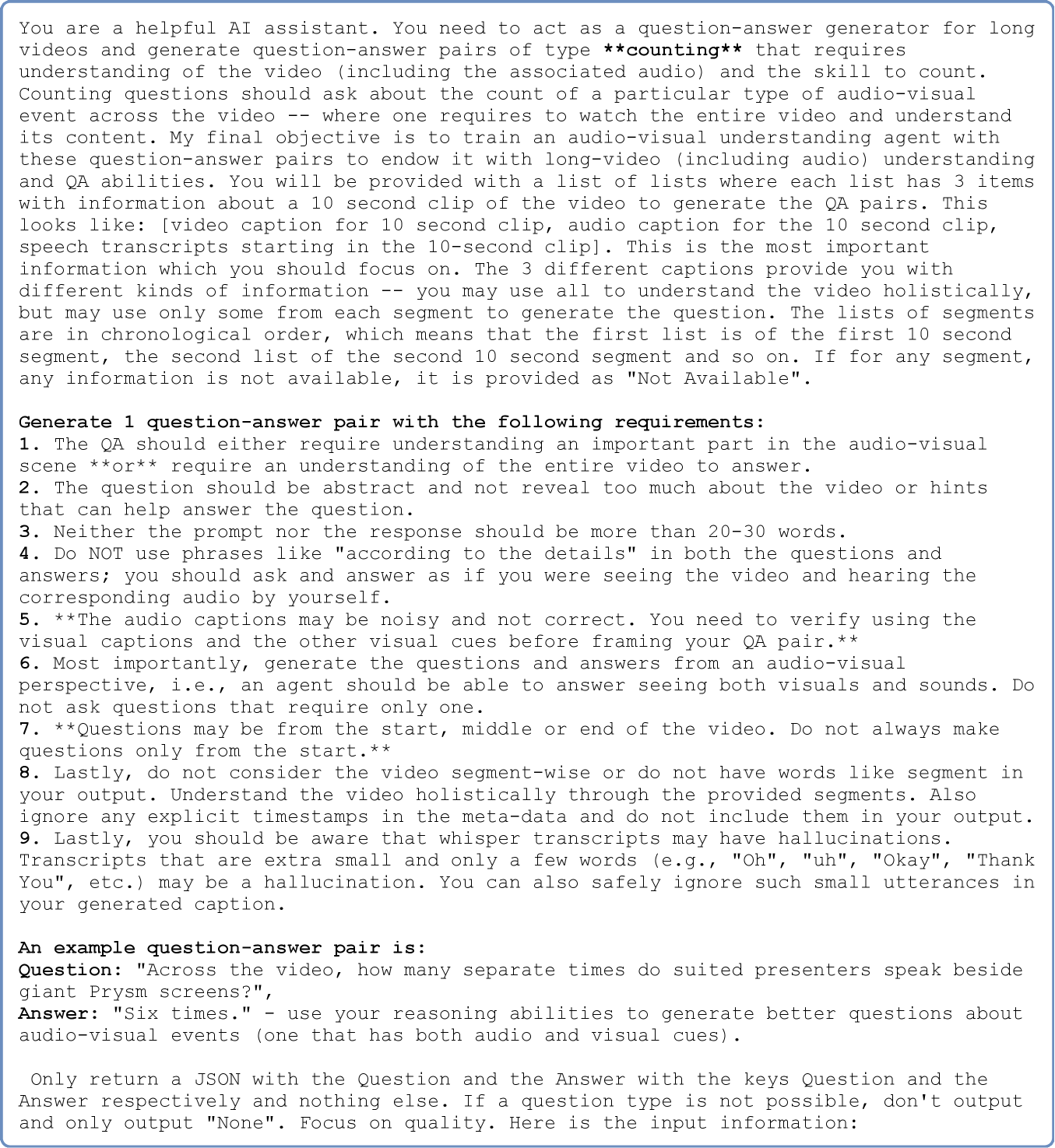}
\caption{\small \textbf{Prompt for Counting QA generation.} Instructs the LLM to produce a question requiring the model to count occurrences of a specific audio-visual event across the full video, verifying noisy audio captions against visual cues.}
  \label{fig:counting}
\vspace{-2mm}
\end{figure*}

\begin{figure*}[t]
  \centering
  \includegraphics[width=\textwidth]{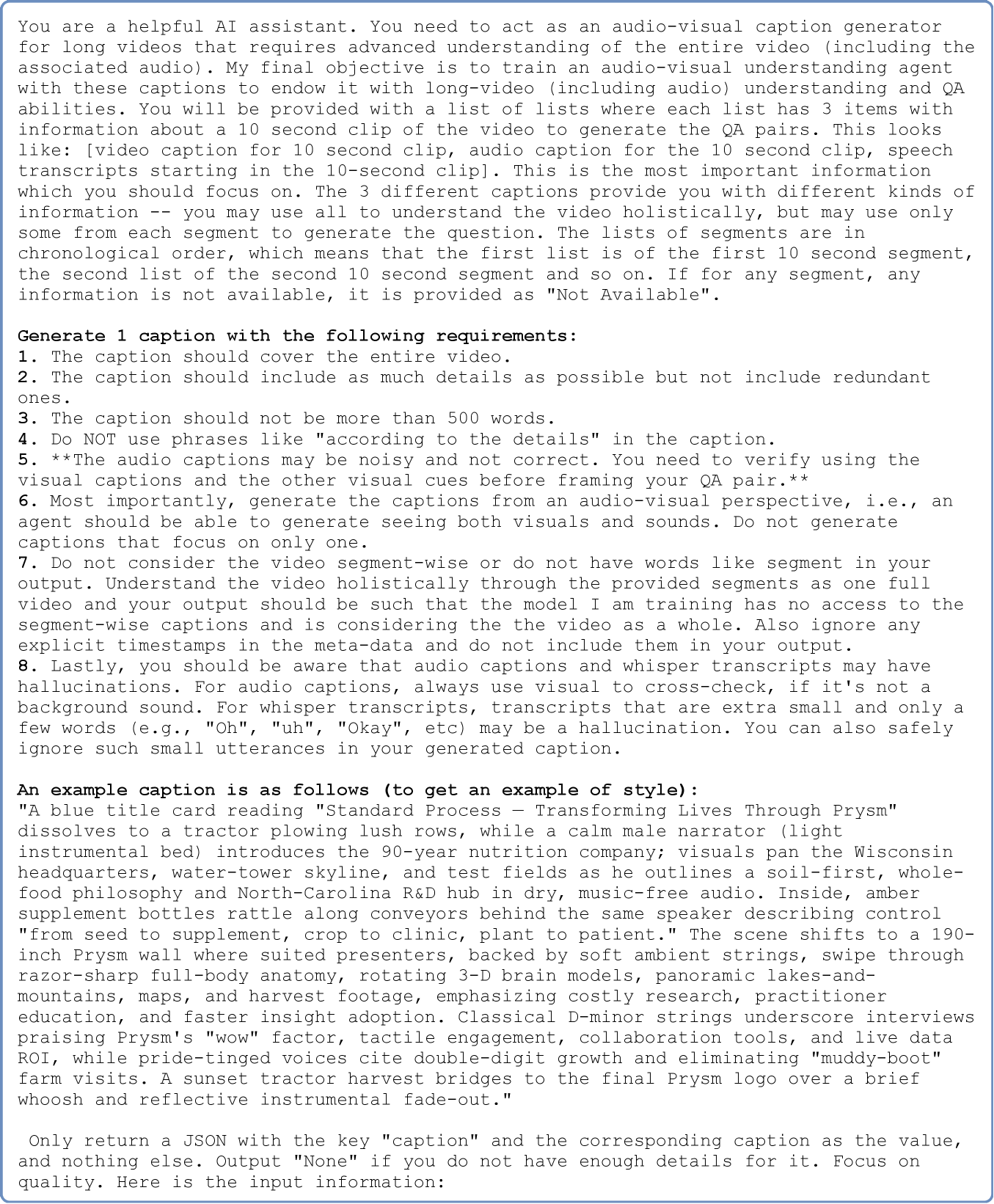}
\caption{\small \textbf{Prompt for Detailed Captioning.} Instructs the LLM to produce a single audio-visual caption (up to 500 words) summarizing the full video from a joint perceptual perspective, without segment-wise structure or explicit timestamps.}
  \label{fig:detailed_caption}
\vspace{-2mm}
\end{figure*}

\begin{figure*}[t]
  \centering
  \includegraphics[width=\textwidth]{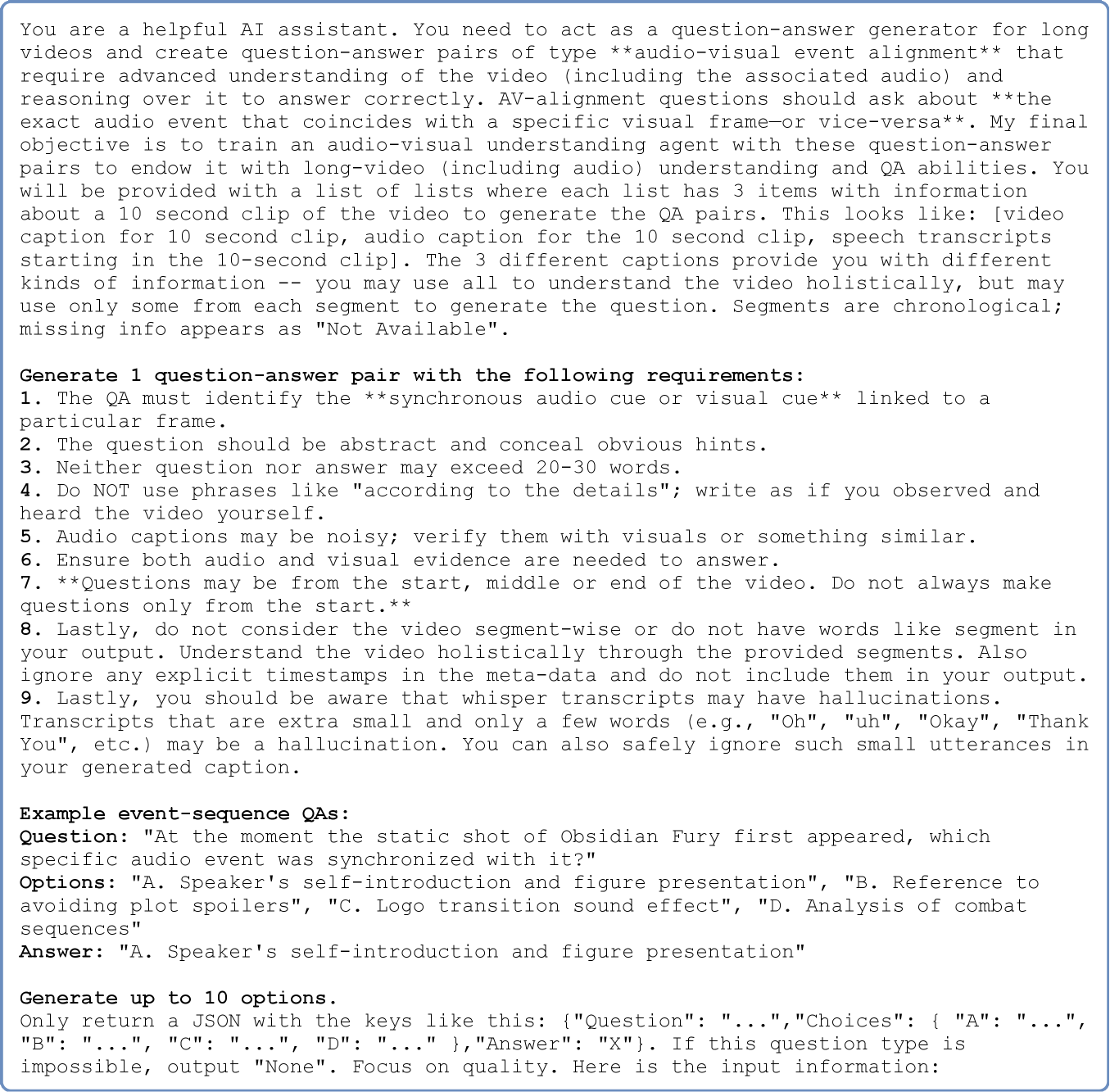}
\caption{\small \textbf{Prompt for Audio-Visual Event Alignment QA generation.} Instructs the LLM to produce a multiple-choice question identifying the exact audio cue that coincides with a specific visual frame, or vice versa, requiring tight cross-modal synchronization.}
  \label{fig:event_alignment}
\vspace{-2mm}
\end{figure*}

\begin{figure*}[t]
  \centering
  \includegraphics[width=\textwidth]{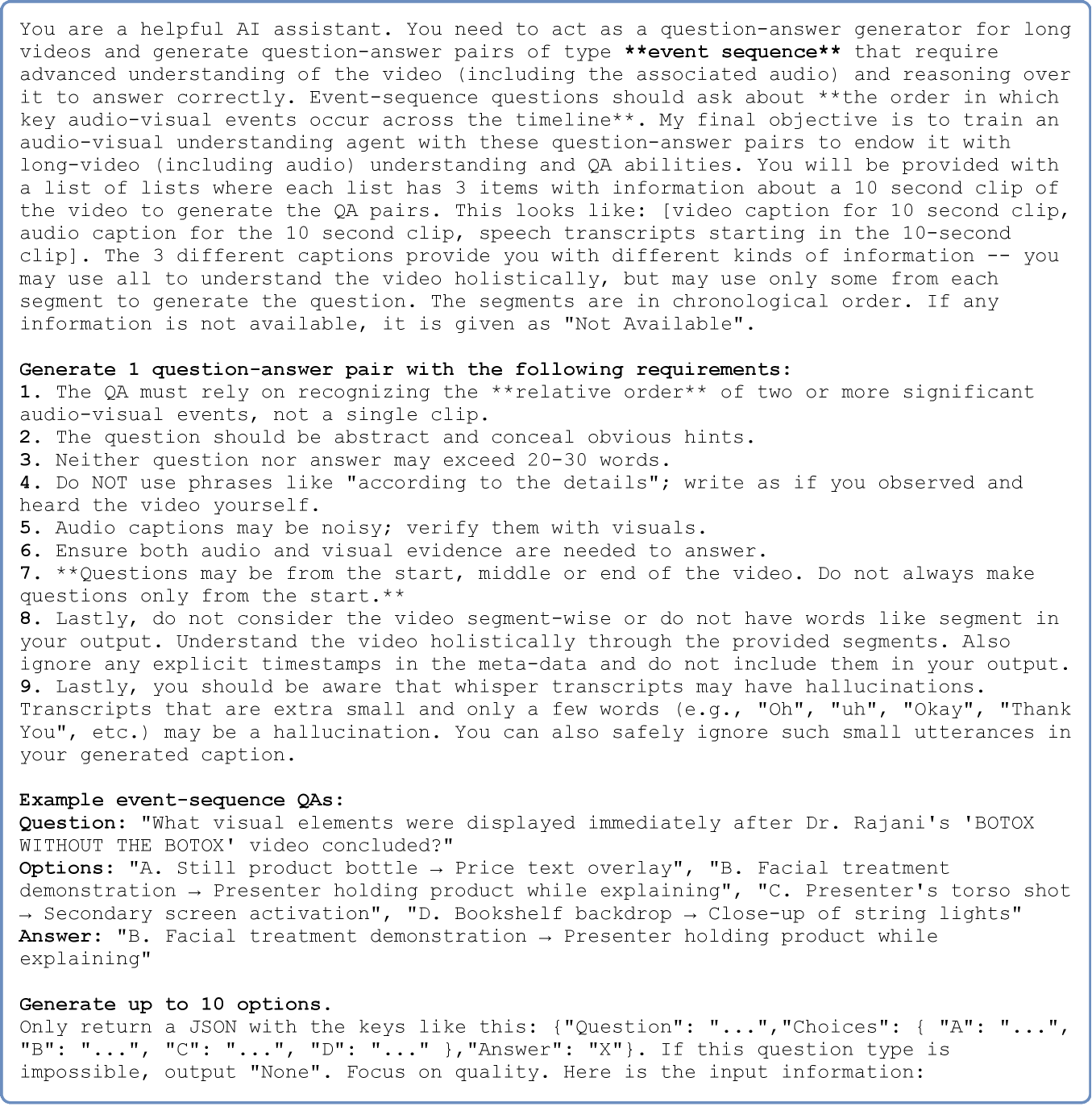}
\caption{\small \textbf{Prompt for Event Sequence QA generation.} Instructs the LLM to produce a multiple-choice question about the relative temporal ordering of two or more significant audio-visual events spanning the video timeline.}
  \label{fig:event_sequence}
\vspace{-2mm}
\end{figure*}

\begin{figure*}[t]
  \centering
  \includegraphics[width=\textwidth]{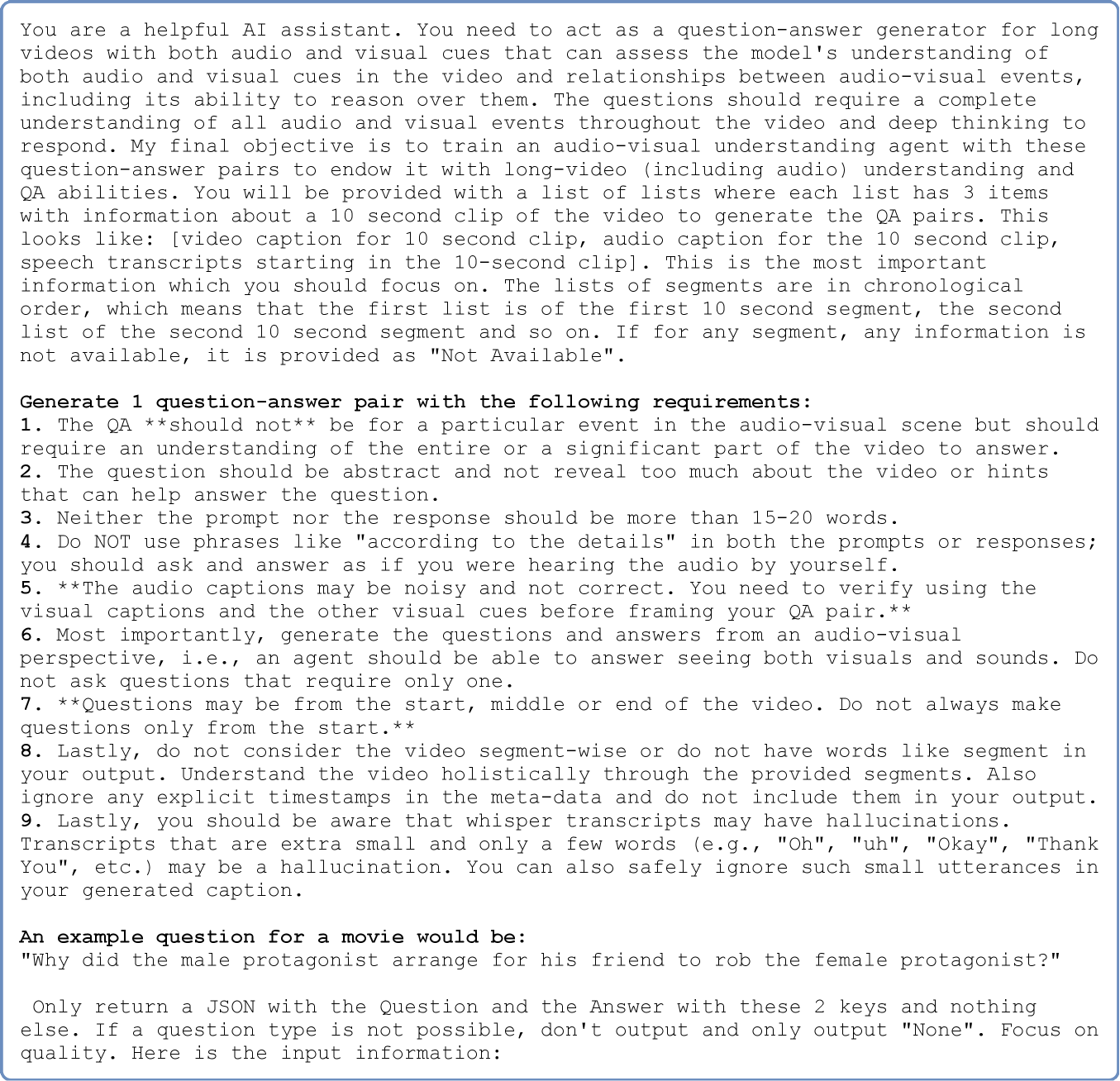}
\caption{\small \textbf{Prompt for Holistic Reasoning QA generation.} Instructs the LLM to produce an open-ended question that requires integrating audio-visual evidence distributed across a substantial portion of the video, rather than any single localized moment.}
  \label{fig:holistic_reasoning}
\vspace{-2mm}
\end{figure*}

\begin{figure*}[t]
  \centering
  \includegraphics[width=\textwidth]{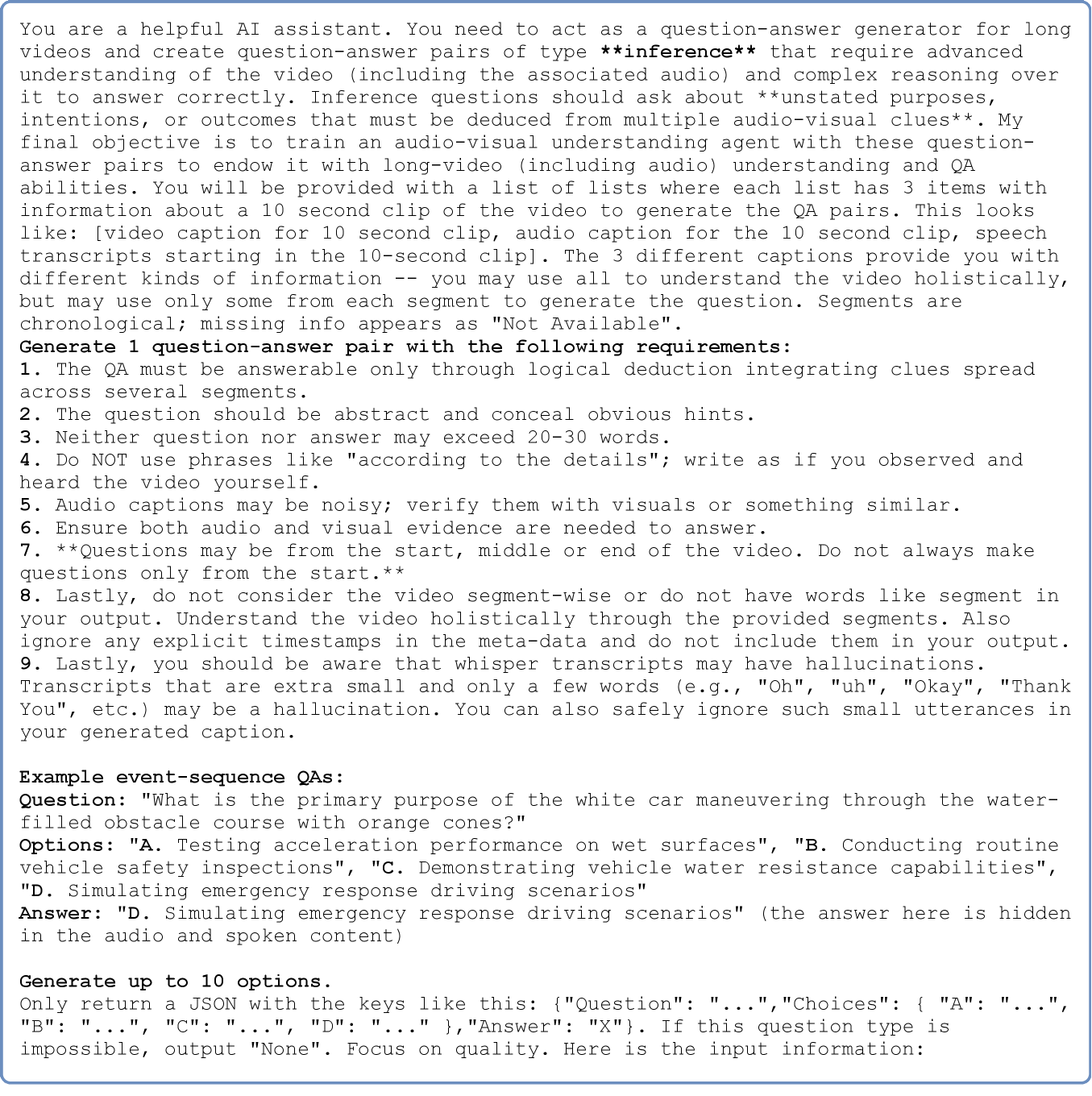}
\caption{\small \textbf{Prompt for Inference QA generation.} Instructs the LLM to produce a multiple-choice question whose answer can only be deduced by combining multiple implicit audio-visual clues spread across the video, with the correct option deliberately not stated directly.}
  \label{fig:inference}
\vspace{-2mm}
\end{figure*}

\begin{figure*}[t]
  \centering
  \includegraphics[width=\textwidth]{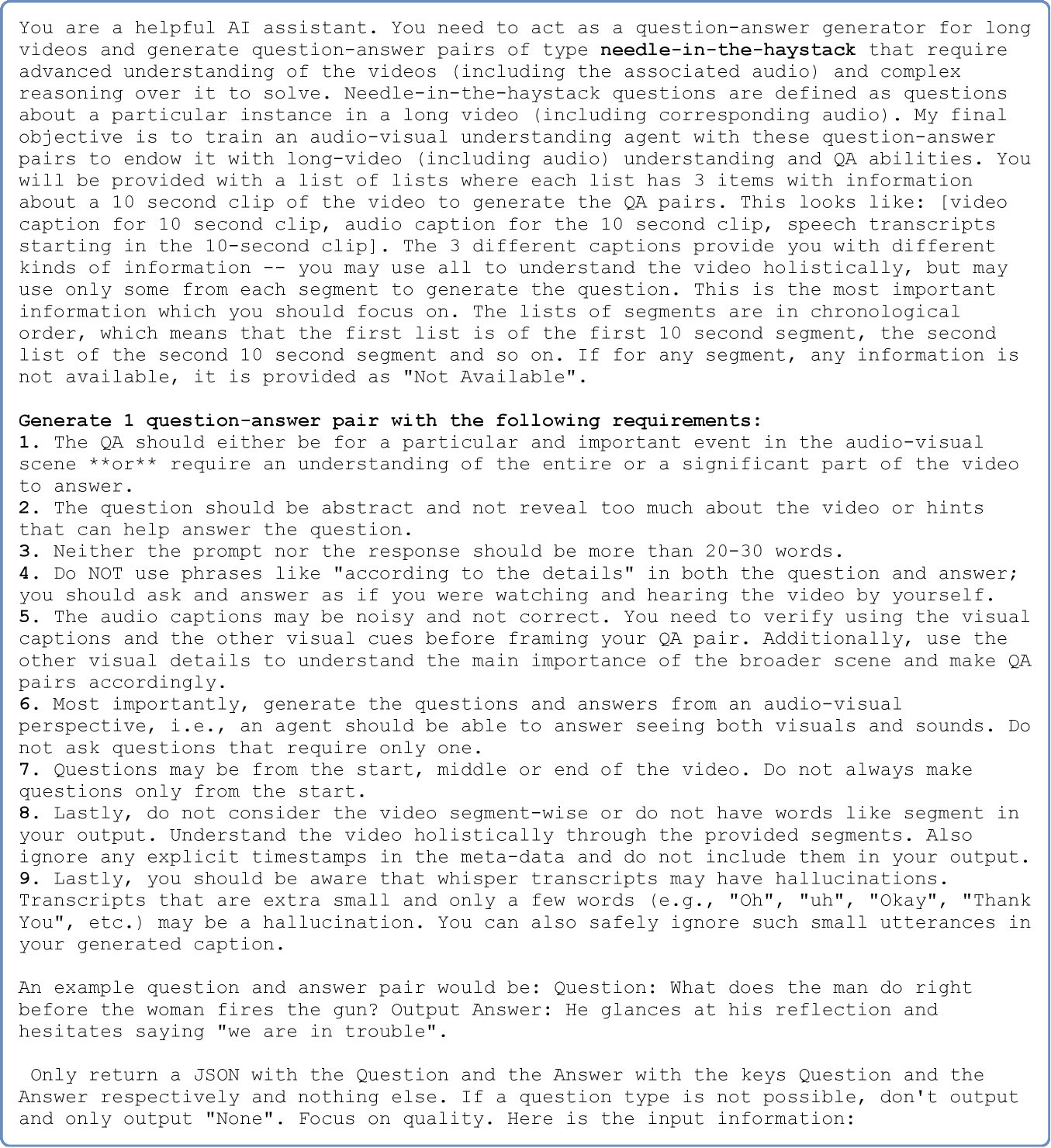}
\caption{\small \textbf{Prompt for Needle-in-the-Haystack QA generation.} Instructs the LLM to produce an open-ended question targeting a specific but important moment in a long video, where broader audio-visual context is still required to identify it correctly.}
  \label{fig:needle}
\vspace{-2mm}
\end{figure*}

\begin{figure*}[t]
  \centering
  \includegraphics[width=\textwidth]{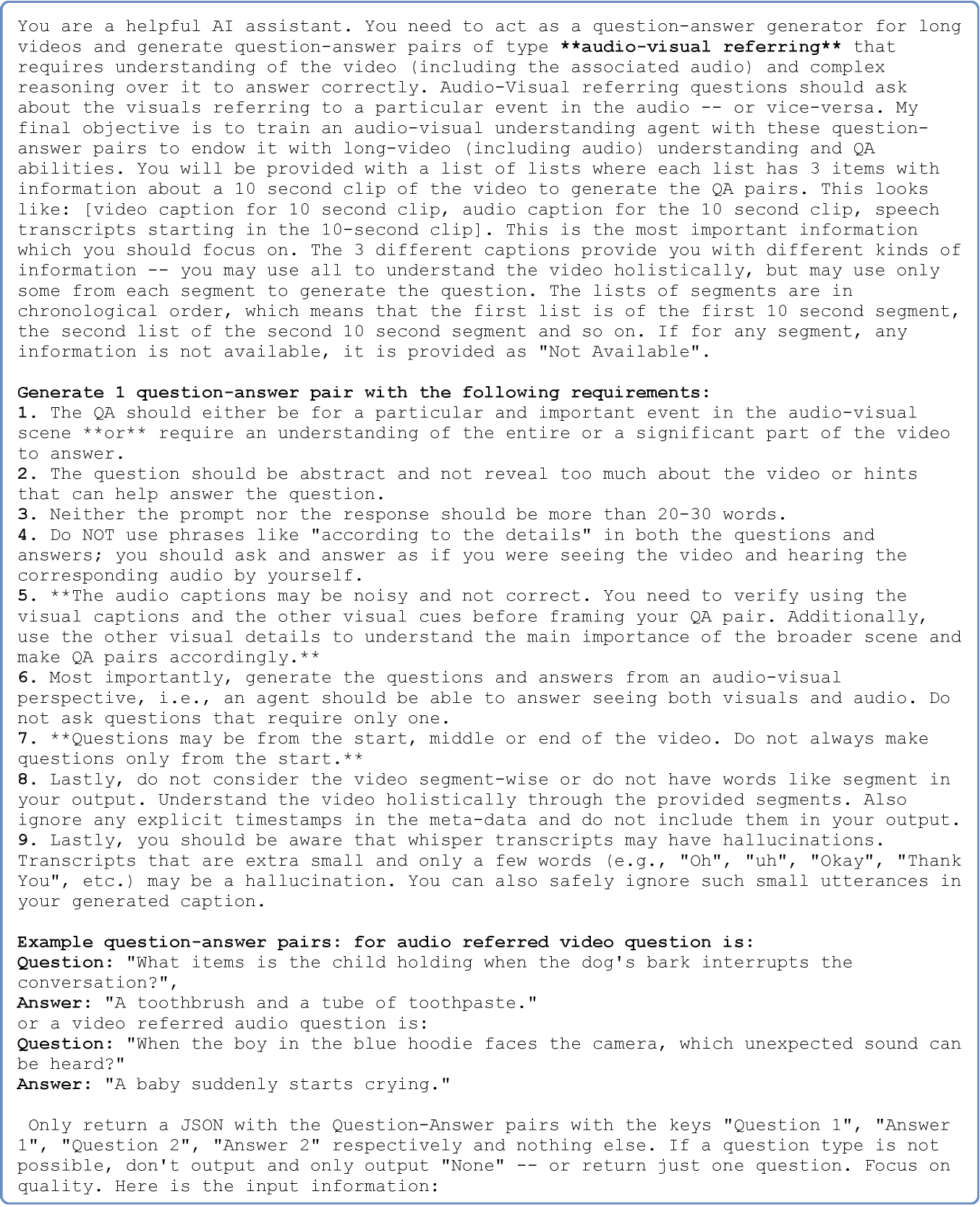}
\caption{\small \textbf{Prompt for Audio-Visual Referring QA generation.} Instructs the LLM to produce up to two question--answer pairs that link a visual event to its corresponding sound, or a sound to its corresponding visual scene, exercising cross-modal grounding.}
  \label{fig:referring}
\vspace{-2mm}
\end{figure*}

\begin{figure*}[t]
  \centering
  \includegraphics[width=\textwidth]{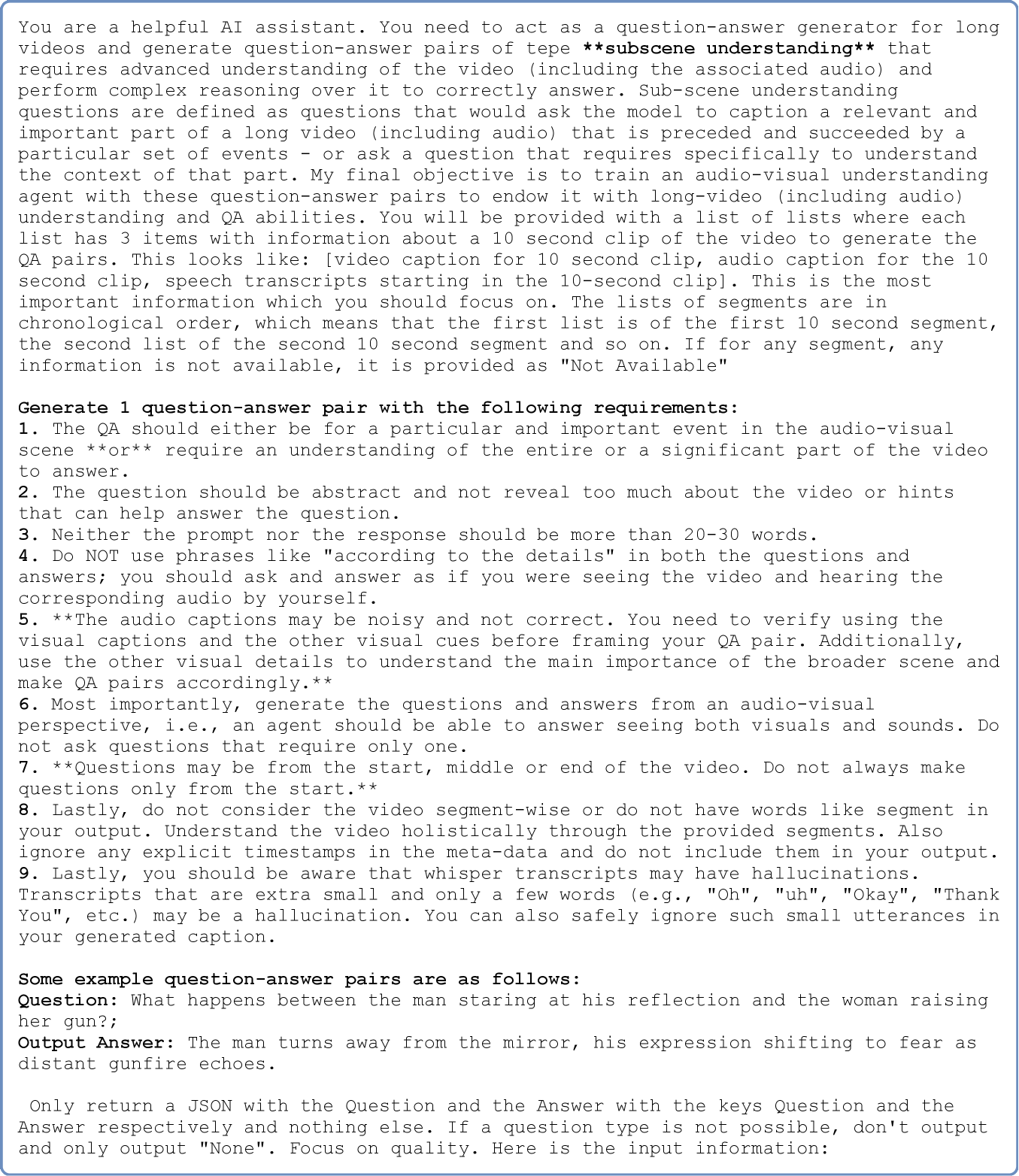}
\caption{\small \textbf{Prompt for Sub-scene Understanding QA generation.} Instructs the LLM to produce an open-ended question about a meaningful intermediate portion of a long video, defined by the events that precede and follow it.}
  \label{fig:subscene}
\vspace{-2mm}
\end{figure*}

\begin{figure*}[t]
  \centering
  \includegraphics[width=\textwidth]{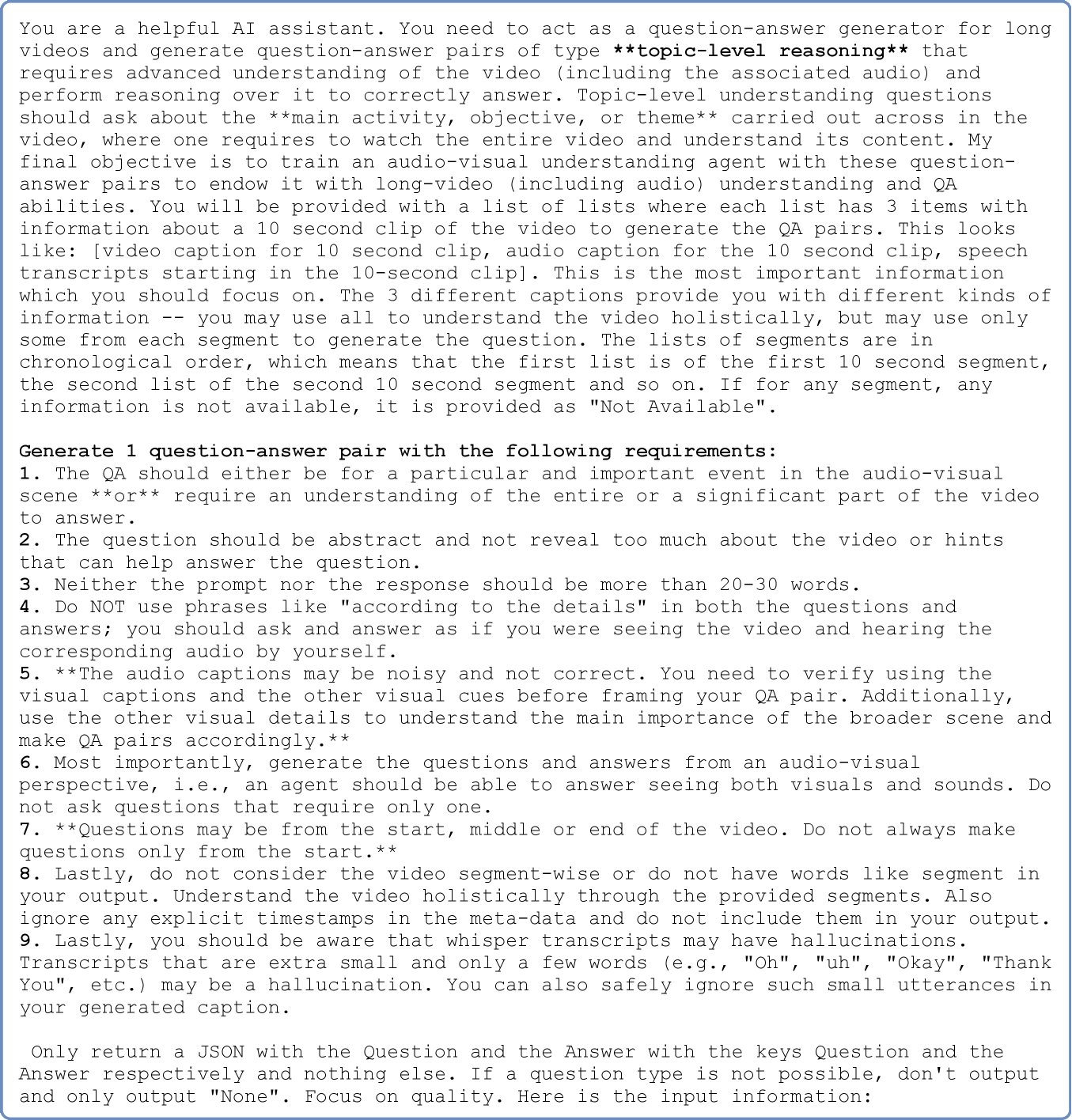}
\caption{\small \textbf{Prompt for Topic-Level Reasoning QA generation.} Instructs the LLM to produce an open-ended question about the main activity, objective, or overall theme of the video, requiring synthesis across the full timeline.}
      \label{fig:topic_level}
\vspace{-2mm}
\end{figure*}

For AV-Skills-Long datasets, we curate synthetic QA pairs and options for questions from various categories using GPT. The category-specific AV-Skills-Long prompt templates are shown in Figures~\ref{fig:comparative}--\ref{fig:topic_level}. Figure~\ref{fig:prompt_avthink} shows the full prompt used to construct the question--answer--reasoning triplets in AV-Think.

\begin{figure*}[t]
  \centering
  \includegraphics[width=\textwidth]{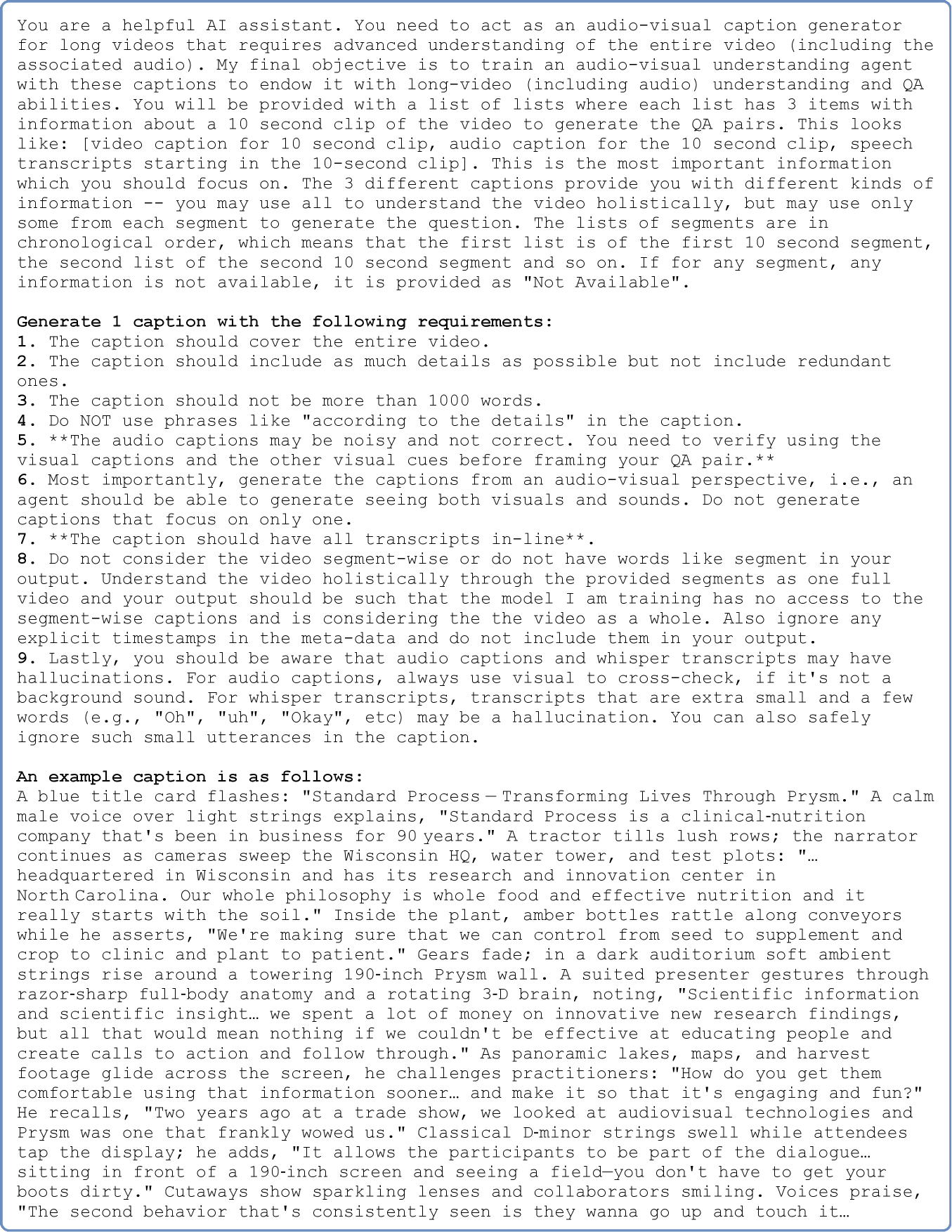}
  \caption{\small \textbf{Prompt for Detailed Captioning with Transcripts. (Part 1 of 2)}}
\end{figure*}

\begin{figure*}[t]
  \ContinuedFloat
  \centering
  \includegraphics[width=\textwidth]{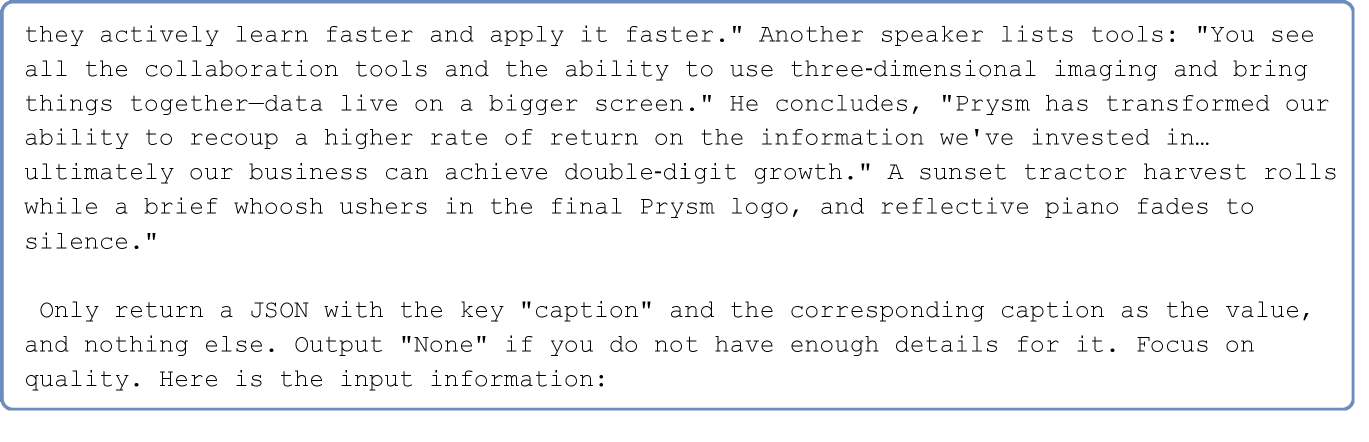}
  \caption{\small \textbf{Prompt for Detailed Captioning with Transcripts. (Part 2 of 2)} Instructs the LLM to produce a single comprehensive audio-visual caption (up to 1000 words) that integrates speech transcripts in-line with descriptions of visuals, sounds, and scene dynamics across the entire video.}
  \label{fig:detailed_caption_w_transcript}
\end{figure*}

\begin{figure*}[t]
  \centering
  \includegraphics[width=\textwidth]{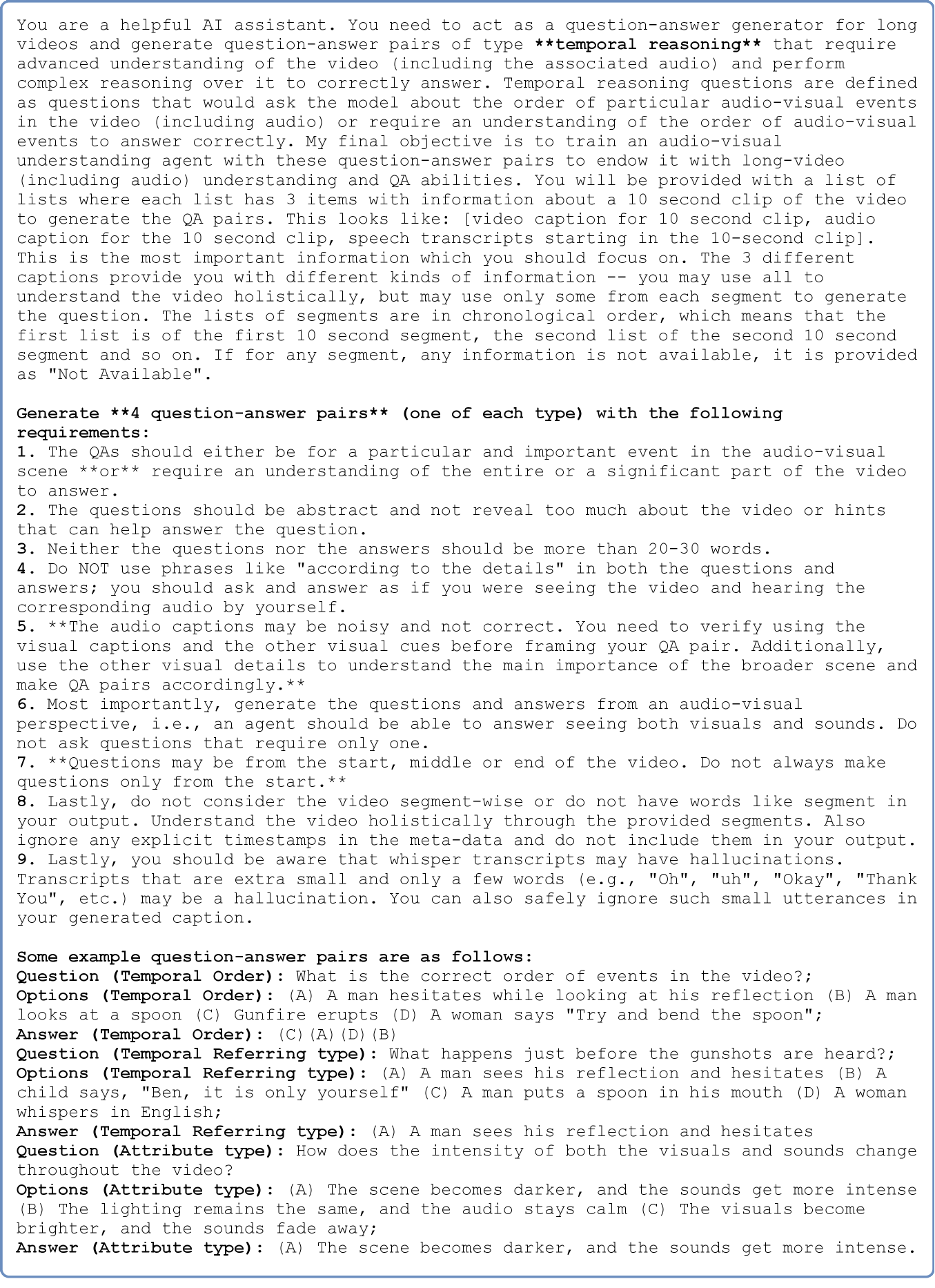}
  \caption{\small \textbf{Prompt for Temporal Reasoning QA generation. (Part 1 of 2).}}
\end{figure*}

\begin{figure*}[t]
  \ContinuedFloat
  \centering
  \includegraphics[width=\textwidth]{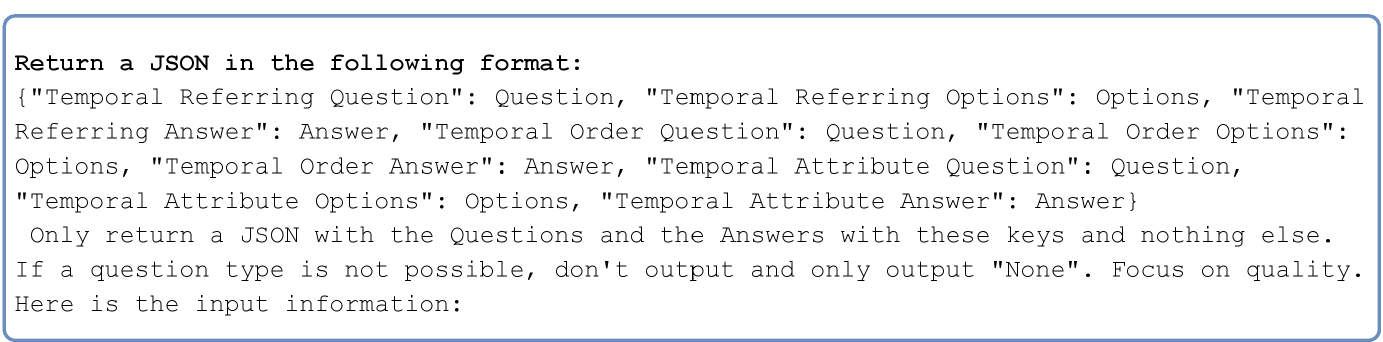}
  \caption{\small \textbf{Prompt for Temporal Reasoning QA generation. (Part 2 of 2)} Instructs the LLM to produce four question--answer pairs covering Temporal Order, Temporal Referring, and Temporal Attribute reasoning over the chronological structure and evolving dynamics of the video.}
  \label{fig:temporal_reasoning}
\end{figure*}

\begin{figure*}[t]
  \centering
  \includegraphics[width=\textwidth]{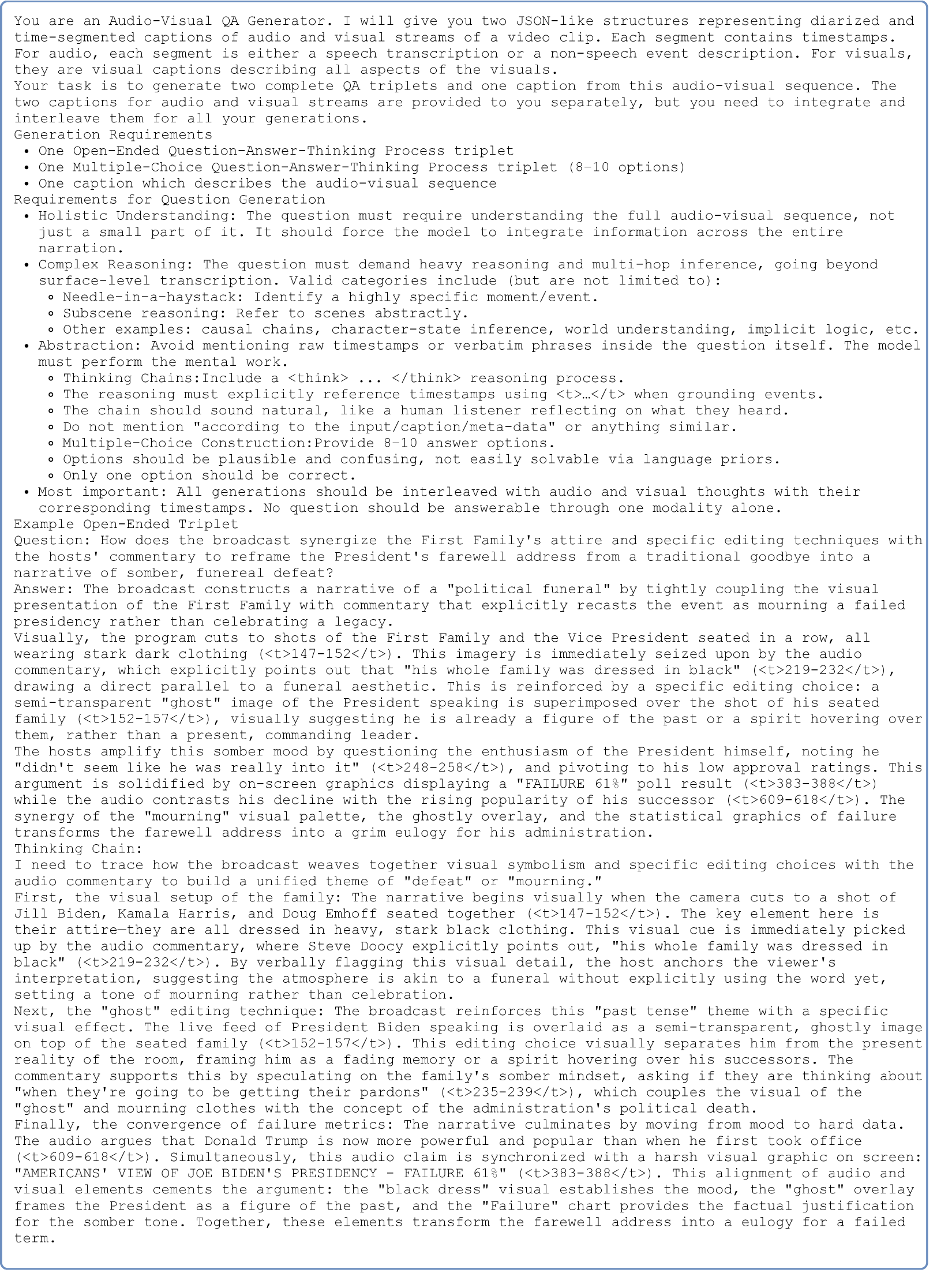}
  \caption{\small Prompt used for synthesizing AV-Think reasoning triplets (Part 1 of 2).}
\end{figure*}

\begin{figure*}[t]
  \ContinuedFloat
  \centering
  \includegraphics[width=\textwidth]{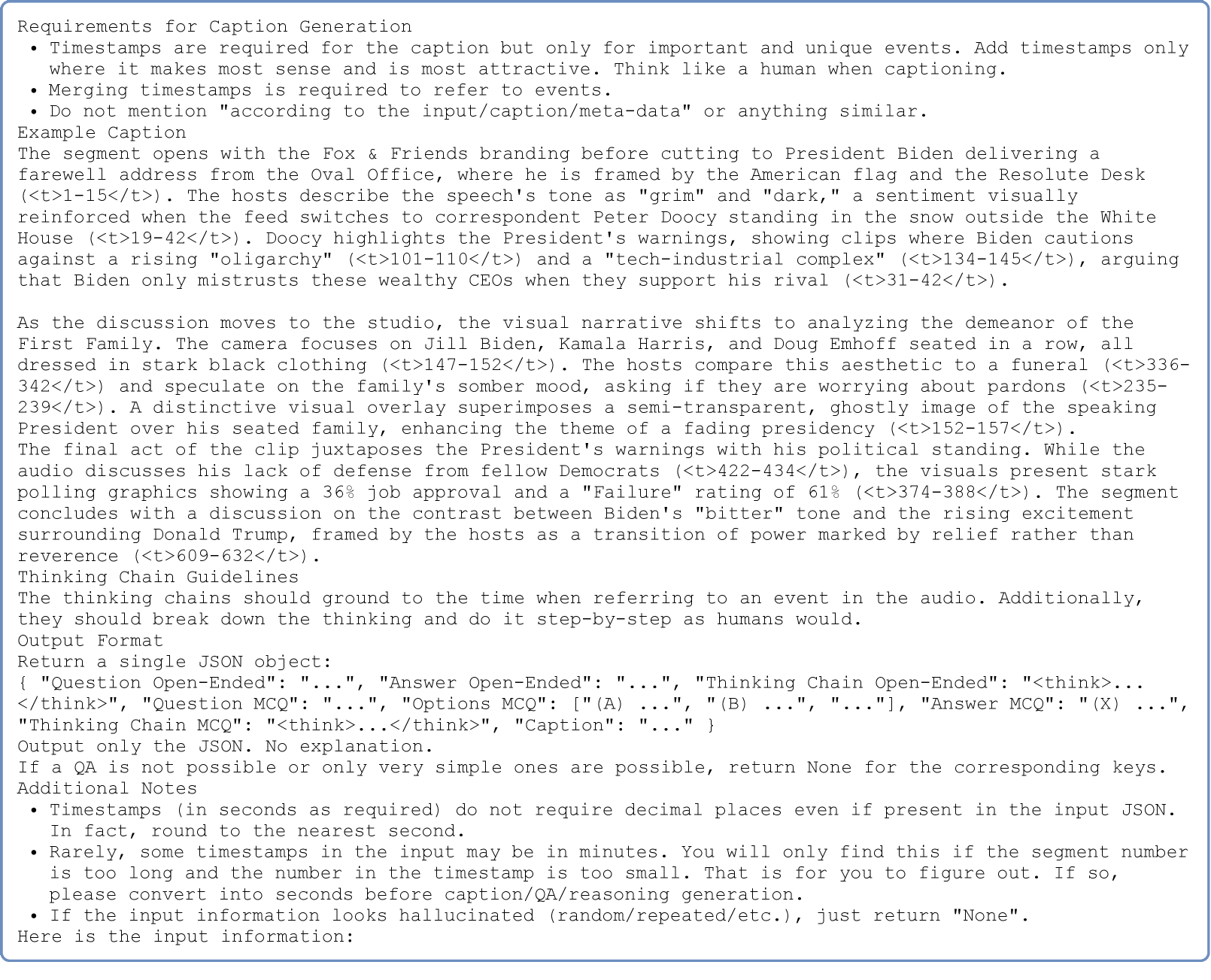}
  \caption{\small \textbf{Prompt used for synthesizing AV-Think reasoning triplets (Part 2 of 2).} Given timestamped audio and visual captions extracted from a long-form video, we prompt an LLM to generate a question--answer--reasoning triplet along with an integrated audio-visual caption. The prompt instructs the model to (i) require holistic understanding across both modalities, (ii) demand multi-hop temporal reasoning rather than surface-level transcription, and (iii) ground every intermediate thought to specific timestamps using \texttt{<t>...</t>} tags. We additionally enforce that no question be answerable from a single modality alone, ensuring that the resulting reasoning chains genuinely interleave audio and visual evidence.}
  \label{fig:prompt_avthink}
\end{figure*}

\section{Assets and Licenses}
\label{app:licenses}

Table~\ref{tab:licenses} lists the datasets and model checkpoints used in this paper, along with their associated licenses or distribution terms. For web-video datasets distributed as metadata, captions, or URLs, the listed terms refer to the released dataset artifacts; underlying source media remain subject to the original platform and copyright-holder terms.
The Aidatatang corpus was retracted by its distributor after our training runs were completed; we retain it in Table~\ref{tab:licenses} for transparency about the data used to train the released checkpoints, and it is not redistributed as part of any AVF release artifact.

\begin{table}[h]
\centering
\caption{Licenses for all datasets and model checkpoints used in AV-Flamingo.}
\vspace{2pt}
\label{tab:licenses}
\resizebox{\columnwidth}{!}{%
\begin{tabular}{llll}
\toprule
\textbf{Asset} & \textbf{Type} & \textbf{License} & \textbf{Reference} \\
\midrule
\multicolumn{4}{l}{\textit{Model Checkpoints}} \\
OmniVinci        & Init. checkpoint  & NVIDIA OneWay Noncommercial License & \citet{ye2025omnivinci} \\
AF-Whisper       & Audio encoder     & NVIDIA OneWay Noncommercial License & \citet{goel2025audio} \\
Qwen2.5-7B       & Base LLM          & Apache 2.0                          & \citet{qwen2025qwen25technicalreport} \\
\midrule
\multicolumn{4}{l}{\textit{Training Datasets}} \\
YouTube-8M       & Video             & CC BY 4.0                           & \citet{abu2016youtube} \\
HD-VILA          & Video             & Microsoft Research License (non-commercial) & \citet{xue2022advancinghighresolutionvideolanguagerepresentation} \\
InternVid        & Video             & CC BY-NC-SA 4.0                     & \citet{wang2024internvidlargescalevideotextdataset} \\
VidChapters-7M   & Video             & MIT License; YouTube/API terms for source media & \citet{yang2023vidchapters7mvideochaptersscale} \\
HarmonySet       & Video             & Not specified; source-video terms apply & \citet{zhou2025harmonyset} \\
LSMDC            & Video             & Research use only                   & \citet{rohrbach2016moviedescription} \\
MMTrail          & Video             & R-UDA 1.0 (non-commercial research) & \citet{chi2024mmtrailmultimodaltrailervideo} \\
MovieCLIP        & Video             & Not specified; source-media terms apply & \citet{bose2022movieclipvisualscenerecognition} \\
MiraData         & Video             & GPL-3.0 metadata; source-video terms apply & \citet{ju2024miradatalargescalevideodataset} \\
AF3 training mix & Audio             & NVIDIA OneWay Noncommercial License & \citet{goel2025audio} \\
AudioSkills-XL   & Audio             & NVIDIA OneWay Noncommercial License & \citet{goel2025audio} \\
MF-Skills        & Audio             & NVIDIA OneWay Noncommercial License & \citet{ghosh2025music} \\
OmniVinci image-text mix & Image-text & Mixed public sources; see source licenses & \citet{ye2025omnivinci} \\
OmniVinci video-text mix & Video-text & Mixed public sources; see source licenses & \citet{ye2025omnivinci} \\
LibriSpeech      & Speech            & CC BY 4.0                           & \citet{panayotov2015librispeech} \\
CHiME            & Speech            & Research use/source corpus terms    & \citet{vincent2013second} \\
AliMeeting       & Speech            & Research/challenge use terms        & \citet{Yu2022M2MeT} \\
CoVoST           & Speech            & CC0 (Common Voice-based)            & \citet{wang2020covost} \\
GigaSpeech       & Speech            & Apache 2.0                          & \citet{Chen_2021} \\
EMILIA           & Speech            & CC BY-NC 4.0                        & \citet{he2024emilia} \\
Aishell-1        & Speech            & Apache 2.0                          & \citet{aishell_2017} \\
Aidatatang       & Speech            & CC BY-NC-ND 4.0 (retracted by distributor)         & \citet{aidatatang_200zh} \\
Multi-talker Switchboard & Speech    & LDC User Agreement                  & \citet{godfrey1993switchboard} \\
MuST-C           & Speech            & CC BY-NC-ND 4.0                     & \citet{cattoni2021must} \\
AV-Safety QA     & AV safety         & AVF release terms                   & Ours \\
Audio-Think-Time & Audio reasoning   & AVF release terms                   & Ours \\
Video-Think-Time & Video reasoning   & AVF release terms                   & Ours \\
AV-Think         & AV reasoning      & AVF release terms                   & Ours \\
\midrule
\multicolumn{4}{l}{\textit{Baseline Models (Evaluation Only)}} \\
Gemini 1.5/2.0/2.5 & AV Model       & Proprietary (API Terms of Service)  & \citet{team2024gemini,comanici2025gemini25pushingfrontier} \\
GPT-4o           & LLM               & Proprietary (API Terms of Service)  & \citet{hurst2024gpt} \\
Qwen2.5-Omni     & AV Model          & Apache 2.0                          & \citet{xu2025qwen25omnitechnicalreport} \\
Qwen3-Omni       & AV Model          & Apache 2.0                          & \citet{xu2025qwen3} \\
Audio Flamingo 3 & Audio Model       & NVIDIA OneWay Noncommercial License & \citet{goel2025audio} \\
Phi-4-mm         & AV Model          & MIT License                         & \citet{abouelenin2025phi} \\
NVILA            & VLM               & CC BY-NC-SA 4.0 for weights; Apache 2.0 for code & \citet{liu2025nvilaefficientfrontiervisual} \\
\bottomrule
\end{tabular}%
}
\end{table}

\section{Broader Impacts}
\label{app:broader_impacts}

AV-Flamingo advances open audio-visual intelligence, with positive societal applications in accessibility (e.g., automated audio description for the visually impaired), education (e.g., lecture and documentary understanding), and media analysis (e.g., automated captioning and content moderation). By releasing the model, training code, and AV-Skills dataset openly, we also lower the barrier for researchers without access to large-scale proprietary resources, supporting a more inclusive research ecosystem.

However, models capable of joint audio-visual understanding carry potential risks. They could be misused to automate surveillance, assist in the generation or verification of deepfakes, or enable multimodal disinformation at scale by jointly analyzing and synthesizing audio and visual content. To mitigate these risks, we release AV-Flamingo and AV-Skills for non-commercial research use only, under licenses that explicitly prohibit harmful applications. In addition, we include AV-Safety QA (Section~\ref{sec.avf_training_data}) during training to encourage safe refusal behavior on harmful audio-visual requests. We encourage the community to continue developing safeguards, evaluation protocols, and governance frameworks for open audio-visual systems.
\end{document}